\documentclass{aa}  

\usepackage{graphicx}
\usepackage{txfonts}
\usepackage[]{hyperref}%
\usepackage[usenames,dvipsnames]
{xcolor}
\usepackage{multirow}
\usepackage{booktabs}
\usepackage{placeins}

\newcommand{\vpeak}{V_{\rm peak}}

\newcommand{\vvir}{V_{\rm vir}}
\newcommand{\vmax}{V_{\rm max}}

\newcommand{\rth}{r_{200c}}

\newcommand{\sig}{\sigma_{8}}
\newcommand{\OmM}{\Omega_\mathrm{m}}

\newcommand{\ns}{{n_{\rm s}}}

\newcommand{\ihMpcC}{ h^{3}{\rm Mpc}^{-3}}
\newcommand{\ihMpc}{ h^{-1}{\rm Mpc} }
\newcommand{\ihGpc}{ h^{-1}{\rm Gpc} }
\newcommand{\hMpc}{ h^{-1}{\rm Mpc} }
\newcommand{\ihGpcC}{ h^{3}{\rm Gpc}^{-3}}
\newcommand{\hMsun}{ h^{-1}{\rm M_{ \odot}}}
\newcommand{\Msun}{{\rm M_{ \odot}}}

\usepackage{datatool, filecontents}
\DTLsetseparator{,}

\DTLloaddb[noheader, keys={thekey,thevalue}]{TNG_SHAMe}{Files/TNG_SHAMe.dat}
\newcommand{\TNG}[1]{\DTLfetch{TNG_SHAMe}{thekey}{#1}{thevalue}}

\DTLloaddb[noheader, keys={thekey,thevalue}]{TNG_ref}{Files/TNG_ref.dat}
\newcommand{\TNGref}[1]{\DTLfetch{TNG_ref}{thekey}{#1}{thevalue}}

\DTLloaddb[noheader, keys={thekey,thevalue}]{TNGSAM_SHAMe}{Files/TNGSAM_SHAMe.dat}
\newcommand{\TNGSAM}[1]{\DTLfetch{TNGSAM_SHAMe}{thekey}{#1}{thevalue}}

\DTLloaddb[noheader, keys={thekey,thevalue}]{TNGSAM_ref}{Files/TNGSAM_ref.dat}
\newcommand{\TNGSAMref}[1]{\DTLfetch{TNGSAM_ref}{thekey}{#1}{thevalue}}

\DTLloaddb[noheader, keys={thekey,thevalue}]{DESIl_SHAMe}{Files/DESIl_SHAMe.dat}
\newcommand{\DESIl}[1]{\DTLfetch{DESIl_SHAMe}{thekey}{#1}{thevalue}}

\DTLloaddb[noheader, keys={thekey,thevalue}]{DESIh_SHAMe}{Files/DESIh_SHAMe.dat}
\newcommand{\DESIh}[1]{\DTLfetch{DESIh_SHAMe}{thekey}{#1}{thevalue}}

\begin{document}
\begin{filecontents*}{Files/TNG_SHAMe.dat}
sat,30.7
satup,10.5
satdown,3.4
sat2,24.2
satup2,3.6
satdown2,0.3
consat,8.4
consatdown,6.6
consatup,1.2
consat2,12.7
consatdown2,4.3
consatup2,1.7
consat_outrmin,5.9
consatdown_outrmin,4.8
consatup_outrmin,1.0
consat2_outrmin,12.2
consatdown2_outrmin,4.3
consatup2_outrmin,1.3
gconsat,247.6
gconsatdown,174.6
gconsatup,22.7
gconsat2,323.2
gconsatdown2,85.6
gconsatup2,28.0
m200cen,12.1
m200cendown,0.2
m200cenup,0.1
m200cenerr,0.2
m200cen2,12.1
m200cendown2,0.2
m200cenup2,0.1
m200cenerr2,0.2
m200sat,13.0
m200satdown,0.2
m200satup,0.1
m200saterr,0.2
m200sat2,13.0
m200satdown2,0.2
m200satup2,0.1
m200saterr2,0.2
fmonosat,80.8
fmonosatdown,3.3
fmonosatup,1.0
fmonosat2,83.3
fmonosatdown2,0.9
fmonosatup2,1.3
fmultisat,19.2
fmultisatdown,1.0
fmultisatup,3.3
fmultisat2,16.7
fmultisatdown2,1.3
fmultisatup2,0.9
fconfcentral,3.2
fconfcentraldown,2.1
fconfcentralup,0.3
fconfcentral2,8.4
fconfcentraldown2,0.6
fconfcentralup2,11.1
fnoconfcentral,70.7
fnoconfcentraldown,9.7
fnoconfcentralup,1.6
fnoconfcentral2,69.9
fnoconfcentraldown2,10.7
fnoconfcentralup2,1.6
shuffled_fnoconfcentral,99.3
shuffled_fnoconfcentraldown,0.4
shuffled_fnoconfcentralup,0.1
shuffled_fnoconfcentral2,99.0
shuffled_fnoconfcentraldown2,0.4
shuffled_fnoconfcentralup2,0.2
mixed_fnoconfcentral,96.8
mixed_fnoconfcentraldown,0.3
mixed_fnoconfcentralup,2.1
mixed_fnoconfcentral2,91.6
mixed_fnoconfcentraldown2,11.1
mixed_fnoconfcentralup2,0.6
m200cenwithsat,12.5
m200cenwithsatdown,0.3
m200cenwithsatup,0.2
m200cenwithsat2,12.3
m200cenwithsatdown2,0.4
m200cenwithsatup2,0.1
m200cennosat,12.1
m200cennosatdown,0.2
m200cennosatup,0.1
m200cennosat2,12.1
m200cennosatdown2,0.2
m200cennosatup2,0.1
m200consat,12.5
m200consatdown,0.3
m200consatup,0.3
m200consat2,13.4
m200consatdown2,0.4
m200consatup2,0.1
m200noconsat,13.1
m200noconsatdown,0.3
m200noconsatup,0.1
m200noconsat2,13.0
m200noconsatdown2,0.2
m200noconsatup2,0.0
m200monoconsat,12.5
m200monoconsatdown,0.3
m200monoconsatup,0.1
m200monoconsat2,12.7
m200monoconsatdown2,0.0
m200monoconsatup2,0.3
m200mononoconsat,12.7
m200mononoconsatdown,0.1
m200mononoconsatup,0.0
m200mononoconsat2,12.7
m200mononoconsatdown2,0.0
m200mononoconsatup2,0.0
m200multiconsat,12.7
m200multiconsatdown,0.1
m200multiconsatup,0.5
m200multiconsat2,14.0
m200multiconsatdown2,0.5
m200multiconsatup2,0.0
m200multinoconsat,13.6
m200multinoconsatdown,0.5
m200multinoconsatup,0.1
m200multinoconsat2,13.6
m200multinoconsatdown2,0.5
m200multinoconsatup2,0.1
fmonosat_outrmin,84.1
fmonosatdown_outrmin,2.7
fmonosatup_outrmin,1.2
fmonosat2,85.5
fmonosatdown2,1.1
fmonosatup2,4.2
fmultisat_outrmin,15.9
fmultisatdown_outrmin,1.2
fmultisatup_outrmin,2.7
fmultisat2_outrmin,14.5
fmultisatdown2_outrmin,4.2
fmultisatup2_outrmin,1.1
fconfcentral_outrmin,3.2
fconfcentraldown_outrmin,2.1
fconfcentralup_outrmin,0.3
fconfcentral2_outrmin,8.4
fconfcentraldown2_outrmin,0.6
fconfcentralup2_outrmin,11.1
m200cenwithsat_outrmin,12.5
m200cenwithsatdown_outrmin,0.3
m200cenwithsatup_outrmin,0.2
m200cenwithsat2_outrmin,12.3
m200cenwithsatdown2_outrmin,0.4
m200cenwithsatup2_outrmin,0.1
m200cennosat_outrmin,12.1
m200cennosatdown_outrmin,0.2
m200cennosatup_outrmin,0.1
m200cennosat2_outrmin,12.1
m200cennosatdown2_outrmin,0.2
m200cennosatup2_outrmin,0.1
m200consat_outrmin,12.6
m200consatdown_outrmin,0.3
m200consatup_outrmin,0.4
m200consat2_outrmin,13.7
m200consatdown2_outrmin,0.5
m200consatup2_outrmin,0.1
m200noconsat_outrmin,13.3
m200noconsatdown_outrmin,0.4
m200noconsatup_outrmin,0.1
m200noconsat2_outrmin,13.2
m200noconsatdown2_outrmin,0.4
m200noconsatup2_outrmin,0.1
m200monoconsat_outrmin,12.6
m200monoconsatdown_outrmin,0.3
m200monoconsatup_outrmin,0.2
m200monoconsat2_outrmin,13.0
m200monoconsatdown2_outrmin,0.1
m200monoconsatup2_outrmin,0.2
m200mononoconsat_outrmin,12.8
m200mononoconsatdown_outrmin,0.1
m200mononoconsatup_outrmin,0.1
m200mononoconsat2_outrmin,12.9
m200mononoconsatdown2_outrmin,0.1
m200mononoconsatup2_outrmin,0.1
m200multiconsat_outrmin,12.7
m200multiconsatdown_outrmin,0.1
m200multiconsatup_outrmin,0.9
m200multiconsat2_outrmin,14.2
m200multiconsatdown2_outrmin,0.6
m200multiconsatup2_outrmin,0.0
m200multinoconsat_outrmin,13.9
m200multinoconsatdown_outrmin,0.7
m200multinoconsatup_outrmin,0.0
m200multinoconsat2_outrmin,14.0
m200multinoconsatdown2_outrmin,0.6
m200multinoconsatup2_outrmin,0.0
fhalos_onlysat,24.3
fhalos_onlysatdown,2.0
fhalos_onlysatup,12.8
fhalos_onlysat2,21.3
fhalos_onlysatup2,1.8
fhalos_onlysatdown2,4.6
fhalos_conformal,4.9
fhalos_conformalup,3.5
fhalos_conformaldown,0.5
fhalos_conformal2,9.9
fhalos_conformalup2,1.0
fhalos_conformaldown2,7.0
shuffled_fhalos_onlysat,24.4
shuffled_fhalos_onlysatdown,2.1
shuffled_fhalos_onlysatup,11.7
shuffled_fhalos_onlysatdown2,21.4
shuffled_fhalos_onlysatup2,1.7
shuffled_fhalos_onlysatdown2,4.1
shuffled_fhalos_conformal,3.1
shuffled_fhalos_conformalup,1.0
shuffled_fhalos_conformaldown,0.3
shuffled_fhalos_conformal2,3.9
shuffled_fhalos_conformalup2,0.3
shuffled_fhalos_conformaldown2,0.3
mixed_fhalos_onlysat,24.5
mixed_fhalos_onlysatdown,2.0
mixed_fhalos_onlysatup,11.7
mixed_fhalos_onlysatdown2,21.6
mixed_fhalos_onlysatup2,1.8
mixed_fhalos_onlysatdown2,4.5
mixed_fhalos_conformal,4.8
mixed_fhalos_conformalup,2.5
mixed_fhalos_conformaldown,0.5
mixed_fhalos_onlysat2,9.3
mixed_fhalos_onlysatup2,0.8
mixed_fhalos_onlysatdown2,7.6
fhalos_multisat,2.4
fhalos_multisatdown,0.2
fhalos_multisatup,2.0
fhalos_multisat2,1.9
fhalos_multisatdown2,0.2
fhalos_multisatup2,0.4
fsat_inm13m14,3.5
fsat_inm13m14down,1.4
fsat_inm13m14up,7.6
fsat_inm12p5m13,20.5
fsat_inm12p5m13down,7.5
fsat_inm12p5m13up,1.5
fsat_inm12m12p5,10.8
fsat_inm12m12p5down,1.2
fsat_inm12m12p5up,10.2
fsat_under12,4.2
fsat_under12down,0.9
fsat_under12up,12.7
fsat_inr200,57.0
fsat_inr200down,20.9
fsat_inr200up,2.2
\end{filecontents*}

\begin{filecontents*}{Files/TNG_ref.dat}
sat,36.9
satup,0.4
satdown,0.5
sat2,25.7
satup2,0.4
satdown2,0.5
consat,5.9
consatdown,0.5
consatup,0.1
consat2,11.8
consatdown2,0.6
consatup2,0.3
consat_outrmin,4.0
consatdown_outrmin,0.4
consatup_outrmin,0.4
consat2_outrmin,11.3
consatdown2_outrmin,0.5
consatup2_outrmin,0.8
gconsat,217.3
gconsatdown,19.9
gconsatup,4.4
gconsat2,303.2
gconsatdown2,12.3
gconsatup2,12.4
m200cen,11.9
m200cendown,0.0
m200cenup,0.0
m200cenerr,0.0
m200cen2,11.9
m200cendown2,0.0
m200cenup2,0.0
m200cenerr2,0.0
m200sat,12.9
m200satdown,0.0
m200satup,0.0
m200saterr,0.0
m200sat2,12.9
m200satdown2,0.0
m200satup2,0.0
m200saterr2,0.0
fmonosat,78.9
fmonosatdown,0.6
fmonosatup,1.0
fmonosat2,83.8
fmonosatdown2,1.0
fmonosatup2,0.5
fmultisat,21.1
fmultisatdown,1.0
fmultisatup,0.6
fmultisat2,16.2
fmultisatdown2,0.5
fmultisatup2,1.0
fconfcentral,3.3
fconfcentraldown,0.3
fconfcentralup,0.0
fconfcentral2,17.8
fconfcentraldown2,0.4
fconfcentralup2,0.3
fnoconfcentral,63.6
fnoconfcentraldown,0.3
fnoconfcentralup,0.6
fnoconfcentral2,62.2
fnoconfcentraldown2,0.3
fnoconfcentralup2,0.6
shuffled_fnoconfcentral,96.7
shuffled_fnoconfcentraldown,0.0
shuffled_fnoconfcentralup,0.3
shuffled_fnoconfcentral2,82.2
shuffled_fnoconfcentraldown2,0.3
shuffled_fnoconfcentralup2,0.4
mixed_fnoconfcentral,96.7
mixed_fnoconfcentraldown,0.0
mixed_fnoconfcentralup,0.3
mixed_fnoconfcentral2,82.2
mixed_fnoconfcentraldown2,0.3
mixed_fnoconfcentralup2,0.4
m200cenwithsat,12.2
m200cenwithsatdown,0.0
m200cenwithsatup,0.0
m200cenwithsat2,11.8
m200cenwithsatdown2,0.0
m200cenwithsatup2,0.0
m200cennosat,11.9
m200cennosatdown,0.0
m200cennosatup,0.0
m200cennosat2,11.9
m200cennosatdown2,0.0
m200cennosatup2,0.0
m200consat,12.2
m200consatdown,0.0
m200consatup,0.0
m200consat2,13.1
m200consatdown2,0.1
m200consatup2,0.1
m200noconsat,13.0
m200noconsatdown,0.0
m200noconsatup,0.0
m200noconsat2,12.9
m200noconsatdown2,0.0
m200noconsatup2,0.0
m200monoconsat,12.2
m200monoconsatdown,0.0
m200monoconsatup,0.0
m200monoconsat2,12.9
m200monoconsatdown2,0.1
m200monoconsatup2,0.0
m200mononoconsat,12.8
m200mononoconsatdown,0.0
m200mononoconsatup,0.0
m200mononoconsat2,12.8
m200mononoconsatdown2,0.0
m200mononoconsatup2,0.0
m200multiconsat,12.4
m200multiconsatdown,0.1
m200multiconsatup,0.0
m200multiconsat2,13.4
m200multiconsatdown2,0.2
m200multiconsatup2,0.1
m200multinoconsat,13.3
m200multinoconsatdown,0.1
m200multinoconsatup,0.0
m200multinoconsat2,13.2
m200multinoconsatdown2,0.0
m200multinoconsatup2,0.1
fmonosat_outrmin,79.8
fmonosatdown_outrmin,1.1
fmonosatup_outrmin,1.0
fmonosat2,84.8
fmonosatdown2,0.9
fmonosatup2,1.8
fmultisat_outrmin,20.2
fmultisatdown_outrmin,1.0
fmultisatup_outrmin,1.1
fmultisat2_outrmin,15.2
fmultisatdown2_outrmin,1.8
fmultisatup2_outrmin,0.9
fconfcentral_outrmin,3.3
fconfcentraldown_outrmin,0.3
fconfcentralup_outrmin,0.0
fconfcentral2_outrmin,17.8
fconfcentraldown2_outrmin,0.4
fconfcentralup2_outrmin,0.3
m200cenwithsat_outrmin,12.2
m200cenwithsatdown_outrmin,0.0
m200cenwithsatup_outrmin,0.0
m200cenwithsat2_outrmin,11.8
m200cenwithsatdown2_outrmin,0.0
m200cenwithsatup2_outrmin,0.0
m200cennosat_outrmin,11.9
m200cennosatdown_outrmin,0.0
m200cennosatup_outrmin,0.0
m200cennosat2_outrmin,11.9
m200cennosatdown2_outrmin,0.0
m200cennosatup2_outrmin,0.0
m200consat_outrmin,12.2
m200consatdown_outrmin,0.0
m200consatup_outrmin,0.0
m200consat2_outrmin,13.3
m200consatdown2_outrmin,0.1
m200consatup2_outrmin,0.1
m200noconsat_outrmin,13.0
m200noconsatdown_outrmin,0.0
m200noconsatup_outrmin,0.0
m200noconsat2_outrmin,13.0
m200noconsatdown2_outrmin,0.0
m200noconsatup2_outrmin,0.0
m200monoconsat_outrmin,12.2
m200monoconsatdown_outrmin,0.0
m200monoconsatup_outrmin,0.0
m200monoconsat2_outrmin,13.2
m200monoconsatdown2_outrmin,0.1
m200monoconsatup2_outrmin,0.1
m200mononoconsat_outrmin,12.9
m200mononoconsatdown_outrmin,0.0
m200mononoconsatup_outrmin,0.0
m200mononoconsat2_outrmin,12.9
m200mononoconsatdown2_outrmin,0.0
m200mononoconsatup2_outrmin,0.0
m200multiconsat_outrmin,12.4
m200multiconsatdown_outrmin,0.2
m200multiconsatup_outrmin,0.1
m200multiconsat2_outrmin,13.6
m200multiconsatdown2_outrmin,0.2
m200multiconsatup2_outrmin,0.1
m200multinoconsat_outrmin,13.4
m200multinoconsatdown_outrmin,0.1
m200multinoconsatup_outrmin,0.0
m200multinoconsat2_outrmin,13.4
m200multinoconsatdown2_outrmin,0.1
m200multinoconsatup2_outrmin,0.0
fhalos_onlysat,32.1
fhalos_onlysatdown,0.4
fhalos_onlysatup,0.3
fhalos_onlysat2,23.0
fhalos_onlysatup2,0.4
fhalos_onlysatdown2,0.1
fhalos_conformal,4.4
fhalos_conformalup,0.4
fhalos_conformaldown,0.0
fhalos_conformal2,16.6
fhalos_conformalup2,0.4
fhalos_conformaldown2,0.2
shuffled_fhalos_onlysat,33.1
shuffled_fhalos_onlysatdown,0.6
shuffled_fhalos_onlysatup,0.4
shuffled_fhalos_onlysatdown2,23.9
shuffled_fhalos_onlysatup2,0.5
shuffled_fhalos_onlysatdown2,0.2
shuffled_fhalos_conformal,3.2
shuffled_fhalos_conformalup,0.1
shuffled_fhalos_conformaldown,0.1
shuffled_fhalos_conformal2,15.6
shuffled_fhalos_conformalup2,0.4
shuffled_fhalos_conformaldown2,0.2
mixed_fhalos_onlysat,33.3
mixed_fhalos_onlysatdown,0.7
mixed_fhalos_onlysatup,0.2
mixed_fhalos_onlysatdown2,24.2
mixed_fhalos_onlysatup2,0.2
mixed_fhalos_onlysatdown2,0.3
mixed_fhalos_conformal,3.1
mixed_fhalos_conformalup,0.1
mixed_fhalos_conformaldown,0.1
mixed_fhalos_onlysat2,15.3
mixed_fhalos_onlysatup2,0.2
mixed_fhalos_onlysatdown2,0.2
fhalos_multisat,3.6
fhalos_multisatdown,0.1
fhalos_multisatup,0.2
fhalos_multisat2,2.0
fhalos_multisatdown2,0.1
fhalos_multisatup2,0.1
fsat_inm13m14,16.3
fsat_inm13m14down,0.4
fsat_inm13m14up,0.6
fsat_inm12p5m13,21.0
fsat_inm12p5m13down,0.1
fsat_inm12p5m13up,0.4
fsat_inm12m12p5,22.9
fsat_inm12m12p5down,1.0
fsat_inm12m12p5up,0.4
fsat_under12,7.1
fsat_under12down,0.5
fsat_under12up,0.1
fsat_inr200,31.7
fsat_inr200down,0.3
fsat_inr200up,0.6
\end{filecontents*}

\begin{filecontents*}{Files/TNGSAM_SHAMe.dat}
sat,11.2
satup,2.3
satdown,2.2
sat2,4.4
satup2,1.2
satdown2,1.1
consat,11.9
consatdown,2.9
consatup,2.2
consat2,15.0
consatdown2,2.6
consatup2,2.6
consat_outrmin,11.9
consatdown_outrmin,3.0
consatup_outrmin,2.3
consat2_outrmin,15.6
consatdown2_outrmin,3.8
consatup2_outrmin,3.4
gconsat,132.0
gconsatdown,18.3
gconsatup,14.0
gconsat2,60.1
gconsatdown2,7.9
gconsatup2,13.7
m200cen,12.0
m200cendown,0.1
m200cenup,0.0
m200cenerr,0.1
m200cen2,11.9
m200cendown2,0.1
m200cenup2,0.0
m200cenerr2,0.1
m200sat,12.6
m200satdown,0.2
m200satup,0.1
m200saterr,0.2
m200sat2,12.6
m200satdown2,0.2
m200satup2,0.2
m200saterr2,0.2
fmonosat,95.5
fmonosatdown,2.8
fmonosatup,1.5
fmonosat2,98.4
fmonosatdown2,1.3
fmonosatup2,0.7
fmultisat,4.5
fmultisatdown,1.5
fmultisatup,2.8
fmultisat2,1.6
fmultisatdown2,0.7
fmultisatup2,1.3
fconfcentral,1.5
fconfcentraldown,0.2
fconfcentralup,0.2
fconfcentral2,8.9
fconfcentraldown2,1.9
fconfcentralup2,1.0
fnoconfcentral,87.7
fnoconfcentraldown,2.3
fnoconfcentralup,2.2
fnoconfcentral2,87.5
fnoconfcentraldown2,2.3
fnoconfcentralup2,2.3
shuffled_fnoconfcentral,100.0
shuffled_fnoconfcentraldown,0.1
shuffled_fnoconfcentralup,0.0
shuffled_fnoconfcentral2,99.9
shuffled_fnoconfcentraldown2,0.1
shuffled_fnoconfcentralup2,0.0
mixed_fnoconfcentral,98.5
mixed_fnoconfcentraldown,0.2
mixed_fnoconfcentralup,0.2
mixed_fnoconfcentral2,91.1
mixed_fnoconfcentraldown2,1.0
mixed_fnoconfcentralup2,1.9
m200cenwithsat,12.2
m200cenwithsatdown,0.1
m200cenwithsatup,0.1
m200cenwithsat2,11.9
m200cenwithsatdown2,0.1
m200cenwithsatup2,0.0
m200cennosat,11.9
m200cennosatdown,0.1
m200cennosatup,0.0
m200cennosat2,11.9
m200cennosatdown2,0.1
m200cennosatup2,0.0
m200consat,12.2
m200consatdown,0.1
m200consatup,0.1
m200consat2,12.4
m200consatdown2,0.1
m200consatup2,0.2
m200noconsat,12.7
m200noconsatdown,0.2
m200noconsatup,0.1
m200noconsat2,12.6
m200noconsatdown2,0.2
m200noconsatup2,0.2
m200monoconsat,12.2
m200monoconsatdown,0.1
m200monoconsatup,0.1
m200monoconsat2,12.4
m200monoconsatdown2,0.1
m200monoconsatup2,0.1
m200mononoconsat,12.7
m200mononoconsatdown,0.2
m200mononoconsatup,0.1
m200mononoconsat2,12.6
m200mononoconsatdown2,0.2
m200mononoconsatup2,0.2
m200multiconsat,12.3
m200multiconsatdown,0.1
m200multiconsatup,0.1
m200multiconsat2,12.4
m200multiconsatdown2,0.1
m200multiconsatup2,0.3
m200multinoconsat,13.0
m200multinoconsatdown,0.3
m200multinoconsatup,0.2
m200multinoconsat2,12.6
m200multinoconsatdown2,0.2
m200multinoconsatup2,0.6
fmonosat_outrmin,95.6
fmonosatdown_outrmin,2.5
fmonosatup_outrmin,1.4
fmonosat2,98.7
fmonosatdown2,1.0
fmonosatup2,0.7
fmultisat_outrmin,4.4
fmultisatdown_outrmin,1.4
fmultisatup_outrmin,2.5
fmultisat2_outrmin,1.3
fmultisatdown2_outrmin,0.7
fmultisatup2_outrmin,1.0
fconfcentral_outrmin,1.5
fconfcentraldown_outrmin,0.2
fconfcentralup_outrmin,0.2
fconfcentral2_outrmin,8.9
fconfcentraldown2_outrmin,1.9
fconfcentralup2_outrmin,1.0
m200cenwithsat_outrmin,12.2
m200cenwithsatdown_outrmin,0.1
m200cenwithsatup_outrmin,0.1
m200cenwithsat2_outrmin,11.9
m200cenwithsatdown2_outrmin,0.1
m200cenwithsatup2_outrmin,0.0
m200cennosat_outrmin,11.9
m200cennosatdown_outrmin,0.1
m200cennosatup_outrmin,0.0
m200cennosat2_outrmin,11.9
m200cennosatdown2_outrmin,0.1
m200cennosatup2_outrmin,0.0
m200consat_outrmin,12.2
m200consatdown_outrmin,0.1
m200consatup_outrmin,0.1
m200consat2_outrmin,12.5
m200consatdown2_outrmin,0.1
m200consatup2_outrmin,0.2
m200noconsat_outrmin,12.7
m200noconsatdown_outrmin,0.2
m200noconsatup_outrmin,0.1
m200noconsat2_outrmin,12.7
m200noconsatdown2_outrmin,0.3
m200noconsatup2_outrmin,0.2
m200monoconsat_outrmin,12.2
m200monoconsatdown_outrmin,0.1
m200monoconsatup_outrmin,0.1
m200monoconsat2_outrmin,12.5
m200monoconsatdown2_outrmin,0.2
m200monoconsatup2_outrmin,0.2
m200mononoconsat_outrmin,12.7
m200mononoconsatdown_outrmin,0.2
m200mononoconsatup_outrmin,0.1
m200mononoconsat2_outrmin,12.7
m200mononoconsatdown2_outrmin,0.3
m200mononoconsatup2_outrmin,0.2
m200multiconsat_outrmin,12.3
m200multiconsatdown_outrmin,0.1
m200multiconsatup_outrmin,0.1
m200multiconsat2_outrmin,12.5
m200multiconsatdown2_outrmin,0.2
m200multiconsatup2_outrmin,0.6
m200multinoconsat_outrmin,13.0
m200multinoconsatdown_outrmin,0.3
m200multinoconsatup_outrmin,0.2
m200multinoconsat2_outrmin,12.7
m200multinoconsatdown2_outrmin,0.2
m200multinoconsatup2_outrmin,0.7
fhalos_onlysat,9.6
fhalos_onlysatdown,1.7
fhalos_onlysatup,2.2
fhalos_onlysat2,3.6
fhalos_onlysatup2,0.7
fhalos_onlysatdown2,1.6
fhalos_conformal,2.6
fhalos_conformalup,0.4
fhalos_conformaldown,0.3
fhalos_conformal2,9.2
fhalos_conformalup2,1.8
fhalos_conformaldown2,0.9
shuffled_fhalos_onlysat,10.0
shuffled_fhalos_onlysatdown,1.8
shuffled_fhalos_onlysatup,2.0
shuffled_fhalos_onlysatdown2,3.8
shuffled_fhalos_onlysatup2,0.7
shuffled_fhalos_onlysatdown2,1.5
shuffled_fhalos_conformal,1.0
shuffled_fhalos_conformalup,0.2
shuffled_fhalos_conformaldown,0.2
shuffled_fhalos_conformal2,0.5
shuffled_fhalos_conformalup2,0.1
shuffled_fhalos_conformaldown2,0.3
mixed_fhalos_onlysat,10.1
mixed_fhalos_onlysatdown,1.9
mixed_fhalos_onlysatup,2.1
mixed_fhalos_onlysatdown2,3.8
mixed_fhalos_onlysatup2,0.7
mixed_fhalos_onlysatdown2,1.6
mixed_fhalos_conformal,2.1
mixed_fhalos_conformalup,0.3
mixed_fhalos_conformaldown,0.2
mixed_fhalos_onlysat2,9.0
mixed_fhalos_onlysatup2,1.8
mixed_fhalos_onlysatdown2,0.9
fhalos_multisat,0.3
fhalos_multisatdown,0.1
fhalos_multisatup,0.2
fhalos_multisat2,0.0
fhalos_multisatdown2,0.0
fhalos_multisatup2,0.0
fsat_inm13m14,7.3
fsat_inm13m14down,3.5
fsat_inm13m14up,4.0
fsat_inm12p5m13,17.3
fsat_inm12p5m13down,3.5
fsat_inm12p5m13up,3.0
fsat_inm12m12p5,35.5
fsat_inm12m12p5down,7.2
fsat_inm12m12p5up,4.0
fsat_under12,22.1
fsat_under12down,3.8
fsat_under12up,8.4
fsat_inr200,14.8
fsat_inr200down,2.6
fsat_inr200up,4.3
\end{filecontents*}

\begin{filecontents*}{Files/TNGSAM_ref.dat}
sat,14.5
satup,0.2
satdown,0.2
sat2,3.8
satup2,0.1
satdown2,0.1
consat,7.2
consatdown,0.7
consatup,0.5
consat2,13.3
consatdown2,1.3
consatup2,1.8
consat_outrmin,6.9
consatdown_outrmin,0.8
consatup_outrmin,0.7
consat2_outrmin,13.2
consatdown2_outrmin,1.2
consatup2_outrmin,2.2
gconsat,103.7
gconsatdown,9.6
gconsatup,7.5
gconsat2,51.6
gconsatdown2,5.7
gconsatup2,6.2
m200cen,11.9
m200cendown,0.0
m200cenup,0.0
m200cenerr,0.0
m200cen2,11.9
m200cendown2,0.0
m200cenup2,0.0
m200cenerr2,0.0
m200sat,12.9
m200satdown,0.0
m200satup,0.0
m200saterr,0.0
m200sat2,12.9
m200satdown2,0.0
m200satup2,0.0
m200saterr2,0.0
fmonosat,91.6
fmonosatdown,0.6
fmonosatup,0.9
fmonosat2,97.6
fmonosatdown2,1.0
fmonosatup2,0.7
fmultisat,8.4
fmultisatdown,0.9
fmultisatup,0.6
fmultisat2,2.4
fmultisatdown2,0.7
fmultisatup2,1.0
fconfcentral,1.2
fconfcentraldown,0.1
fconfcentralup,0.1
fconfcentral2,12.1
fconfcentraldown2,0.1
fconfcentralup2,0.3
fnoconfcentral,85.0
fnoconfcentraldown,0.2
fnoconfcentralup,0.2
fnoconfcentral2,84.5
fnoconfcentraldown2,0.3
fnoconfcentralup2,0.2
shuffled_fnoconfcentral,98.5
shuffled_fnoconfcentraldown,0.3
shuffled_fnoconfcentralup,0.2
shuffled_fnoconfcentral2,98.5
shuffled_fnoconfcentraldown2,0.3
shuffled_fnoconfcentralup2,0.2
mixed_fnoconfcentral,98.8
mixed_fnoconfcentraldown,0.1
mixed_fnoconfcentralup,0.1
mixed_fnoconfcentral2,87.9
mixed_fnoconfcentraldown2,0.3
mixed_fnoconfcentralup2,0.1
m200cenwithsat,12.3
m200cenwithsatdown,0.1
m200cenwithsatup,0.1
m200cenwithsat2,11.7
m200cenwithsatdown2,0.0
m200cenwithsatup2,0.0
m200cennosat,11.9
m200cennosatdown,0.0
m200cennosatup,0.0
m200cennosat2,11.9
m200cennosatdown2,0.0
m200cennosatup2,0.0
m200consat,12.3
m200consatdown,0.1
m200consatup,0.1
m200consat2,13.0
m200consatdown2,0.1
m200consatup2,0.0
m200noconsat,12.9
m200noconsatdown,0.0
m200noconsatup,0.0
m200noconsat2,12.9
m200noconsatdown2,0.1
m200noconsatup2,0.1
m200monoconsat,12.3
m200monoconsatdown,0.1
m200monoconsatup,0.1
m200monoconsat2,12.9
m200monoconsatdown2,0.1
m200monoconsatup2,0.1
m200mononoconsat,12.8
m200mononoconsatdown,0.0
m200mononoconsatup,0.0
m200mononoconsat2,12.8
m200mononoconsatdown2,0.0
m200mononoconsatup2,0.1
m200multiconsat,12.4
m200multiconsatdown,0.0
m200multiconsatup,0.1
m200multiconsat2,13.1
m200multiconsatdown2,0.0
m200multiconsatup2,0.3
m200multinoconsat,13.3
m200multinoconsatdown,0.1
m200multinoconsatup,0.1
m200multinoconsat2,13.4
m200multinoconsatdown2,0.5
m200multinoconsatup2,0.2
fmonosat_outrmin,91.4
fmonosatdown_outrmin,0.6
fmonosatup_outrmin,1.0
fmonosat2,97.5
fmonosatdown2,0.7
fmonosatup2,0.9
fmultisat_outrmin,8.6
fmultisatdown_outrmin,1.0
fmultisatup_outrmin,0.6
fmultisat2_outrmin,2.5
fmultisatdown2_outrmin,0.9
fmultisatup2_outrmin,0.7
fconfcentral_outrmin,1.2
fconfcentraldown_outrmin,0.1
fconfcentralup_outrmin,0.1
fconfcentral2_outrmin,12.1
fconfcentraldown2_outrmin,0.1
fconfcentralup2_outrmin,0.3
m200cenwithsat_outrmin,12.3
m200cenwithsatdown_outrmin,0.1
m200cenwithsatup_outrmin,0.1
m200cenwithsat2_outrmin,11.7
m200cenwithsatdown2_outrmin,0.0
m200cenwithsatup2_outrmin,0.0
m200cennosat_outrmin,11.9
m200cennosatdown_outrmin,0.0
m200cennosatup_outrmin,0.0
m200cennosat2_outrmin,11.9
m200cennosatdown2_outrmin,0.0
m200cennosatup2_outrmin,0.0
m200consat_outrmin,12.3
m200consatdown_outrmin,0.1
m200consatup_outrmin,0.1
m200consat2_outrmin,13.0
m200consatdown2_outrmin,0.1
m200consatup2_outrmin,0.0
m200noconsat_outrmin,12.9
m200noconsatdown_outrmin,0.0
m200noconsatup_outrmin,0.0
m200noconsat2_outrmin,12.9
m200noconsatdown2_outrmin,0.1
m200noconsatup2_outrmin,0.1
m200monoconsat_outrmin,12.3
m200monoconsatdown_outrmin,0.1
m200monoconsatup_outrmin,0.1
m200monoconsat2_outrmin,13.0
m200monoconsatdown2_outrmin,0.1
m200monoconsatup2_outrmin,0.1
m200mononoconsat_outrmin,12.8
m200mononoconsatdown_outrmin,0.0
m200mononoconsatup_outrmin,0.0
m200mononoconsat2_outrmin,12.9
m200mononoconsatdown2_outrmin,0.0
m200mononoconsatup2_outrmin,0.1
m200multiconsat_outrmin,12.4
m200multiconsatdown_outrmin,0.0
m200multiconsatup_outrmin,0.1
m200multiconsat2_outrmin,13.1
m200multiconsatdown2_outrmin,0.0
m200multiconsatup2_outrmin,0.3
m200multinoconsat_outrmin,13.3
m200multinoconsatdown_outrmin,0.1
m200multinoconsatup_outrmin,0.1
m200multinoconsat2_outrmin,13.5
m200multinoconsatdown2_outrmin,0.6
m200multinoconsatup2_outrmin,0.1
fhalos_onlysat,12.9
fhalos_onlysatdown,0.2
fhalos_onlysatup,0.2
fhalos_onlysat2,3.8
fhalos_onlysatup2,0.1
fhalos_onlysatdown2,0.2
fhalos_conformal,2.1
fhalos_conformalup,0.2
fhalos_conformaldown,0.2
fhalos_conformal2,12.2
fhalos_conformalup2,0.1
fhalos_conformaldown2,0.2
shuffled_fhalos_onlysat,13.0
shuffled_fhalos_onlysatdown,0.1
shuffled_fhalos_onlysatup,0.2
shuffled_fhalos_onlysatdown2,3.9
shuffled_fhalos_onlysatup2,0.2
shuffled_fhalos_onlysatdown2,0.1
shuffled_fhalos_conformal,2.3
shuffled_fhalos_conformalup,0.3
shuffled_fhalos_conformaldown,0.3
shuffled_fhalos_conformal2,2.0
shuffled_fhalos_conformalup2,0.2
shuffled_fhalos_conformaldown2,0.2
mixed_fhalos_onlysat,13.0
mixed_fhalos_onlysatdown,0.1
mixed_fhalos_onlysatup,0.3
mixed_fhalos_onlysatdown2,3.9
mixed_fhalos_onlysatup2,0.1
mixed_fhalos_onlysatdown2,0.1
mixed_fhalos_conformal,1.9
mixed_fhalos_conformalup,0.1
mixed_fhalos_conformaldown,0.1
mixed_fhalos_onlysat2,12.1
mixed_fhalos_onlysatup2,0.1
mixed_fhalos_onlysatdown2,0.2
fhalos_multisat,0.6
fhalos_multisatdown,0.1
fhalos_multisatup,0.0
fhalos_multisat2,0.0
fhalos_multisatdown2,0.0
fhalos_multisatup2,0.0
fsat_inm13m14,16.6
fsat_inm13m14down,0.6
fsat_inm13m14up,0.3
fsat_inm12p5m13,23.8
fsat_inm12p5m13down,0.2
fsat_inm12p5m13up,0.3
fsat_inm12m12p5,36.6
fsat_inm12m12p5down,0.3
fsat_inm12m12p5up,1.6
fsat_under12,19.4
fsat_under12down,1.0
fsat_under12up,0.6
fsat_inr200,2.4
fsat_inr200down,0.4
fsat_inr200up,0.2
\end{filecontents*}

\begin{filecontents*}{Files/DESIl_SHAMe.dat}
sat,34.0
satup,1.5
satdown,1.9
sat2,25.5
satup2,3.5
satdown2,3.8
consat,3.9
consatdown,1.1
consatup,1.3
consat2,5.4
consatdown2,2.0
consatup2,2.4
consat_outrmin,5.0
consatdown_outrmin,1.3
consatup_outrmin,0.8
consat2_outrmin,5.7
consatdown2_outrmin,1.9
consatup2_outrmin,2.3
gconsat,136.8
gconsatdown,40.2
gconsatup,39.0
gconsat2,134.0
gconsatdown2,37.1
gconsatup2,28.9
m200cen,11.7
m200cendown,0.1
m200cenup,0.1
m200cenerr,0.1
m200cen2,11.7
m200cendown2,0.1
m200cenup2,0.1
m200cenerr2,0.1
m200sat,12.5
m200satdown,0.2
m200satup,0.2
m200saterr,0.2
m200sat2,12.5
m200satdown2,0.2
m200satup2,0.2
m200saterr2,0.2
fmonosat,92.3
fmonosatdown,2.1
fmonosatup,1.5
fmonosat2,94.4
fmonosatdown2,1.9
fmonosatup2,0.9
fmultisat,7.7
fmultisatdown,1.5
fmultisatup,2.1
fmultisat2,5.6
fmultisatdown2,0.9
fmultisatup2,1.9
fconfcentral,2.0
fconfcentraldown,0.6
fconfcentralup,0.5
fconfcentral2,12.5
fconfcentraldown2,2.7
fconfcentralup2,4.8
fnoconfcentral,65.3
fnoconfcentraldown,1.1
fnoconfcentralup,1.8
fnoconfcentral2,64.9
fnoconfcentraldown2,1.1
fnoconfcentralup2,1.8
shuffled_fnoconfcentral,99.9
shuffled_fnoconfcentraldown,0.5
shuffled_fnoconfcentralup,0.1
shuffled_fnoconfcentral2,99.8
shuffled_fnoconfcentraldown2,0.4
shuffled_fnoconfcentralup2,0.2
mixed_fnoconfcentral,98.0
mixed_fnoconfcentraldown,0.5
mixed_fnoconfcentralup,0.6
mixed_fnoconfcentral2,87.5
mixed_fnoconfcentraldown2,4.8
mixed_fnoconfcentralup2,2.7
m200cenwithsat,12.0
m200cenwithsatdown,0.2
m200cenwithsatup,0.2
m200cenwithsat2,11.6
m200cenwithsatdown2,0.2
m200cenwithsatup2,0.2
m200cennosat,11.7
m200cennosatdown,0.1
m200cennosatup,0.1
m200cennosat2,11.7
m200cennosatdown2,0.1
m200cennosatup2,0.1
m200consat,12.0
m200consatdown,0.2
m200consatup,0.2
m200consat2,12.4
m200consatdown2,0.2
m200consatup2,0.7
m200noconsat,12.4
m200noconsatdown,0.1
m200noconsatup,0.3
m200noconsat2,12.4
m200noconsatdown2,0.1
m200noconsatup2,0.2
m200monoconsat,12.0
m200monoconsatdown,0.2
m200monoconsatup,0.2
m200monoconsat2,12.4
m200monoconsatdown2,0.2
m200monoconsatup2,0.4
m200mononoconsat,12.4
m200mononoconsatdown,0.1
m200mononoconsatup,0.1
m200mononoconsat2,12.4
m200mononoconsatdown2,0.1
m200mononoconsatup2,0.1
m200multiconsat,12.2
m200multiconsatdown,0.3
m200multiconsatup,0.2
m200multiconsat2,12.5
m200multiconsatdown2,0.1
m200multiconsatup2,1.4
m200multinoconsat,12.7
m200multinoconsatdown,0.2
m200multinoconsatup,0.7
m200multinoconsat2,12.6
m200multinoconsatdown2,0.2
m200multinoconsatup2,0.7
fmonosat_outrmin,94.9
fmonosatdown_outrmin,2.2
fmonosatup_outrmin,1.4
fmonosat2,97.0
fmonosatdown2,2.1
fmonosatup2,0.7
fmultisat_outrmin,5.1
fmultisatdown_outrmin,1.4
fmultisatup_outrmin,2.2
fmultisat2_outrmin,3.0
fmultisatdown2_outrmin,0.7
fmultisatup2_outrmin,2.1
fconfcentral_outrmin,2.0
fconfcentraldown_outrmin,0.6
fconfcentralup_outrmin,0.5
fconfcentral2_outrmin,12.5
fconfcentraldown2_outrmin,2.7
fconfcentralup2_outrmin,4.8
m200cenwithsat_outrmin,12.0
m200cenwithsatdown_outrmin,0.2
m200cenwithsatup_outrmin,0.2
m200cenwithsat2_outrmin,11.6
m200cenwithsatdown2_outrmin,0.2
m200cenwithsatup2_outrmin,0.2
m200cennosat_outrmin,11.7
m200cennosatdown_outrmin,0.1
m200cennosatup_outrmin,0.1
m200cennosat2_outrmin,11.7
m200cennosatdown2_outrmin,0.1
m200cennosatup2_outrmin,0.1
m200consat_outrmin,12.0
m200consatdown_outrmin,0.2
m200consatup_outrmin,0.2
m200consat2_outrmin,12.5
m200consatdown2_outrmin,0.3
m200consatup2_outrmin,0.9
m200noconsat_outrmin,12.6
m200noconsatdown_outrmin,0.2
m200noconsatup_outrmin,0.3
m200noconsat2_outrmin,12.6
m200noconsatdown2_outrmin,0.2
m200noconsatup2_outrmin,0.3
m200monoconsat_outrmin,12.0
m200monoconsatdown_outrmin,0.2
m200monoconsatup_outrmin,0.2
m200monoconsat2_outrmin,12.5
m200monoconsatdown2_outrmin,0.3
m200monoconsatup2_outrmin,0.5
m200mononoconsat_outrmin,12.5
m200mononoconsatdown_outrmin,0.1
m200mononoconsatup_outrmin,0.2
m200mononoconsat2_outrmin,12.5
m200mononoconsatdown2_outrmin,0.1
m200mononoconsatup2_outrmin,0.2
m200multiconsat_outrmin,12.1
m200multiconsatdown_outrmin,0.3
m200multiconsatup_outrmin,0.3
m200multiconsat2_outrmin,12.8
m200multiconsatdown2_outrmin,0.3
m200multiconsatup2_outrmin,1.5
m200multinoconsat_outrmin,12.8
m200multinoconsatdown_outrmin,0.3
m200multinoconsatup_outrmin,0.9
m200multinoconsat2_outrmin,12.8
m200multinoconsatdown2_outrmin,0.3
m200multinoconsatup2_outrmin,1.0
fhalos_onlysat,32.1
fhalos_onlysatdown,2.5
fhalos_onlysatup,1.3
fhalos_onlysat2,24.6
fhalos_onlysatup2,3.8
fhalos_onlysatdown2,3.0
fhalos_conformal,2.7
fhalos_conformalup,0.8
fhalos_conformaldown,0.8
fhalos_conformal2,10.7
fhalos_conformalup2,2.7
fhalos_conformaldown2,5.1
shuffled_fhalos_onlysat,32.5
shuffled_fhalos_onlysatdown,2.5
shuffled_fhalos_onlysatup,1.5
shuffled_fhalos_onlysatdown2,24.8
shuffled_fhalos_onlysatup2,4.4
shuffled_fhalos_onlysatdown2,3.0
shuffled_fhalos_conformal,1.1
shuffled_fhalos_conformalup,0.5
shuffled_fhalos_conformaldown,0.6
shuffled_fhalos_conformal2,1.4
shuffled_fhalos_conformalup2,0.5
shuffled_fhalos_conformaldown2,0.7
mixed_fhalos_onlysat,32.5
mixed_fhalos_onlysatdown,2.5
mixed_fhalos_onlysatup,1.4
mixed_fhalos_onlysatdown2,25.0
mixed_fhalos_onlysatup2,4.1
mixed_fhalos_onlysatdown2,2.9
mixed_fhalos_conformal,2.2
mixed_fhalos_conformalup,0.6
mixed_fhalos_conformaldown,0.6
mixed_fhalos_onlysat2,10.3
mixed_fhalos_onlysatup2,2.5
mixed_fhalos_onlysatdown2,5.2
fhalos_multisat,1.2
fhalos_multisatdown,0.2
fhalos_multisatup,0.4
fhalos_multisat2,0.8
fhalos_multisatdown2,0.2
fhalos_multisatup2,0.2
fsat_inm13m14,1.0
fsat_inm13m14down,0.6
fsat_inm13m14up,4.1
fsat_inm12p5m13,10.7
fsat_inm12p5m13down,3.5
fsat_inm12p5m13up,2.6
fsat_inm12m12p5,22.0
fsat_inm12m12p5down,3.3
fsat_inm12m12p5up,3.6
fsat_under12,15.2
fsat_under12down,2.3
fsat_under12up,7.8
fsat_inr200,48.8
fsat_inr200down,12.5
fsat_inr200up,7.7
\end{filecontents*}

\begin{filecontents*}{Files/DESIh_SHAMe.dat}
sat,27.6
satup,3.2
satdown,3.6
sat2,19.5
satup2,3.4
satdown2,2.2
consat,3.9
consatdown,1.6
consatup,0.9
consat2,5.1
consatdown2,1.9
consatup2,1.2
consat_outrmin,4.2
consatdown_outrmin,1.5
consatup_outrmin,0.8
consat2_outrmin,5.6
consatdown2_outrmin,2.0
consatup2_outrmin,1.4
gconsat,106.0
gconsatdown,48.7
gconsatup,22.4
gconsat2,95.2
gconsatdown2,20.9
gconsatup2,22.9
m200cen,11.9
m200cendown,0.1
m200cenup,0.1
m200cenerr,0.1
m200cen2,11.9
m200cendown2,0.1
m200cenup2,0.1
m200cenerr2,0.1
m200sat,12.5
m200satdown,0.2
m200satup,0.1
m200saterr,0.2
m200sat2,12.6
m200satdown2,0.2
m200satup2,0.1
m200saterr2,0.2
fmonosat,95.3
fmonosatdown,3.2
fmonosatup,0.6
fmonosat2,96.8
fmonosatdown2,3.9
fmonosatup2,1.0
fmultisat,4.7
fmultisatdown,0.6
fmultisatup,3.2
fmultisat2,3.2
fmultisatdown2,1.0
fmultisatup2,3.9
fconfcentral,1.4
fconfcentraldown,0.7
fconfcentralup,0.3
fconfcentral2,12.3
fconfcentraldown2,5.6
fconfcentralup2,3.3
fnoconfcentral,71.0
fnoconfcentraldown,2.0
fnoconfcentralup,4.6
fnoconfcentral2,70.8
fnoconfcentraldown2,2.3
fnoconfcentralup2,4.5
shuffled_fnoconfcentral,99.9
shuffled_fnoconfcentraldown,0.2
shuffled_fnoconfcentralup,0.1
shuffled_fnoconfcentral2,99.8
shuffled_fnoconfcentraldown2,0.2
shuffled_fnoconfcentralup2,0.1
mixed_fnoconfcentral,98.6
mixed_fnoconfcentraldown,0.3
mixed_fnoconfcentralup,0.7
mixed_fnoconfcentral2,87.7
mixed_fnoconfcentraldown2,3.3
mixed_fnoconfcentralup2,5.6
m200cenwithsat,12.3
m200cenwithsatdown,0.1
m200cenwithsatup,0.1
m200cenwithsat2,11.8
m200cenwithsatdown2,0.1
m200cenwithsatup2,0.1
m200cennosat,11.9
m200cennosatdown,0.1
m200cennosatup,0.1
m200cennosat2,11.9
m200cennosatdown2,0.1
m200cennosatup2,0.1
m200consat,12.3
m200consatdown,0.1
m200consatup,0.1
m200consat2,12.4
m200consatdown2,0.1
m200consatup2,0.6
m200noconsat,12.5
m200noconsatdown,0.1
m200noconsatup,0.2
m200noconsat2,12.4
m200noconsatdown2,0.1
m200noconsatup2,0.2
m200monoconsat,12.3
m200monoconsatdown,0.1
m200monoconsatup,0.1
m200monoconsat2,12.4
m200monoconsatdown2,0.1
m200monoconsatup2,0.2
m200mononoconsat,12.4
m200mononoconsatdown,0.1
m200mononoconsatup,0.1
m200mononoconsat2,12.4
m200mononoconsatdown2,0.1
m200mononoconsatup2,0.1
m200multiconsat,12.5
m200multiconsatdown,0.2
m200multiconsatup,0.1
m200multiconsat2,12.6
m200multiconsatdown2,0.2
m200multiconsatup2,1.4
m200multinoconsat,12.6
m200multinoconsatdown,0.2
m200multinoconsatup,0.8
m200multinoconsat2,12.6
m200multinoconsatdown2,0.1
m200multinoconsatup2,0.8
fmonosat_outrmin,96.3
fmonosatdown_outrmin,4.0
fmonosatup_outrmin,0.8
fmonosat2,98.1
fmonosatdown2,2.7
fmonosatup2,0.8
fmultisat_outrmin,3.7
fmultisatdown_outrmin,0.8
fmultisatup_outrmin,4.0
fmultisat2_outrmin,1.9
fmultisatdown2_outrmin,0.8
fmultisatup2_outrmin,2.7
fconfcentral_outrmin,1.4
fconfcentraldown_outrmin,0.7
fconfcentralup_outrmin,0.3
fconfcentral2_outrmin,12.3
fconfcentraldown2_outrmin,5.6
fconfcentralup2_outrmin,3.3
m200cenwithsat_outrmin,12.3
m200cenwithsatdown_outrmin,0.1
m200cenwithsatup_outrmin,0.1
m200cenwithsat2_outrmin,11.8
m200cenwithsatdown2_outrmin,0.1
m200cenwithsatup2_outrmin,0.1
m200cennosat_outrmin,11.9
m200cennosatdown_outrmin,0.1
m200cennosatup_outrmin,0.1
m200cennosat2_outrmin,11.9
m200cennosatdown2_outrmin,0.1
m200cennosatup2_outrmin,0.1
m200consat_outrmin,12.3
m200consatdown_outrmin,0.2
m200consatup_outrmin,0.1
m200consat2_outrmin,12.5
m200consatdown2_outrmin,0.1
m200consatup2_outrmin,0.7
m200noconsat_outrmin,12.5
m200noconsatdown_outrmin,0.1
m200noconsatup_outrmin,0.3
m200noconsat2_outrmin,12.5
m200noconsatdown2_outrmin,0.1
m200noconsatup2_outrmin,0.3
m200monoconsat_outrmin,12.3
m200monoconsatdown_outrmin,0.2
m200monoconsatup_outrmin,0.2
m200monoconsat2_outrmin,12.5
m200monoconsatdown2_outrmin,0.1
m200monoconsatup2_outrmin,0.4
m200mononoconsat_outrmin,12.5
m200mononoconsatdown_outrmin,0.1
m200mononoconsatup_outrmin,0.1
m200mononoconsat2_outrmin,12.5
m200mononoconsatdown2_outrmin,0.1
m200mononoconsatup2_outrmin,0.2
m200multiconsat_outrmin,12.5
m200multiconsatdown_outrmin,0.2
m200multiconsatup_outrmin,0.2
m200multiconsat2_outrmin,12.7
m200multiconsatdown2_outrmin,0.1
m200multiconsatup2_outrmin,1.5
m200multinoconsat_outrmin,12.7
m200multinoconsatdown_outrmin,0.2
m200multinoconsatup_outrmin,1.0
m200multinoconsat2_outrmin,12.8
m200multinoconsatdown2_outrmin,0.2
m200multinoconsatup2_outrmin,1.0
fhalos_onlysat,27.0
fhalos_onlysatdown,4.4
fhalos_onlysatup,2.5
fhalos_onlysat2,18.8
fhalos_onlysatup2,2.1
fhalos_onlysatdown2,3.1
fhalos_conformal,2.1
fhalos_conformalup,0.9
fhalos_conformaldown,0.4
fhalos_conformal2,11.0
fhalos_conformalup2,4.7
fhalos_conformaldown2,2.8
shuffled_fhalos_onlysat,27.1
shuffled_fhalos_onlysatdown,4.7
shuffled_fhalos_onlysatup,2.5
shuffled_fhalos_onlysatdown2,18.8
shuffled_fhalos_onlysatup2,2.1
shuffled_fhalos_onlysatdown2,3.1
shuffled_fhalos_conformal,1.0
shuffled_fhalos_conformalup,0.3
shuffled_fhalos_conformaldown,0.3
shuffled_fhalos_conformal2,1.0
shuffled_fhalos_conformalup2,0.2
shuffled_fhalos_conformaldown2,0.3
mixed_fhalos_onlysat,27.1
mixed_fhalos_onlysatdown,4.7
mixed_fhalos_onlysatup,2.5
mixed_fhalos_onlysatdown2,19.0
mixed_fhalos_onlysatup2,2.1
mixed_fhalos_onlysatdown2,2.9
mixed_fhalos_conformal,1.9
mixed_fhalos_conformalup,0.7
mixed_fhalos_conformaldown,0.4
mixed_fhalos_onlysat2,10.7
mixed_fhalos_onlysatup2,4.4
mixed_fhalos_onlysatdown2,2.9
fhalos_multisat,0.7
fhalos_multisatdown,0.1
fhalos_multisatup,0.2
fhalos_multisat2,0.3
fhalos_multisatdown2,0.1
fhalos_multisatup2,0.3
fsat_inm13m14,1.0
fsat_inm13m14down,0.6
fsat_inm13m14up,1.9
fsat_inm12p5m13,14.6
fsat_inm12p5m13down,5.4
fsat_inm12p5m13up,4.5
fsat_inm12m12p5,24.5
fsat_inm12m12p5down,7.4
fsat_inm12m12p5up,4.7
fsat_under12,16.9
fsat_under12down,7.1
fsat_under12up,9.0
fsat_inr200,38.3
fsat_inr200down,7.6
fsat_inr200up,16.4
\end{filecontents*}

   \title{Investigating the galaxy-halo connection of DESI Emission-Line Galaxies with SHAMe-SF}

   \author{Sara Ortega-Martinez
      \inst{1,2}\fnmsep\thanks{email: sara.ortega@dipc.org}
          \and
          Sergio Contreras\inst{1}
          \and
          Raul E. Angulo \inst{1,3}
          \and
          Jon\'as Chaves-Montero\inst{4}
          }

   \institute{Donostia International Physics Center (DIPC), Donostia-San Sebastian, Spain
         \and
             University of the Basque Country UPV/EHU, Department of Theoretical Physics, Bilbao, E-48080, Spain
        \and
            IKERBASQUE, Basque Foundation for Science, 48013, Bilbao, Spain
        \and
            Institut de F\'{\i}sica d'Altes Energies (IFAE), The Barcelona Institute of Science and Technology, 08193 Bellaterra (Barcelona), Spain
             }

   \date{Received November, 2024; accepted YYYYY YY, YYYY}

 
  \abstract
   {The Dark Energy Spectroscopic Instrument (DESI) survey is mapping the large-scale distribution of millions of Emission Line Galaxies (ELGs) over vast cosmic volumes to measure the growth history of the Universe. However, compared to Luminous Red Galaxies (LRGs), very little is known about the connection of ELGs with the underlying matter field.}
   {In this paper, we employ a novel theoretical model, SHAMe-SF, to infer the connection between ELGs and their host dark matter subhaloes. SHAMe-SF is a version of subhalo abundance matching that incorporates prescriptions for multiple processes, including star formation, tidal stripping, environmental correlations, and quenching.}
   {We analyse the public measurements of the projected and redshift-space ELGs correlation functions at $z=1.0$ and $z=1.3$ from DESI One Percent data release, which we fit over a broad range of scales $r \in [0.1, 30]/\hMpc$ to within the statistical uncertainties of the data. We also validate the inference pipeline using two mock DESI ELG catalogues built from hydrodynamical (TNG300) and semi-analytical galaxy formation models (\texttt{L-Galaxies}).}
   {SHAMe-SF is able to reproduce the clustering of DESI-ELGs and the mock DESI samples within statistical uncertainties. We infer that DESI ELGs typically reside in haloes of $\sim 10^{11.8}\hMsun$ when they are central, and $\sim 10^{12.5}\hMsun$ when they are a satellite, which occurs in $\sim$30 \% of the cases. In addition, compared to the distribution of dark matter within halos, satellite ELGs preferentially reside both in the outskirts and inside haloes, and have a net infall velocity towards the centre. Finally, our results show evidence of assembly bias and conformity. All these findings are in qualitative agreement with the mock DESI catalogues. 
   }
   {These results pave the way for a cosmological interpretation of DESI ELG measurements on small scales using SHAMe-SF.
   }

   \titlerunning{Galaxy-subhalo connection of DESI ELGs with SHAMe-SF}
   \maketitle
%

\section{Introduction}

Current galaxy surveys are mapping the large-scale structure (LSS) of the Universe across a wide redshift range, seeking to offer new insights into the nature of our Universe and its components. In the redshift range $0.8 < z < 1.6$, Emission Line Galaxies (ELGs) are an ideal target for these surveys because of their high number density and characteristic spectral features. The Dark Energy Spectroscopic Instrument (DESI, \citealt{DESI:2016}), one of these surveys, recently released the first 1\% of its data, collecting spectra from $\sim 250\,000$ ELGs identified using the [OII] line doublet \citep{DESI:2023_EDR}. The number of DESI ELGs is already similar to that of eBOSS -- the largest previous ELG survey \cite[e.g.][]{Raichoor:2016}.

To maximally exploit the data provided by surveys like DESI, it is necessary to link the observed galaxies with the underlying dark-matter field. Understanding this connection and the properties of the observed galaxies allows the creation of more realistic mocks and the development of more accurate theoretical models \citep{Cuesta-Lazaro:2023,ChavesMontero:2023,C2023:lensing}. Several analyses on semi-analytical models \citep{GonzalezPerez:2018,Gonzalez-Perez:2020} and hydrodynamical simulations \citep{Hadzhiyska:2021,Yuan:2022b} characterised ELGs as galaxies with high (specific) star formation rates that inhabit intermediate mass haloes, while $\sim 10$\% of them are passive galaxies hosting an AGN. 
This is supported by observational studies at redshifts $0.02<z<0.22$ \citep{Favole:2024} and $0.8<z<1$ \citep{Yuan:2024_conformity}. 

Previous analysis of the galaxy clustering of DESI and other ELG samples modelled the galaxy-halo connection using Halo Occupation Distributions (HOD) and SubHalo Abundance Matching (SHAM). Both are empirical models that select, based on their properties, which halos/subhalos are hosting the galaxies of a given sample (see, e.g. \citealt{Wechsler:2018} for a review). Even within the same family of models, the assumptions made for the galaxy-halo connection can be completely different.
 
HODs \citep{Jing:1998a, Benson:2000, Peacock:2000, Berlind:2003, Zheng:2005, Zheng:2007, Guo:2015a,C23_HOD} model the probability distribution of finding galaxies (central and number of satellites) depending on their host halo mass. To improve their accuracy at describing ELG samples, HODs underwent several extensions and modifications, for instance in adopted functional forms and scatter around the mean \citep{Jimenez:2019, Alam:2020, Rocher2023:DESI, Hadzhiyska:2022_onehalo, VosGines:2024,Garcia-Quitero:2024}, including secondary properties \citep{Hearin:2016, Hadzhiyska:2022_onehalo} or altering the satellite phase-space distribution \citep[e.g.][]{Avila:2020, Rocher2023:DESI}. In addition to these changes, to reproduce the small-scale clustering of ELGs, state-of-the-art HODs also include conditional probabilities for satellites based on the type of their central galaxy \citep{Alam:2020, Yuan:2024_conformity, Reyes-Peraza:2024}.

Standard SHAMs \citep{Vale:2006, Shankar:2006, Conroy:2006, Trujillo-Gomez:2011} assume that the most massive haloes host the most luminous galaxies. In the case of ELGs at redshifts below $z = 2$, this relation does not hold since the highest star-forming galaxies do not usually populate the highest mass subhaloes \citep[e.g.][]{GonzalezPerez:2018}. SHAMs can account for this by changing the subhalo selection criteria in their main property \citep{Favole:2016, Prada:2023, Yu:2024_DESISHAM}. These extended SHAMs also regulate the number of satellites in the sample, using the satellite fraction as a free parameter. Other approaches first perform a SHAM to model stellar mass and then select ELGs using a secondary subhalo property \citep{Favole:2022, Lin:2023} or by selecting from a set of ELG candidates \citep{Gao1:2022, Gao2:2023}. In the latter, the satellite fraction is regulated by an additional free parameter \citep{Gao3:2024} and they include correlations between centrals and their satellites (an effect known as ``conformity''), which improve the modelling of galaxy clustering on scales below $0.3 \, \hMpc$. 

In a previous paper \citep{SOM2024}, we presented SHAMe-SF, a new model for ELGs and star-forming samples, which we extensively validated with a hydrodynamical simulation and a semi-analytical model. Here, we apply the SHAMe-SF model ELG to data from DESI's 1\% Data Release \citep{DESI:2023_EDR}. Our model can accurately reproduce the galaxy clustering on small scales ($r \in [0.1, 30]/\hMpc$). After fitting the clustering, we compute the posterior predictive distribution of the host halo mass distribution, assembly bias, satellite fractions, and phase space distribution. We find that central ELGs are hosted by haloes with average masses of $10^{11.8} \, \hMsun$ if they are central, and $10^{12.5} \, \hMsun$ when they are satellites, and their clustering depends on properties beyond halo mass. Satellites are distributed almost isotropically, and we can divide them between an infalling population (outside the halo boundary with negative radial velocities) and an orbiting population. Even if most satellites inhabit haloes where the central galaxy is not an ELG, some haloes with lower halo masses ($10^{12.2} \, \hMsun$) are conformal. SHAMe-SF's capability to deliver this prediction was validated using two mock DESI samples built using a hydrodynamical simulation and a semi-analytical model. We also compare our findings with other models applied to DESI-ELGs, finding good agreement for all the aspects of the galaxy-halo connection we analysed, except for the satellite fractions. 

The work is structured as follows: A detailed description of the observational data, the choice of the validation mock DESI ELG samples, and the simulations used in this work are given in Section~\ref{sec:gensims}. We describe the SHAMe-SF model and the galaxy clustering statistics in Section~\ref{sec:SHAMemodel}. We use these tools in Section~\ref{sec:clustering} to fit the galaxy clustering of DESI's ELGs in the redshift ranges $0.8<z<1.1$ and $1.1<z<1.6$. Finally, our inference on the galaxy-(sub)halo connection is reported in Section~\ref{sec:DESIinference}. We compare these results with other ELG analyses in Section~\ref{sec:DESIothermodels}. Our conclusions are summarized in Section~\ref{sec:conclusions}.

\section{Observational and mock data}
\label{sec:gensims}
    In this section, we describe the different datasets we employ. First, we provide details on the ELG samples observed by DESI (Sect.~\ref{sec:thDESI}). Then, we describe the TNG300 simulation and the semi-analytical model we will employ for building mock DESI-ELG catalogues (Sect.~\ref{sec:Mocks}). 

\subsection{ELGs in DESI 1\% data release}
\label{sec:thDESI}

    In June 2023, the DESI collaboration released the first 1\% of their data \citep{DESI:2023_EDR} --  the last step of its Survey Validation program \citep{DESI:2023_validation}. The sky area scanned consisted of 140 sq deg, made of 20 non-overlapping rosettes with repeated observations to guarantee high completeness in the number of galaxies with reliable redshift estimations.

    This data contains a sample of ELGs, over the redshift range $0.6 < z < 1.6$, with a selection criteria similar to that planned for the complete DESI survey \citep{DESI:2016, Raichoor:2020_DESIELG, Raichoor:2023_DESIELG}. These galaxies were selected photometrically (via magnitude and colour cuts) and later confirmed spectroscopically by identifying the $[\ion{O}{ii}]$ doublet in their spectra. Overall, the success rate of the target selection reached $\sim 90$\% \citep{DESI:2023_EDR}. Despite the limited volume, this dataset represents a unique opportunity to learn about ELGs at high redshift and their connection with the underlying dark matter.

    Here, we will employ public measurements of the clustering of these ELG galaxies provided by the DESI collaboration \citep{Rocher2023:DESI}. The ELG sample was divided into a low-$z$ and a high-$z$ bin: $z\in[0.8, 1.1]$ and $z \in [1.1, 1.6]$. The clustering measurements cover a wide range of scales, $r \in [0.1, 30]\,\hMpc$, and have been corrected for observational systematic effects such as incompleteness and density inhomogeneities originating from fibre assignment. We will also employ the covariance matrices provided by DESI. As described in \cite{Rocher2023:DESI}; diagonal elements were estimated using a jackknife (JN) method and off-diagonal elements using a suite of HOD catalogues.

    The $[\ion{O}{ii}]$ emission of some galaxies in the DESI-ELG sample can come from AGNs instead of star formation. This fraction is estimated to be very low ($\sim4$\%, \citealt{Lan:2024}); thus, we do not expect this to affect our results significantly.
    
\subsection{Mock DESI catalogues}
\label{sec:Mocks}

\begin{figure*}
		\centering
		\includegraphics[width=0.95\textwidth]{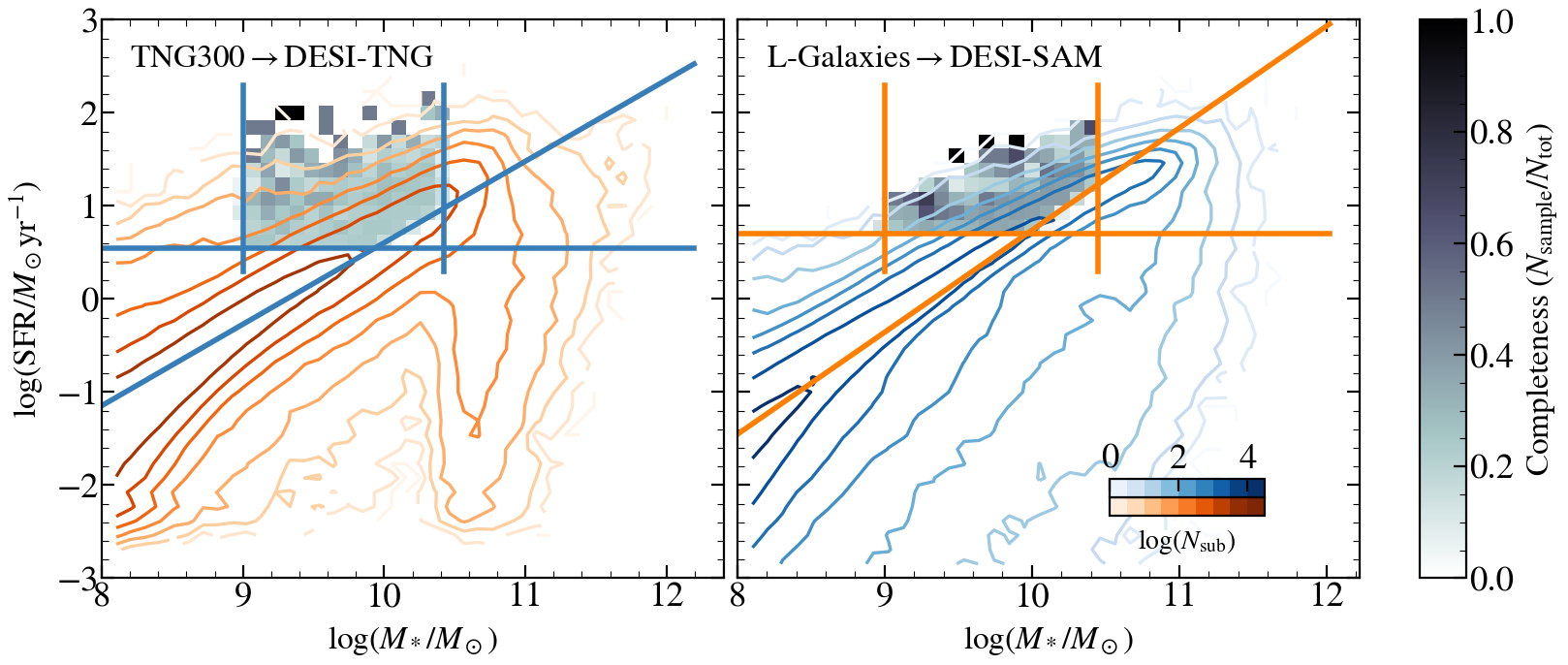}
        \caption{
  Distribution of stellar masses ($M_*$) and star formation rates (SFR) for galaxies at $z=1$ as predicted by the TNG300 hydrodynamical simulation (``DESI-TNG'', left panel) and by the \texttt{L-Galaxies} semi-analytical galaxy formation model (``DESI-SAM'', right panel). Our mock DESI samples are defined by selection criteria in $M_*$, SFR, and the main sequence (similar to a cut in specific SFR), as indicated by solid lines (see Fig. 5 of \citealt{Yuan:2024_conformity} for a similar plot for a crossmatch between observed DESI and COSMOS galaxies). Note that the units for stellar mass are $M_\odot$ (without the $h$ factor). The fraction of selected galaxies for such samples is shown in a greyscale.}
        \label{fig:ELGselection}
\end{figure*}

We have built two mock catalogues based on the TNG300 hydrodynamical simulation and a semi-analytical model to validate and interpret our results. In hydrodynamical simulations, baryons and gravity are jointly evolved. Semi-analytical models (SAM) use the merger trees of dark matter-only simulation to model the evolution of the baryonic components of galaxies (for a review, see \citealt{Baugh:2006,Benson:2010}). 

\subsubsection{Galaxy Formation Simulations}

TNG300 is a large publicly available high-resolution hydrodynamical simulation, which evolves dark matter, gas, stars, and black holes inside a periodic box of $205\,\hMpc$ \citep[$\sim300$ Mpc,][]{Nelson:2017TNGcolors, TNGb, TNGc, TNGd, TNGe}. Compared to observations, TNG300 reproduces the stellar mass function from SDSS \citep{TNGd,Jackson:2020}, as well as the stellar mass - star formation rate main sequence and the scatter on this relation \citep{Donnari:2019}. 

On the other hand, several aspects of the galaxy population are not reproduced by TNG300. Additionally, it has not been calibrated to match the observed properties of ELGs. For these reasons, we employ another galaxy formation model, namely the \texttt{L-Galaxies} semi-analytical model, with which we can estimate the current modelling uncertainties and validate our inferences in two different galaxy formation scenarios.

We employ the catalogues provided by \cite{LGalaxies_Ayromlou:2021}. These catalogues were generated using the \cite{Henriques:2015} version of \texttt{L-Galaxies} on top of the subhalo merger trees extracted from a gravity-only version of the TNG300. As for the TNG300, it has been extensively tested that \texttt{L-Galaxies} reproduces multiple aspects of the observed galaxy population \citep{Guo:2011,Guo:2013,Henriques:2013,Henriques:2020}.  

\subsubsection{ELG selection}
\label{sec:thColours}

DESI employs photometric selections to identify ELG candidates. Using the same criteria to build the mock DESI-ELG catalogues using TNG300 and \texttt{L-Galaxies} implies modelling of additional processes (such as dust obscuration, metallicities or burstiness), to obtain realistic colour-colour distributions. Instead, we define our mock DESI-ELG samples based on the stellar masses and star formation rates of galaxies.

 \cite{Yuan:2024_conformity} estimated the stellar masses and star formation rates of DESI ELGs in the redshift interval $0.8 < z <1.1$ by cross-matching them with galaxies in the COSMOS survey \citep{Weaver:2022_Cosmos}. These authors show that DESI ELGs populate a well-defined region in SFR, $M_*$, and specific SFR (sSFR). Specifically, these galaxies lay above the Main Sequence, have a minimum SFR of $\sim 1\,\Msun\rm{yr}^{-1}$, a minimum stellar mass of $~10^{8.7}\Msun$, and a maximum stellar mass of $~10^{10.5}\Msun$ avoiding massive quenched galaxies. 

Similar to Figure 5 in  \cite{Yuan:2024_conformity}, in Figure \ref{fig:ELGselection}, we show the distribution of star formation rates (SFR) and stellar masses, ($M_*$) for galaxies in the TNG300 simulation and \texttt{L-Galaxies} semi-analytical models catalogues (orange and blue contours, respectively). Note that this is the only case where mass units are $\Msun$ and not $\hMsun$ to ease the comparison with the crossmatch with COSMOS. Although both models are broadly consistent, we can see some differences. TNG300 shows a more narrowly defined main sequence of star formation and a drastic quenching at $\log(M_*/M_\odot)\sim 10.5$. In contrast, quenching is less evident in the SAM for higher halo masses.

To mimic these constraints in our mocks, we first matched COSMOS's cumulative stellar mass and SFR functions with those in either TNG300 or \texttt{L-Galaxies}. We then defined thresholds that would yield the same abundance of galaxies. Subsequently, we fitted the $M_*$-SFR main sequence in each catalogue and discarded galaxies below it. We label the final catalogues as DESI-TNG (for the hydrodynamical simulation) and DESI-SAM (for the semi-analytical model). Finally, we downsampled each catalogue to match DESI's number density ($\bar{n} = 1.047\times 10^{-3}\,\ihMpcC$) (by a factor of $\sim 4$ for DESI-TNG and $\sim 3$ for DESI-SAM). This downsampling mimics the incompleteness of the selection criteria. The final selection thresholds are indicated as blue (DESI-TNG) and orange (DESI-SAM) solid lines in Figure \ref{fig:ELGselection}. The grey colour bar indicates the fraction of galaxies in the ELG catalogues.

\section{SHAMe-SF galaxy population model}
\label{sec:SHAMemodel}

In this section, we recap our model for the clustering of ELG and describe the gravity-only simulation with which we compute its predictions. Then, we describe an emulator we built to accelerate the evaluation of model predictions.

\subsection{SHAMe-SF}
\label{sec:firstSHAMeSF}

The SubHalo Abundance Matching Extended (Star Formation version), SHAMe-SF, is a model based on the SHAM formalism that includes physical prescriptions to link the SFR of a galaxy to the properties of their host subhalo. The model was first presented in \cite{SOM2024}, where we refer the reader to for an extensive discussion of the main ingredients and assumptions. In short, the SFR of a galaxy is assumed to be a function of the peak circular velocity, $\vmax$, of the host DM subhalo, where nuisance parameters control the functional form, scatter, and correlations with the environment. The SFR is also suppressed as a function of the host halo mass and time since accretion, effectively modelling gas stripping and quenching. Here, and in the remainder of the paper, we define the halo mass, $M_h$, as the mass contained within a sphere of an average density 200 times the critical density of the universe and centred in the minimum of the gravitational potential (whose radius is defined as $\rth$). We take the satellite definition from \texttt{FoF} + \texttt{SUBFIND}, which also considers satellites beyond $\rth$. 

For this work, we have extended SHAMe-SF to allow for an increase in SFR for very massive galaxies. Specifically, the SFR is given by:

\begin{equation}
  {\rm SFR}|_{({\vpeak})} \propto
    \begin{cases}
       \left[ \left( \frac{\vpeak}{V_{1}}\right)^{-\beta} + \left( \frac{\vpeak}{V_{1}}\right)^{\gamma} \right]^{-1}, & \text{if $\vpeak \leq V_{1}+\Delta V_1$}\\
       & \\
      \text{SFR}_{V_{1}+\Delta V_1} \left/~  \left( \frac{\vpeak}{V_{1}+\Delta V_1}\right)^{\gamma + \Delta\gamma}  \ \right.\ & \text{if $\vpeak$ > $V_{1}+\Delta V_1$}
    \end{cases}       
    \label{eq:model}
\end{equation}

\noindent where $\Delta V_1$ and $\Delta \gamma$ are two additional free parameters compared to the original version of the model. $\text{SFR}_{V_{1}+\Delta V_1}$ is a normalization factor set to make the function continuous at $\text{SFR}_{V_{1}+\Delta V_1}$. The option of including an additional feature in the functional form for high values of $\vpeak$ was already discussed when we first presented the SHAMe-SF model. This term accounts for the possible connections between SFR and $\vpeak$ that could appear due to different implementations of quenching and environmental effects (e.g., \citealt{C15}, \citealt{Popesso:2015}), as well as galaxies classified as ELG due to AGN emission lines rather than SF-related lines. 

In the first version of SHAMe-SF, the same parameter regulated the secondary dependence on $\vpeak/\vvir$ for centrals and satellites ($f_k$). In this version, we use the same parametrization that \cite{C2023:lensing}, and allow for different dependencies ($f_{k,\rm{cen}}$ and $f_{k,\rm{sat}}$). This separation can describe differences between centrals and satellites depending on concentration that are not already captured by the quenching mechanism dependent on host halo mass and time since peak mass. 

The original version of the model was validated against multiple catalogues of SFR-selected galaxies extracted from the TNG and a SAM model \citep{SOM2024}. Specifically, SHAMe-SF was able to fit the projected and redshift-space clustering of ELGs with number densities over the range $\bar{n} \in \{10^{-2} - 10^{4}\}\,\ihMpcC$ at $z=0$ and $1$. Here, we further validate the extended version of the model against the mock DESI-ELG catalogues described previously.  In Appendix~\ref{app:TNG}, we display our results, showing that SHAMe-SF can successfully fit the clustering statistics and provide accurate inferences about the galaxy-halo connection of mock DESI-ELG galaxies. 

\subsection{Gravity-Only simulations}
\label{sec:thBaccoPlanck}

To suppress the impact of cosmic variance, we computed the predictions of SHAMe-SF on a simulation much larger than that employed for building our mock DESI-ELG catalogues. Specifically, we use three gravity-only simulations of the ``BACCO simulation project'' \citep{Angulo:2021}. The first simulation evolved $1536^3$ particles of $m_p=10^{9.5} \, \hMsun$ on a box of $512 \, \hMpc$ a side, which we will employ to evaluate SHAMe-SF models. We will use the other two simulations (with paired phases), with $3072^3$ particles on a $1024 \, \hMpc$, to construct a SHAMe-SF emulator. The cosmology adopted by both simulations corresponds to the best-fit analysis of the Planck satellite data \citep{Planck2015}: $\OmM$ = 0.3089, $\sig$ = 0.8159, $\ns$ = 0.9667 and $h$ = 0.6774, which is, hence, the fiducial cosmology adopted throughout our analyses. 

The initial conditions of our simulations were computed using 2LPT, where mode amplitudes were fixed to the ensemble average, drastically suppressing variance on large scales \citep{Angulo:2016}. 
Similarly to other BACCO simulations, we identify on-the-fly haloes and subhaloes using \texttt{FoF} and \texttt{SUBFIND} algorithms, respectively \citep{Davis:1985,Springel:2001}, as well as all the dark matter properties that SHAMe-SF requires.

\subsection{A SHAMe-SF emulator}
\label{sec:synfitting}

Following \cite{Angulo:2021}, we speed up the evaluation of SHAMe-SF predictions by building a simple feed-forward neural network emulator \citep[see also][]{Arico:2021,Arico:2020,Pellejero:2023,Zennaro:2023,C2023:lensing,SOM2024}. 

We start by defining the set of SHAMe-SF parameter combinations, randomly distributed according to a Latin Hypercube over the ranges: 

\begin{eqnarray}
        \beta & \in & [0,20]\nonumber \\
        \gamma & \in & [-15,25] \nonumber \\
        \Delta\gamma & \in & [-10,10] \nonumber \\
        V_1 & \in & [10^{1.8},10^{3.2}] \text{ (km/s)} \nonumber \\
        \Delta V_1 & \in & [10^{0.2},10^{1.9}] \text{ (km/s)} \nonumber \\
        \sigma & \in & [0,2]  \\
        f_{k,\rm{(cen+sat)/2}} & \in & [-1,1] \nonumber \\
        f_{k,\rm{(cen-sat)/2}}  & \in & [-1,1] \nonumber \\
        \alpha_0 & \in & [0,8] \nonumber\\
        \alpha_{\rm exp} & \in & [-8,8] \nonumber \\
        M_{\rm crit} & \in & [8.5,15] \ (\log(\hMsun)), \nonumber
        \label{eq:par_range}
\end{eqnarray}

\noindent where $\{\beta, \gamma, \Delta\gamma,\Delta V_1, \sigma \}$ describe the relation between SFR and circular velocity, $\{f_{k,\rm{(cen+sat)/2}}, f_{k,\rm{(cen-sat)/2}}\}$ control the correlation with environment, $\{ \alpha_0, \alpha_{\rm exp}, M_{\rm crit}\}$ regulate the impact of gas stripping in satellites. These ranges are motivated by the analysis of star-forming galaxies from TNG300 and L-Galaxies in \cite{SOM2024} and are wide enough to avoid truncating the posterior distributions. We refer to \cite{SOM2024} for further details on the relation between parameters. 

We then evaluated SHAMe-SF on top of our $1\,\ihGpcC$ gravity-only simulation at 10 redshifts over the range $z \in [0.8, 1.6]$\footnote{$z\in[1.49, 1.38, 1.31, 1.21, 1.14, 1.11, 0.98, 0.95, 0.89, 0.84]$} and for 6 number density thresholds $\bar{n} \in [10^{-4}, 10^{-2.5}]$ (equally spaced in log scale). Each evaluation takes approximately 20 CPU minutes (including calculating the galaxy clustering for the 8 number densities).

For each catalogue, we use \textsc{corrfunc} \citep{Corrfunc1, Corrfunc2} to compute the redshift-space 2D correlation function $\xi(s,\mu)$, where $s^2 = r^2_{\rm p} + r^2_{\pi}$ and $\mu = \cos\left(\hat{s,\rm los} \right)$. To reduce statistical noise, we average the measurements over the three Cartesian axes and for 8 model evaluations adopting different random seeds.

We compute the projected correlation function as follows:   

\begin{equation}
    w_{\rm p} \equiv 2 \times \int_{0}^{\pi_{\rm max}} d\pi\, \xi(\rm p, \pi),
\end{equation}

\noindent using the same separation bins and line-of-sight integration limit ($\pi_{\rm max}=40\ihMpc$) as those used by DESI \citep{Rocher2023:DESI}. Subsequently, we estimate the multipoles of the correlation function as:

\begin{equation}
    \xi_{\ell} \equiv \frac{2 \ell+1}{2} \int^{1}_{-1} d\mu\, \xi(s,\mu)P_\ell(\mu),
\end{equation}
\noindent where $P_{\ell}$ is the $\ell$-th order Legendre polynomial. 

Overall, we obtained 130000 measurements for each clustering statistic. We used this data to train a Neural Network (NN) using the same architecture as \cite{C2023:lensing}: two fully connected hidden layers with 200 neurons for the projected correlation function and the monopole, and three hidden layers and 200 neurons for the quadrupole (since it has a more complex structure). Variations of this architecture do not significantly change our results. We use the TensorFlow library's Keras front-end with the Adam optimisation algorithm, setting the learning rate to 0.001 and choosing a mean squared error loss function. 10\% of the dataset was kept for validation.

\section{Fitting DESI ELG galaxy clustering}
\label{sec:clustering}
\begin{figure*}
		\centering
		\includegraphics[width=0.95\textwidth]{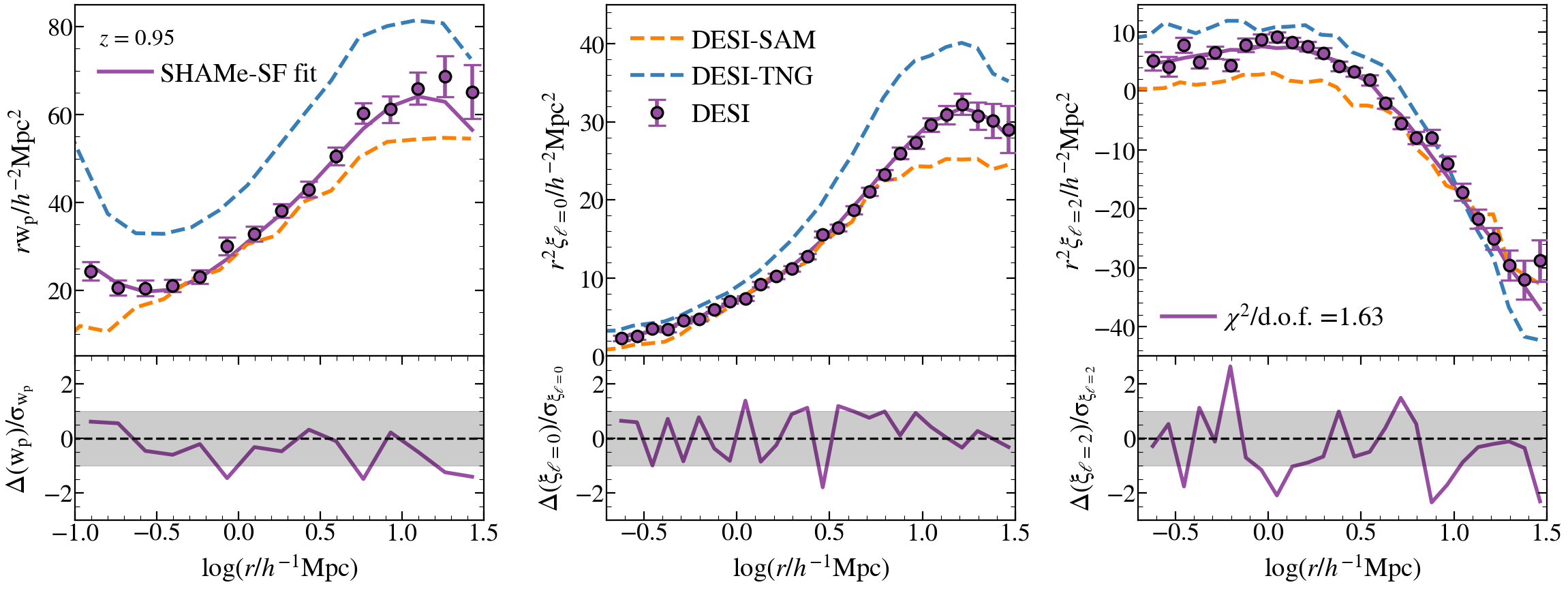}
		\includegraphics[width=0.95\textwidth]{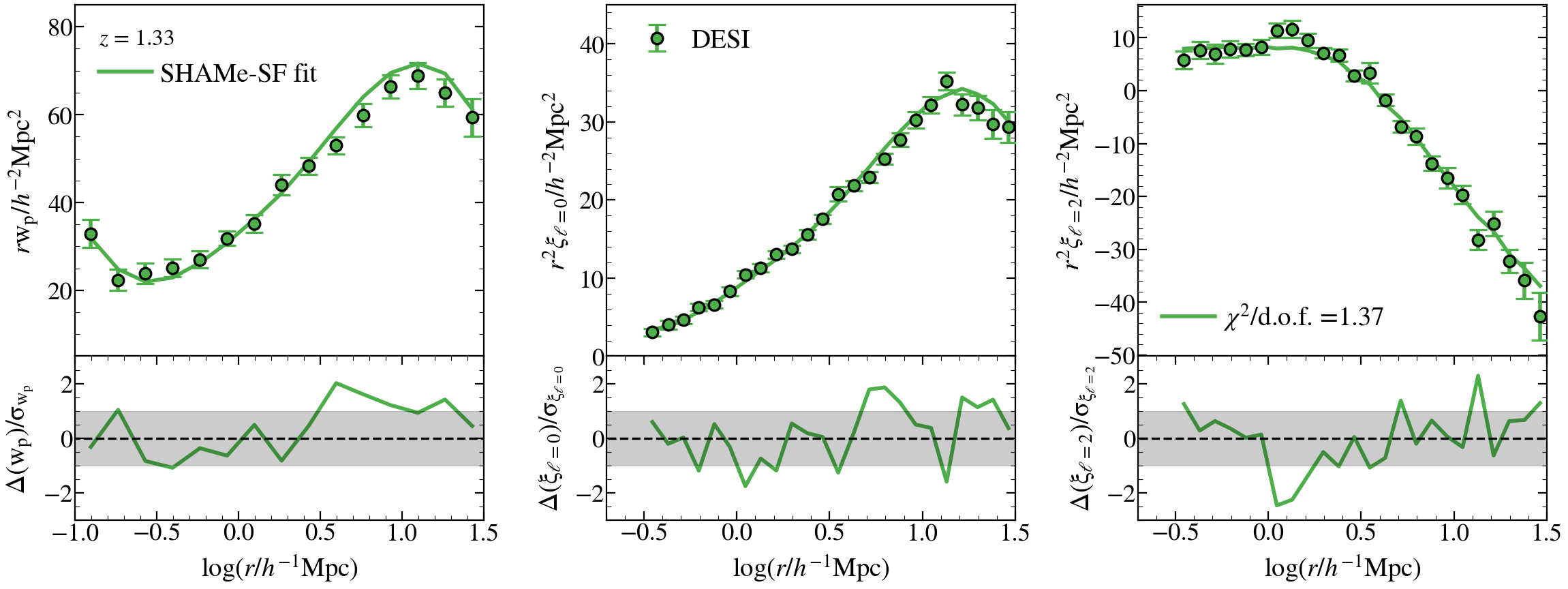}
        
	\caption{Projected correlation function ($w_p$), and monopole ($\xi_{\ell=0}$) and quadrupole ($\xi_{\ell=2}$) of the redshift-space correlation function of ELG galaxies in DESI at $z=0.95$ (top row) and at $z=1.33$ (bottom row), together with the corresponding best fit SHAMe-SF model (purple and green lines). The bottom panel in each plot shows the difference between the data and the fit with SHAMe-SF in units of the diagonal elements of the respective covariance matrix. For comparison, we display the measurements for our DESI-TNG and DESI-SAM mock DESI catalogues at $z=1$ as blue and orange lines, respectively. SHAMe-SF is a reasonably good description of the data for all the statistics and scales considered. Also, note that the data displays fluctuations inconsistent with the error bars, which suggests there could be sources of noise that are not accounted for in the covariance matrices. }
        \label{fig:clusteringDESI}
\end{figure*}

This section discusses the fitting of DESI-ELG clustering measurements using our SHAMe-SF model. 

\subsection{Likelihood, parameter space, and sampling}

We assume that the likelihood of observing a set of clustering statistics: $\vec{d} = \{w_p, \xi_{\ell=0}, \xi_{\ell=2} \}$ is given by a multivariate Gaussian: $\log \mathcal{L}(\vec{\theta}) \propto (\vec{d}-\vec{t})^t \mathcal{C}^{-1} (\vec{d}-\vec{t})$, where $\vec{t}$ is the SHAMe-SF emulator predictions for a given set of parameters $\vec{\theta}$. We will use the data vector, $\vec{d}$, and covariance matrix, $\mathcal{C}^{-1}$, provided by the DESI collaboration (c.f. Sect. \ref{sec:thDESI}). We include the emulator uncertainty (c.f. Sect. \ref{sec:synfitting}) as an additive contribution to the diagonal of the covariance matrix. 

We fix the cosmological parameters to those assumed in our gravity-only simulation (c.f. Sect. \ref{sec:thBaccoPlanck}). Thus, we will only vary parameters that control the relationship between ELGs and the host DM subhaloes. For all of these 11 SHAMe-SF parameters, we adopt flat priors over an interval that coincides with the ranges employed by our emulator. In Appendix A, we show that these ranges comfortably enclose the values expected in the DESI-SAM and DESI-TNG ELG mocks. Thus, we expect these priors to be uninformative and not to affect our results. We note that our SHAMe-SF predictions assume values for \textit{Planck} cosmology that do not match the fiducial cosmology assumed by DESI to transform angular separations and redshifts into distances. In principle, we could apply the so-called Alcock-Paczynski corrections after each model evaluation to account for this. However, the difference in cosmology is small ($\Delta \OmM = 0.006$), and we have checked this negligibly impacts our predictions, and, therefore, we have ignored this effect.

We will separately analyse the low-$z$ and high-$z$ ELG samples. In each case, we evaluate SHAMe-SF at the median redshift of each sample ($0.95$ and $1.33$, respectively). We will only consider separations in the range $r \in [0.1, 30]\,\ihMpc$. The lower limit is set by the minimum scale for which we expect SHAMe-SF to deliver accurate inferences. The upper limit is set by the largest scale included in the public measurements, which also roughly coincides with the largest scale we expect our predictions to be unaffected by the finite size of our gravity-only simulation. Although not shown here, we have checked that none of our results depend strongly on the minimum or maximum scale included in the fit.

We sample the likelihood using an ensemble Markov Chain Monte-Carlo (MCMC) algorithm implemented by {\tt emcee} \citep{emcee}. We employ a configuration consisting of 5000 chains, each with 30000 steps. We note that the high computational efficiency of our emulator allows us to obtain  100,000 model evaluations in under 2 seconds of CPU time. Throughout this work, we will refer to best-fit parameters as those that maximise the likelihood, and $1$ and $2\sigma$ regions will refer to 68 and 95\% of the posterior distribution.

\subsection{Best fit to DESI galaxy clustering}
\label{sec:DESIclus}

In Figure \ref{fig:clusteringDESI}, we display the measured projected correlation function and the monopole and quadrupole of the redshift space correlation function of the low-$z$ and high-$z$ DESI-ELG samples. The best-fit SHAMe-SF model is shown as solid lines.

Firstly, we highlight that our model is an excellent description of the high- and low-$z$ data over the whole range of scales considered. However, we note that the value of the reduced $\chi^2$, displayed in the legend, is somewhat high. Looking at the bottom panels, which display the difference between the data and the model in units of the diagonal elements of the covariance matrix, we can see no large systematic deviations from 0, typically the signature of model errors. Instead, there seem to be large random fluctuations, especially considering that neighbouring data points should exhibit a considerable degree of correlation. This suggests that there could be additional sources of stochastic noise or that the jackknife method (employed by DESI to estimate the covariance matrix) underestimates the actual uncertainty in the data. 

In the case of the low-$z$ sample, for which we have the crossmatch with COSMOS from \cite{Yuan:2024_conformity}, we can directly compare the DESI-ELG clustering with that in our mock DESI-TNG and DESI-SAM catalogues (shown as dashed lines). We can see that for all statistics, the ELG mocks bracket the DESI measurements\footnote{Note that at $r>10\,\hMpc$, we expect finite-box effects to impact the clustering measurements in the mocks. Therefore, we have applied a correction estimated with simulations of two sidelengths, as described in Appendix \ref{app:box}}, which indirectly supports the validity of our procedure to create DESI mocks to deliver plausible predictions for the galaxy-halo connection. The TNG-based sample is about 10\% more biased, whereas the DESI-SAM is 2\% less biased. This naively implies that the DESI-TNG/DESI-SAM predicts more/fewer satellites than measured in DESI and a higher/lower average host halo masses. In subsequent sections, we will explicitly explore this.

\subsection{Constraints on SHAME-SF model parameters}
\label{sec:shame_pars}

\begin{figure}
    \centering
	\includegraphics[width=\columnwidth]{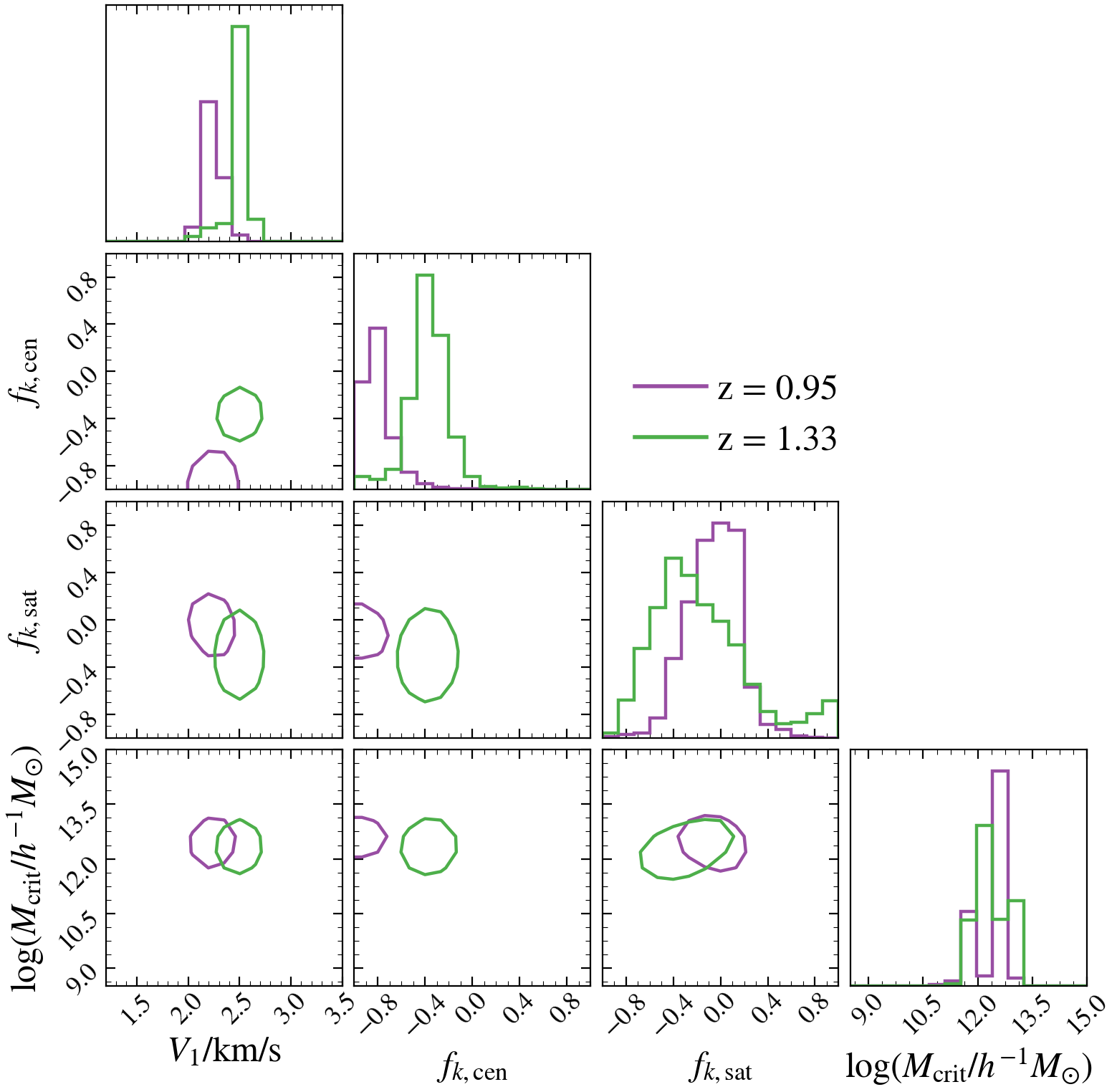}
    \caption{Marginalised $1\sigma$ constraints on the most important SHAMe-SF parameters, obtained from fitting the clustering of DESI-ELGs at $z=0.95$ (purple) and $1.33$ (green).}
    \label{fig:DESIpost}
\end{figure}

By fitting the clustering of DESI-ELGs, we have constrained the free parameters of our SHAMe-SF model. In Figure \ref{fig:DESIpost} we display the posterior distribution function on the four parameters best constrained by the data, namely: $V_1$, the pivot value of $\vmax$ above which galaxies start to be quenched; $f_{k,\rm{cen}}$ and $f_{k,\rm{sat}}$, which control the secondary dependence on subahlo concentration; and $M_{\rm crit}$, the host halo mass where satellites decrease their SFR due to gas stripping. The full posteriors on the 11 parameters are shown in Figure~\ref{fig:allparams} in Appendix \ref{app:posteriors}. 

First, the low and high-$z$ samples prefer statistically consistent values for these SHAMe-SF parameters. This implies that the galaxy formation physics controlling ELGs does not significantly evolve between $z=1.0$ and $1.3$. Specifically, we obtain $\log V_1 [$km/s$]\sim 2.3$: the star formation rate scales with $\vpeak$ until subhalos reach $\vpeak\sim 200$ km/sec, after they start to be quenched for higher $\vpeak$ values. Additionally, from the value of the $M_{\rm crit}$ parameter, we infer that already in haloes of $10^{12}\,\hMsun$, satellite ELGs are affected by stripping. The value of the ordering parameter for centrals ($f_{k,\rm{cen}}$) is negative for both redshift bins. For a fixed value of $\vpeak$, SHAMe-SF preferentially populates subhalos with lower concentrations. This is also true for satellites ($f_{k,\rm{sat}}$), but we find values closer to zero for the lower-z bin (almost random ordering). The following section will explore what this implies for the type of DM structures DESI-ELG galaxies live in. 

\section{Galaxy-halo connection in DESI ELGs}
\label{sec:DESIinference}
In this section, we present the main results of our paper: constraints on the connection between DESI-ELGs and the underlying DM structures. We explore the halo occupation number (Sect.~\ref{sec:DESIHOD}), assembly bias (Sect~\ref{sec:DESIassembly}) and the abundance and distribution of satellites (Sect.~\ref{sec:DESIsat}).

We obtain our inferences by randomly selecting $100$ steps from our MCMC chains after discarding 10000 burn-in steps. We then employ those parameter sets to build SHAMe-SF catalogues from our $512\,\hMpc$ gravity-only simulation. Finally, we compute the statistic of interest and present the median and 16th-84th percentiles of the distributions. We validate our procedure by applying it to our mock DESI catalogues, as shown in Appendix~\ref{app:TNG}.

\subsection{Halo occupation number}
\label{sec:DESIHOD}

\begin{figure*}
	\centering
        \includegraphics[width=0.45\textwidth]{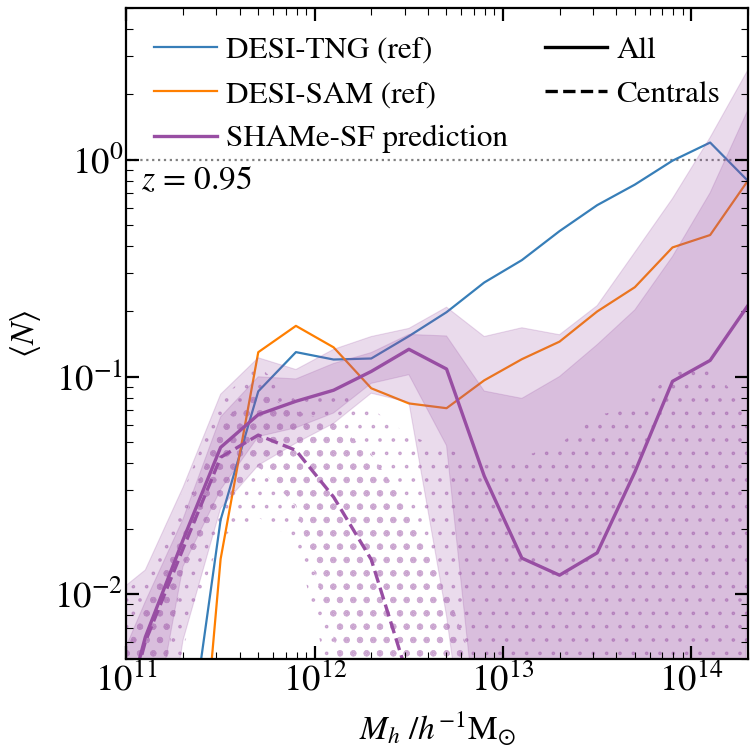}
        \includegraphics[width=0.45\textwidth]{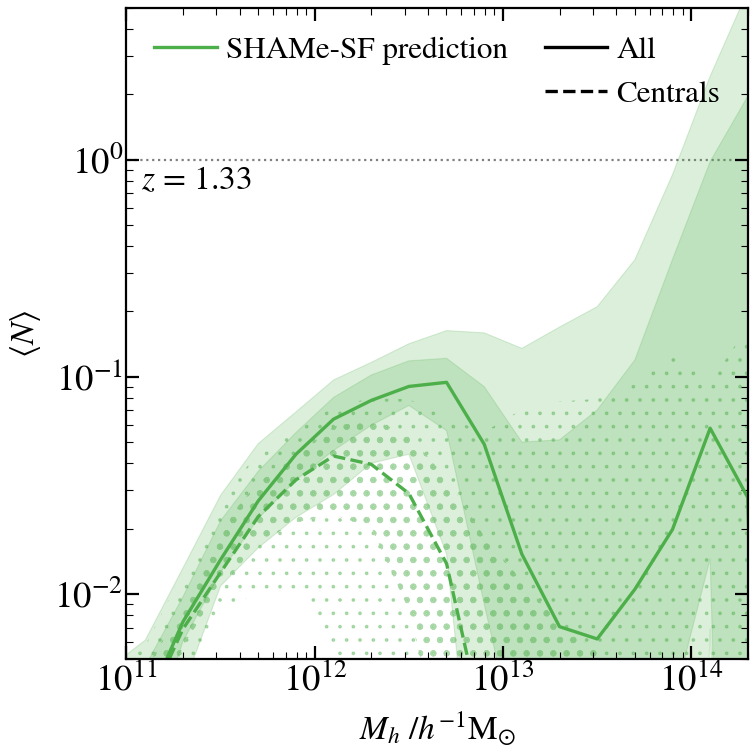}
	    
    \caption{ Inferred halo occupation number for galaxies in the low-$z$ and high-$z$ DESI ELGS samples. The solid (dashed) lines show the median of our model after marginalisation over all the SHAMe-SF parameters, whereas shaded regions (small and large circles) show the $1\sigma$ and $2\sigma$ regions for all galaxies (centrals). We compare the measured occupation distributions in our two DESI ELG mocks. DESI ELGs have similar mean halo masses as our mock DESI catalogues (DESI-TNG and DESI-SAM), but the abundance of satellites is systematically lower. We add as a reference the dotted grey line, indicating an abundance of $\langle N \rangle = 1$.}
	
 \label{fig:HOD}
\end{figure*}

In Figure \ref{fig:HOD}, we present our inferences on the average number of ELG galaxies per halo as a function of halo mass for the low-$z$ (left, purple) and high-$z$ (right, green) samples. The solid lines and shaded regions represent the median and 1- and $2-\sigma$ predictions for all galaxies. The central occupancy is shown using the dashed lines (median), large circled-hatch ($1-\sigma$) and small circled-hatch ($2-\sigma$) regions. With blue/orange solid lines, we display the halo occupation number directly measured in our DESI-ELG mocks (DESI-TNG/DESI-SAM, respectively) for comparison with the low-$z$ sample. 

We find that central ELGs populate on average haloes of mass $\log( M_{\rm h}/\hMsun) = \DESIl{m200cen}\,\pm\,\DESIl{m200cenerr}$ and $\DESIh{m200cen}\,\pm\,\DESIh{m200cenerr}$ in the low-$z$ and high-$z$ bins, respectively. For the mock ELG samples (DESI-TNG and DESI-SAM), we find a similar value in the low-$z$ bin: $\log (M_{\rm h}/\hMsun) \sim$  \TNGref{m200cen}. The probability of finding a central galaxy on a halo is typically well below the unity for all halo masses. In the $1\sigma$ interval, the probability of finding a central ELG falls below $10^{-2.5}$ for galaxies with halo masses above $10^{13}\,\hMsun$. When looking at the $2\sigma$ regions, the model does not constraint the central occupancy for halo masses above $10^{13}\,\hMsun$, but even in this confidence interval, it does not reach the one-central-per-halo expectancy. 

In the case of satellite ELGs, the average halo mass increases by $\sim0.8$dex: $\log(M_{\rm h}/h^{-1}M_\odot)$ = \DESIl{m200sat}$\pm$\DESIl{m200saterr}, and \DESIh{m200sat}$\pm$\DESIh{m200saterr}, for the low and high-$z$ samples. However, we find that satellite ELGs populate lower mass halos than in the DESI-TNG/DESI-SAM mock samples, where $\log (M_{\rm h}/\hMsun) =$ \TNGref{m200sat}. SHAMe-SF places most of the satellites on masses in the interval $\log(M_{\rm h}/h^{-1}M_\odot) \in [11.8,12.8]$. For halo masses above $10^{13}\,\hMsun$, their median abundance drops, and we predict much larger uncertainties (almost two orders of magnitude for $10^{14}\,\hMsun$ in both redshift bins). This scatter between different realizations of the SHAMe-SF model is linked to the low contribution of halos in these mass ranges to the clustering signal (see \citealt{C13}). We do not find the drop in the satellite abundance when comparing the low-$z$ bin with the measurements on the mock DESI ELG samples.

\subsection{Assembly bias}
\label{sec:DESIassembly}

\begin{figure}
		\centering
		\includegraphics[width=0.45\textwidth]{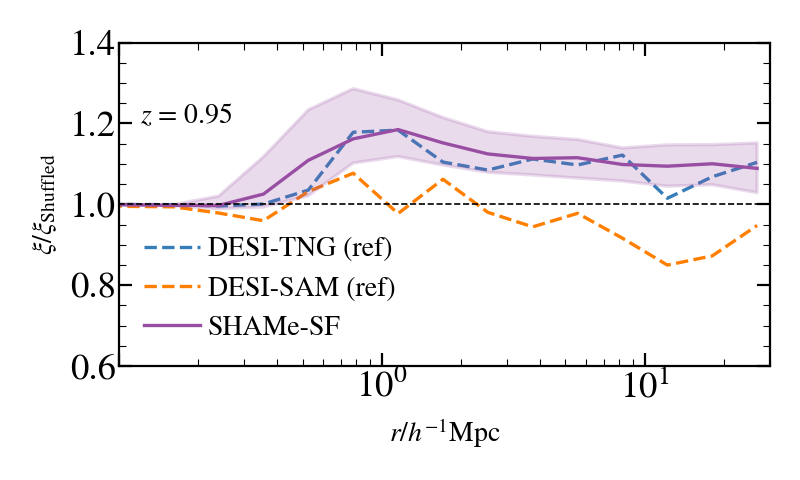}
        \includegraphics[width=0.45\textwidth]{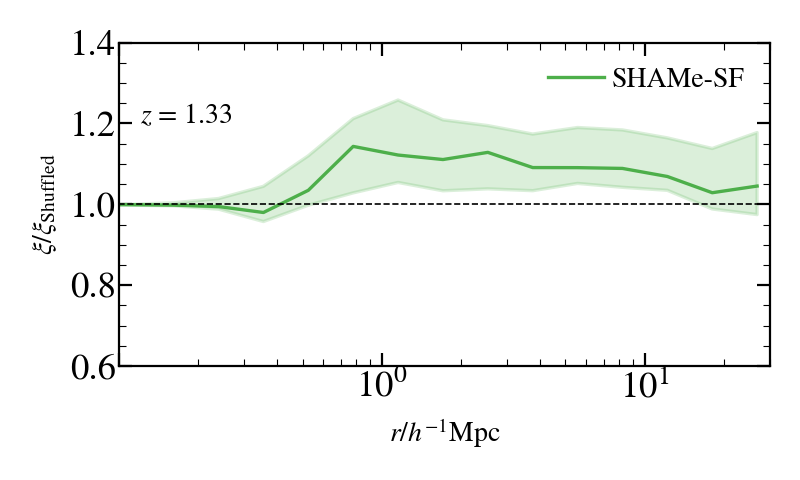}
		\caption{Inferred magnitude of galaxy assembly bias in the low-$z$ and high-$z$ DESI-ELG catalogues. We display the ratio of the correlation function in the best-fit SHAMe catalogue to a version where the position of haloes has been randomly shuffled among haloes of the same mass. Solid lines and shaded regions indicate the mean and $1\sigma$ region after marginalisation over all SHAMe parameters. We display the same quantity estimated in our mock DESI catalogues for comparison (DESI-TNG and DESI-SAM, dashed lines). At a fixed halo mass, DESI ELGs tend to live preferentially in highly biased haloes, which is in qualitative agreement with the predictions of the TNG simulation.}
  \label{fig:DESIassemblybias}
\end{figure}

As observed in simulations, the halo spatial distribution depends on properties beyond halo mass \citep{Sheth:2004, Gao:2005, Gao:2007, Wechsler:2006, Faltenbacher:2010, Angulo:2009, Mao:2018}. Since the evolution of subhaloes and galaxies is bound to that of their host haloes, this dependency (usually called assembly bias) will propagate to galaxies \citep{Croton:2007}. The strength of the assembly bias signal varies depending on the hydrodynamical simulation (or SAM), number density, redshift or sample considered \citep{C19,Contreras:2021bias, Jimenez:2021}.

To quantify the amplitude of the assembly bias signal, we use the technique proposed by \cite{Croton:2007}, where the positions of haloes (including all their satellites) are shuffled among haloes within $0.1$ dex bins in halo mass. Then, assembly bias is estimated by comparing the shuffled and original correlation functions:

\begin{equation}
    b^2 = \xi(r)/\xi_{\rm shuffled}(r).
\end{equation}

We present the results of the inferred assembly bias signal in Figure~\ref{fig:DESIassemblybias} for the low-$z$ (top, purple) and high-$z$ (bottom, green) DESI-ELG samples. The shaded region indicates the 1-$\sigma$ intervals.  We show the mock DESI-TNG (blue) and DESI-SAM (orange) samples as a reference. We find a non-negligible assembly bias signal for both redshift bins (between 10 and 20\% for the low-$z$ bin and $\sim$10\% for the high-$z$ bin). We obtain the same constraints shuffling using halo mass bins of $0.05$ dex and $0.2$ dex. In the case of the low-$z$ bin, we also observe a scale dependence of the signal on large scales, compatible with the findings of \cite{Jimenez:2021} for a semi-analytical model. The authors found different assembly bias signals for different number densities. Thus, we cannot assess the redshift evolution of assembly bias using these samples. \cite{Hadzhiyska:2022_onehalo} point that part of this signal can originate from haloes misplaced as satellites by \texttt{FoF}+\texttt{SUBFIND}. To estimate this effect's impact, we repeat the calculation only shuffling objects within $\rth$. We do not find deviations at large scales. 

\subsection{ELGs as satellites}
\label{sec:DESIsat}
Satellite galaxies dominate the clustering signal on small scales, where baryonic processes are also critical. Their evolution is more challenging to model than central galaxies due to the extensive variety of processes and interactions associated with satellites (gas stripping, strangulation and ram pressure, among others).

Before analysing in depth what is happening to satellites, we measure how present they are in the sample in Sect.~\ref{sec:satfrac}. We take a closer look at the radial phase distribution of these satellites in Sect.~\ref{sec:DESIphasespace}, and dedicate Sect.~\ref{sec:DESIanisotropy} to analyse the presence of angular anisotropies. We finish analysing central-satellite conformity in Sect.~\ref{sec:DESIconformity}.

\subsubsection{Satellite fractions}
\label{sec:satfrac}
In this section, we compute the satellite fractions inferred by the SHAMe-SF model. We previously verified that SHAMe-SF predicted reasonable satellite fractions compared to the DESI-TNG and DESI-SAM mock DESI-ELG catalogues. We discuss further details in Appendix~\ref{sec:synsatfrac}.
Since we are fitting galaxy clustering, labelling a galaxy as a central or a satellite does not change the given spatial distribution. In some cases, this labelling depends on the criteria of the (sub)halo finder: \texttt{FoF}+\texttt{SUBFIND} allows having satellite subhaloes beyond $\rth$, while in \texttt{Rockstar} \citep{Behroozi:2013_Rockstar} these objects would be considered as centrals. This section uses both definitions to distinguish effects near the halo boundary and inside the virialised object. 

We find that $\DESIl{sat2}^{+\DESIl{satup2} }_{- \DESIl{satdown2} }$\% of the galaxies in the lowest redshift bin are satellites (or $\DESIl{sat}^{+\DESIl{satup} }_{- \DESIl{satdown} }$\% when also considering as satellites outside $\rth$). These values are bracketed by the satellite fractions of the DESI-TNG and DESI-SAM mocks (see Table~\ref{tab:mockdata}), as pointed out in Sect.~\ref{sec:DESIclus}.

\subsubsection{Radial phase-space distribution}
\label{sec:DESIphasespace}

\begin{figure*}
		\centering
		\includegraphics[width=0.48\textwidth]{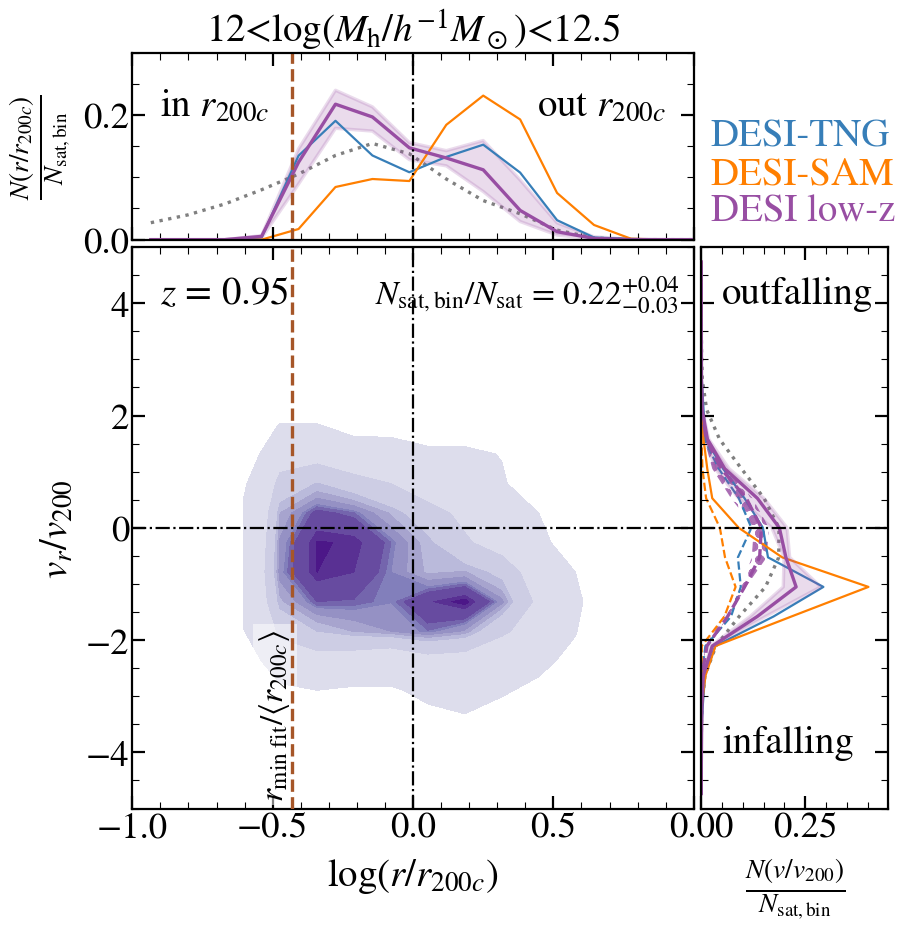}   
        \includegraphics[width=0.48\textwidth]{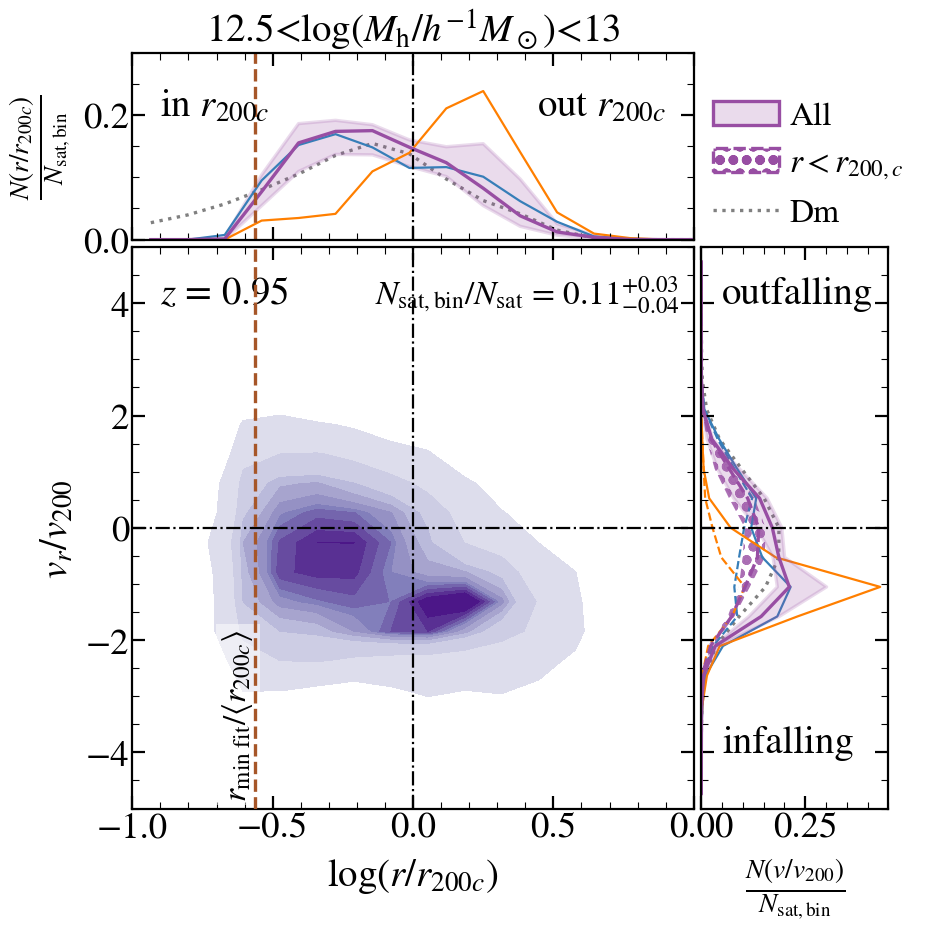}

        \includegraphics[width=0.48\textwidth]{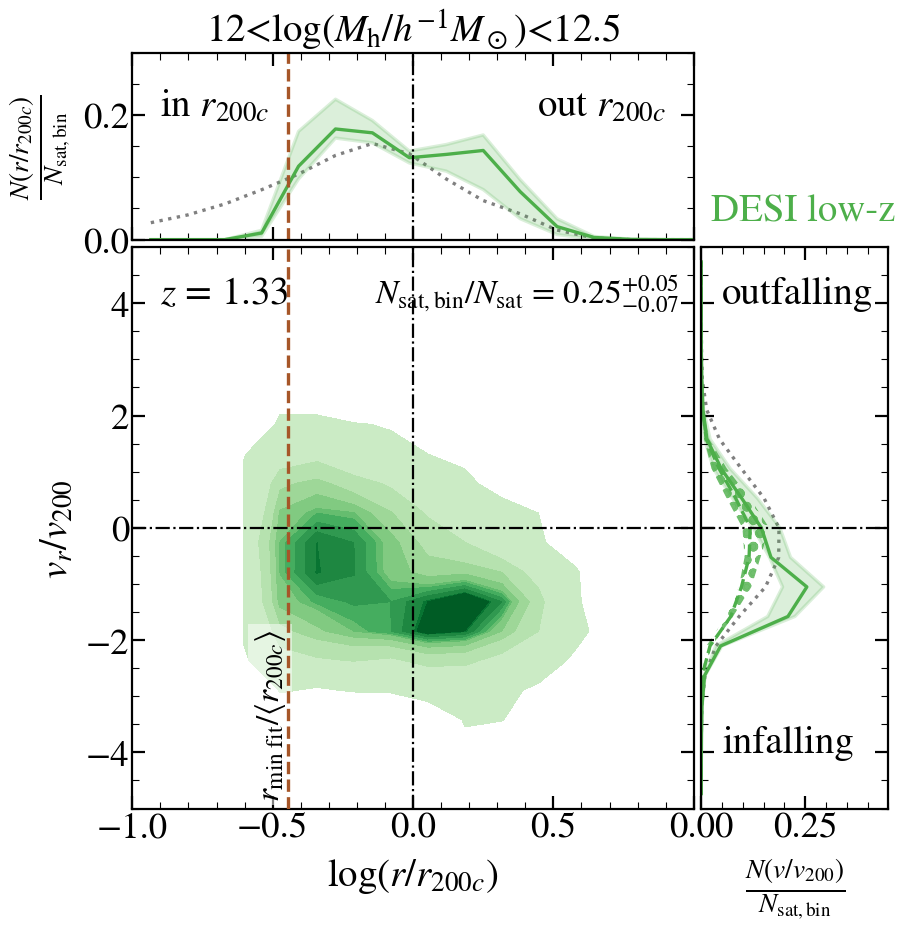}
        \includegraphics[width=0.48\textwidth]{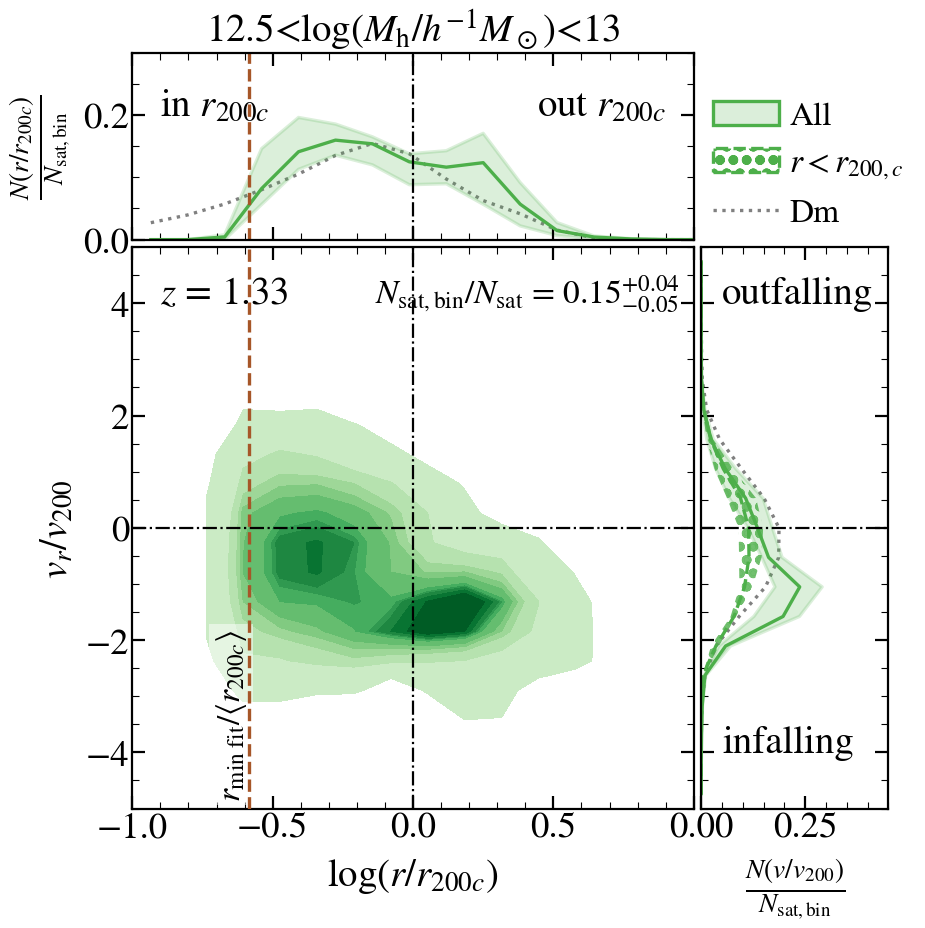} 

		\caption{Inferred phase-space distribution of ELG satellites in the low-$z$ and high-$z$ DESI samples for three halo mass bins for satellites with $r>0.1 \ \hMpc$ (the minimum scale up to where we fit the clustering). The contours mark the regions (from darker to lighter) containing 10\%, 20\%, ..., 90\%, and 99\% of the total satellites. The fraction of all the satellites with $r>0.1 \ \hMpc$ inside mass bins is indicated in each panel. The brown dashed line indicates where this division is located for each halo mass bin. Note that the axes are normalised in units of the host halo $r_{200}$ and $v_{200}$. For comparison, we show the distribution of velocities and halo-centric distances as measured in our mock DESI catalogues (orange and blue lines for DESI-SAM and DESI-TNG, respectively), and for randomly-selected dark matter particles (dotted lines). Most of the satellites are hosted by haloes with $\log(M_{\rm h} \, [\hMsun]) < 13$. For radial velocities, the dashed line and the circled-hatched region represent the median and $1\sigma$ distribution for satellites with $r<\rth$.
        We can distinguish two populations, one infalling outside $\rth$ with a negative infall velocity and a population mainly within the halo boundary with a distribution of radial velocities skewed towards negative values (also infalling).
        }
  	
    \label{fig:DESIphasespace}
\end{figure*}

In this section, we analyse the number density profiles and radial velocities of satellite ELGs. The inferences from the SHAMe-SF model for DESI ELGs are presented in Figure~\ref{fig:DESIphasespace}. We normalise the distances by the host halo $\rth$, and the infall velocity by the host $v_{200}$ (virial velocity at $\rth$). We stack the positions and velocities relative to $\rth$ and $v_{200}$ for satellite subhaloes with $r>0.1\ihMpc$ (otherwise, their positions are not constrained by our fits) for all \texttt{FoF} haloes, and compute the distributions of $r/\rth$ and $v_r/v_{200}$. We analyse the distributions by binning satellites in terms of their host halo mass. The aim is two-fold: i) analyse possible mass dependencies of the profile, and ii) avoid normalization issues originating from removing objects with $r<0.1\ihMpc$ (since this scale is close to $\rth$ for haloes with $\log(M_{\rm h}/\hMsun) < 11.7$). We mark where this scale becomes relevant for each halo mass with the brown dashed line. 

For the high-$z$ and low-$z$ samples in Figure~\ref{fig:DESIphasespace} (green and purple, respectively), we choose two halo mass bins, $\log( M_{\rm h} \, [\hMsun]) \in$ [12, 12.5], [12.5, 13]. We also analyse a higher mass bin in Appendix~\ref{app:DESIhighmassphase} (unlike the DESI-TNG/DESI-SAM samples, in DESI, we find that less than 2\% of satellites are hosted in halos with masses larger than $10^{13}\hMsun$). 

Each diagram in Figure~\ref{fig:DESIphasespace} comprises three panels. The central larger panel shows the 2D distribution of distances and radial velocities of SHAMe-SF galaxies. The 1D projected distributions are shown on the upper and right panels. For comparison, we added the grey dotted line representing the dark matter profile from the simulation. We also include the mock DESI-TNG (blue) and DESI-SAM (orange) distributions for the same halo bins for the low-$z$ sample. Appendix~\ref{sec:synphasespace} provides a detailed analysis of those samples. 

We focus first on the ELGs within $\rth$ (i.e. log(r/$\rth$) < 0).  We represent these haloes in the radial velocity histogram using the dashed line and dotted hatch region. For all mass bins and redshifts, their velocity distribution leans towards more negative values (infalling) when compared to the randomly selected particles, as pointed out by \cite{Orsi:2018}. However, we do not find a significant preference towards the outskirts of the halo compared to the randomly chosen particles dark matter particles for the two lower mass bins shown. In these mass bins, satellite ELGs seem closer to the halo centre than the underlying halo dark matter distribution. 

However, the most interesting feature comes from scales beyond $\rth$. We can distinguish between two populations, a continuation of the profile inside $\rth$ (with a distribution with infalling and outfalling velocities, but also tilted towards infalling) and a separately infalling population, clearly visible 2D histograms, where we observe two peaks in the distribution. As already analysed by \cite{Hadzhiyska:2022_onehalo} for ELGs in MillenniumTNG simulation (MTNG, \citealt{HernandezAguayo2022,Pakmor2022,Barrera2022,Kannan2022,Delgado2022,Ferlito2022,C22a}), this globally infalling population may be a product of the \texttt{FoF} identifiers, which place two haloes on the same \texttt{FoF} object. 

The relative importance of the infalling and orbiting populations for a given mass bin differs between the mock DESI samples. The quenching inside haloes of DESI-SAM is stronger than that of DESI-TNG, which causes "infalling ELGs" to be dominant with respect to the orbiting ones. This is consistent with the findings of \cite{Orsi:2018}, but our DESI inference prefers a milder quenching inside haloes, similar to what we observe in TNG-DESI. Remarkably, this implies that ELG correlation functions can be used in the future to place constraints on ram pressure and quenching inside haloes.

\subsubsection{Angular anisotropy}
\label{sec:DESIanisotropy}
\begin{figure}
		\centering
		\includegraphics[width=0.45\textwidth]{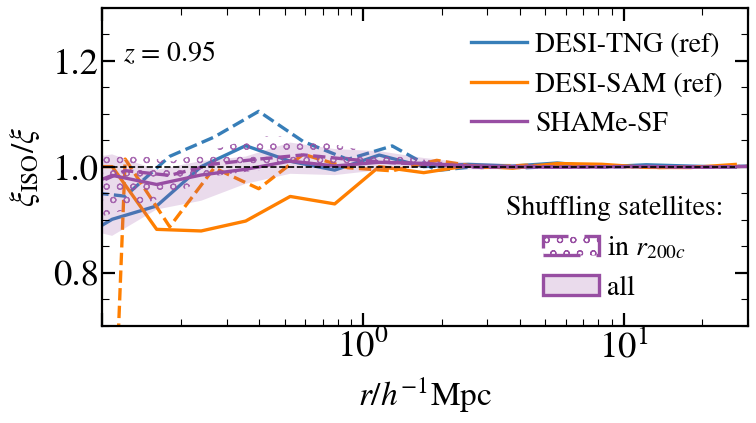}
        \includegraphics[width=0.45\textwidth]{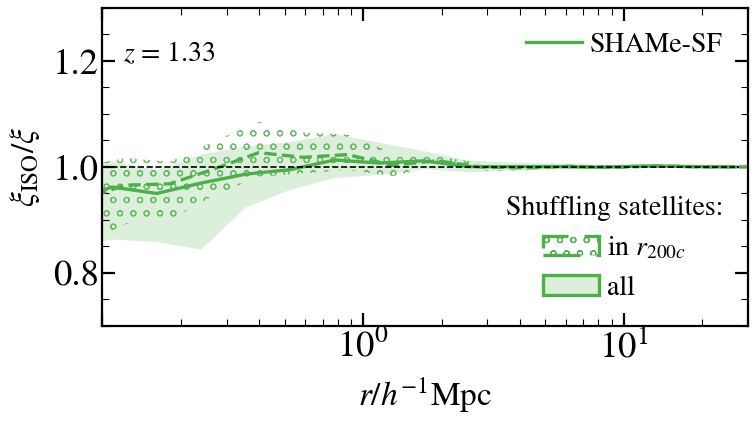}
		\caption{Inferred magnitude of satellite anisotropy in the low-$z$ and high-$z$ DESI-ELG catalogues. We display the ratio of the correlation function in the best-fit SHAMe catalogue to a version where the angular position of haloes has been randomly shuffled among each halo. Solid (dashed) lines and shaded (circle-hatched) regions indicate the mean and $1\sigma$ region after marginalisation over all SHAMe parameters when shuffling all satellites (only satellites within $\rth$). For comparison, we display the same quantity estimated in our mock DESI catalogues. }
   \label{fig:DESIiso}
\end{figure}

Even if the main focus of satellite modelling in HODs leans towards radial distributions, \cite{Hadzhiyska:2022_onehalo} highlight the importance of anisotropy within haloes in their analysis of the MTNG simulation. The authors compare the satellite distributions to different HOD prescriptions, finding an improvement in the two-point correlation function when considering non-radially symmetric satellite distributions (in their case, parameterizing the probability to assign a satellite to another satellite instead of always to the central subhalo).

Subhaloes are generally accreted via filaments, resulting in an excess distribution along the semi-major axis of haloes \citep{Yang:2006,Mezini:2024}. However, it has been found that galaxies along the semi-major axis have a higher quenching fraction \citep{Martin-Navarro:2021, Karp:2023}, which would isotropize the distribution of blue galaxies.  This difference also appears in the alignment for central red/blue galaxies on larger scales (e.g. \citealt{Rodriguez:2024}).

As in the case of assembly bias, we only explore whether this effect is detectable in our catalogues. Following \cite{Hadzhiyska:2022_onehalo}, we keep the central-satellite distance $r$ constant but randomly vary the angular distribution sampling new values of the spherical coordinates ($\phi,\theta$) and compute the ratio between the original and isotropized correlation functions. We present our inferences in Figure~\ref{fig:DESIiso} for the low-$z$ (purple) and high-$z$ (green) redshift bins. For comparison, we added the DESI-TNG and DESI-SAM results as dashed blue and orange lines for the low-$z$ bin. We show the validation of anisotropy predictions for these mock DESI samples in Appendix~\ref{sec:synanisotropy}. The dashed line and circle-hatched region represent the median and $1\sigma$ intervals when only subhaloes within $\rth$ are randomised. Solid lines display our results when considering all \texttt{SUBFIND} satellite subhaloes. 

Even if both redshift bins are consistent with an isotropic distribution on satellites, we find some deviations for the high-$z$ bin (up to 5\%).
Anisotropy can be sourced by satellite pairs or the interaction between neighbouring haloes. The percentage haloes hosting more than one satellite is $\DESIl{fhalos_multisat}^{+\DESIl{fhalos_multisatup}}_{-\DESIl{fhalos_multisatdown}}$\% and $\DESIh{fhalos_multisat}^{+\DESIh{fhalos_multisatup}}_{-\DESIh{fhalos_multisatdown}}$\% for the low-$z$ and high-$z$ bins, respectively. Pairs of satellites with small angular separations can also contribute to the higher signal on small scales observed in the data without necessarily adding central-satellite conformity. 

\subsubsection{Conformity}
\label{sec:DESIconformity}

\begin{figure*}
		\centering
        \includegraphics[width=0.95\textwidth]{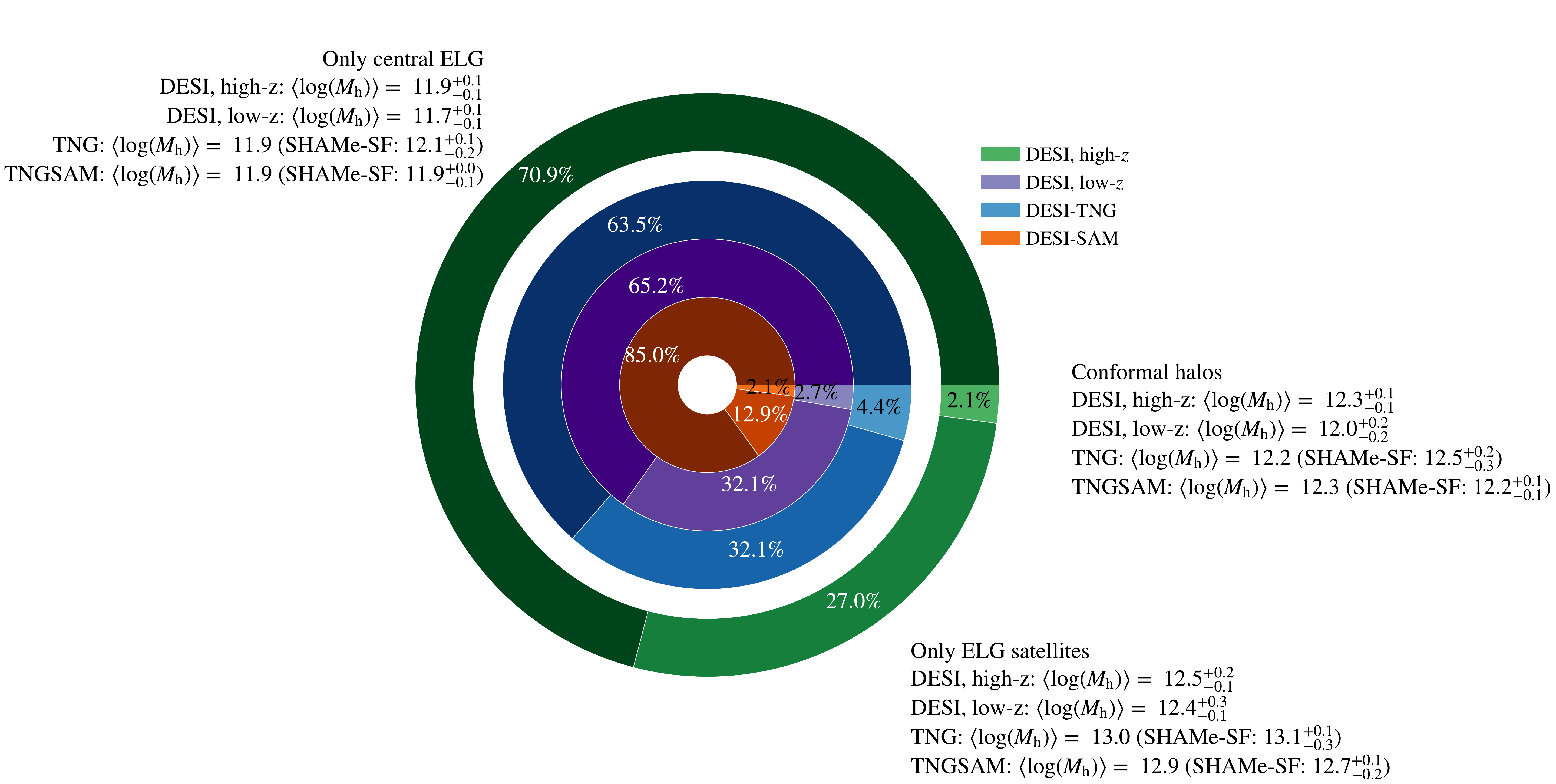}
    \caption{Classification of the occupation of the host haloes of DESI high-$z$ (green) and low-$z$ (purple) samples, as well as the DESI-TNG and DESI-SAM mock samples at $z\sim1$ (blue and orange). We distinguish between haloes containing only a central ELG (top left), only ELG satellites (bottom right) and both ELG centrals and satellites (right, conformal haloes). We also provide the average halo masses in each case. All halo masses are expressed in $\hMsun$ units. In all cases, the inference for DESI is bracketed between the DESI-TNG and DESI-SAM mock DESI catalogues.}
   \label{fig:DESIconf}
\end{figure*}

Conformity refers to correlations between the properties of nearby galaxies \citep[e.g.,][]{Weinmann:2006,Kauffmann:2013,Illustrisb,Hearin:2015,Kauffmann:2015,Bray:2016,Lacerna:2018,Calderon:2018}. One of the most common examples is star-forming conformity: galaxies tend to be quenched around massive quenched centrals. Since ELGs are generally star-forming galaxies, we would expect them to be middle mass centrals (as discussed in Section~\ref{sec:DESIHOD}), satellites near star-forming centrals or satellites in the outskirts of large and quenched galaxies. Due to the mass resolution of our simulation, our fits extend to scales of $r > 0.1\, \hMpc$. However, conformity has been detected observationally and in empirical models up to a few $\hMpc$ \citep[e.g.,][]{Lacerna:2022,Ayromlou:2023}. Since these scales fall within our range, we would expect to detect some amount of conformity in our fits. As discussed in Section~\ref{sec:compconformity}, conformity is included in DESI models as a correlation between the probability of hosting satellite ELGs if the central galaxy is also an ELG, thus central-satellite conformity. 

To analyse the central-satellite conformity in DESI's ELGs, we compute the percentages of haloes in the sample that have only a central ELG (without ELG satellites), haloes with only satellites, and haloes with both centrals and satellites (thus, haloes exhibiting central-satellite conformity). The percentages and average halo masses are shown in Figure~\ref{fig:DESIconf} for the high-$z$ and low-$z$ DESI fit inferences (green and purple), and the DESI-TNG and DESI-SAM mock samples at $z = 1$ (blue and orange). We discuss the model validation using these mock samples in Appendix~\ref{sec:synconformity}. 

From all the selected haloes, only $\DESIl{fhalos_conformal}^{+\DESIl{fhalos_conformalup}}_{-\DESIl{fhalos_conformaldown}}$\% ($\DESIh{fhalos_conformal}^{+\DESIh{fhalos_conformalup}}_{-\DESIh{fhalos_conformaldown}}$\% ) of them host both ELG central and satellites in the low-$z$ (high-$z$) sample. These quantities are bracketed by the values of our mock DESI-ELG samples for the low-$z$ bin (\TNGref{fhalos_conformal}\% for DESI-TNG and \TNGSAMref{fhalos_conformal}\% for DESI-SAM). SHAMe-SF predicts that these conformal halos have lower average halo masses compared to non-conformal satellites $\log( M_{\rm h}/\hMsun) =\DESIl{m200cenwithsat}^{+\DESIl{m200cenwithsatup}}_{-\DESIl{m200cenwithsatdown}}$ ($\DESIh{m200cenwithsat}^{+\DESIh{m200cenwithsatup}}_{-\DESIh{m200cenwithsatdown}} $ for the high-$z$ sample)
.

We now explore whether this is caused by conformity. To do so, we shuffle satellites hosted by haloes within bins of 0.1 dex in halo mass. If the percentage of conformal haloes remains constant, then having an ELG central hosting ELG satellites would be similar within that halo mass bin. For the low-$z$ and high-$z$ bins, respectively, we obtain 
$\DESIl{shuffled_fhalos_conformal}^{+\DESIl{shuffled_fhalos_conformalup}}_{-\DESIl{shuffled_fhalos_conformaldown}}$\%
and $\DESIh{shuffled_fhalos_conformal}^{+\DESIh{shuffled_fhalos_conformalup}}_{-\DESIh{shuffled_fhalos_conformaldown}}$\%. Both percentages are reduced by more than half: even if the number of conformal ELG haloes is small in our sample, the type of central slightly conditions the presence of ELG satellites. This confirms the presence of central-satellite conformity in our ELG sample. The behaviour is similar for DESI-TNG and DESI-SAM mock DESI samples in the low-$z$ bin, which is discussed in Appendix~\ref{sec:synconformity}.

\section{Comparison with other analyses}
\label{sec:DESIothermodels}

In this section, we compare our inferences from Section~\ref{sec:DESIinference} with the findings of other analyses of DESI ELGs. Specifically, we explore the halo occupation number (Sect.~\ref{sec:compHOD}), assembly bias (Sect.~\ref{sec:compassemblybias}), and satellite fractions (Sect.~\ref{sec:comsatfrac}), radial phase space (Sect.~\ref{sec:compphasespace}) and conformity (Sect.~\ref{sec:compconformity}).

\begin{figure*}
		\centering
        \includegraphics[width=0.95\textwidth]{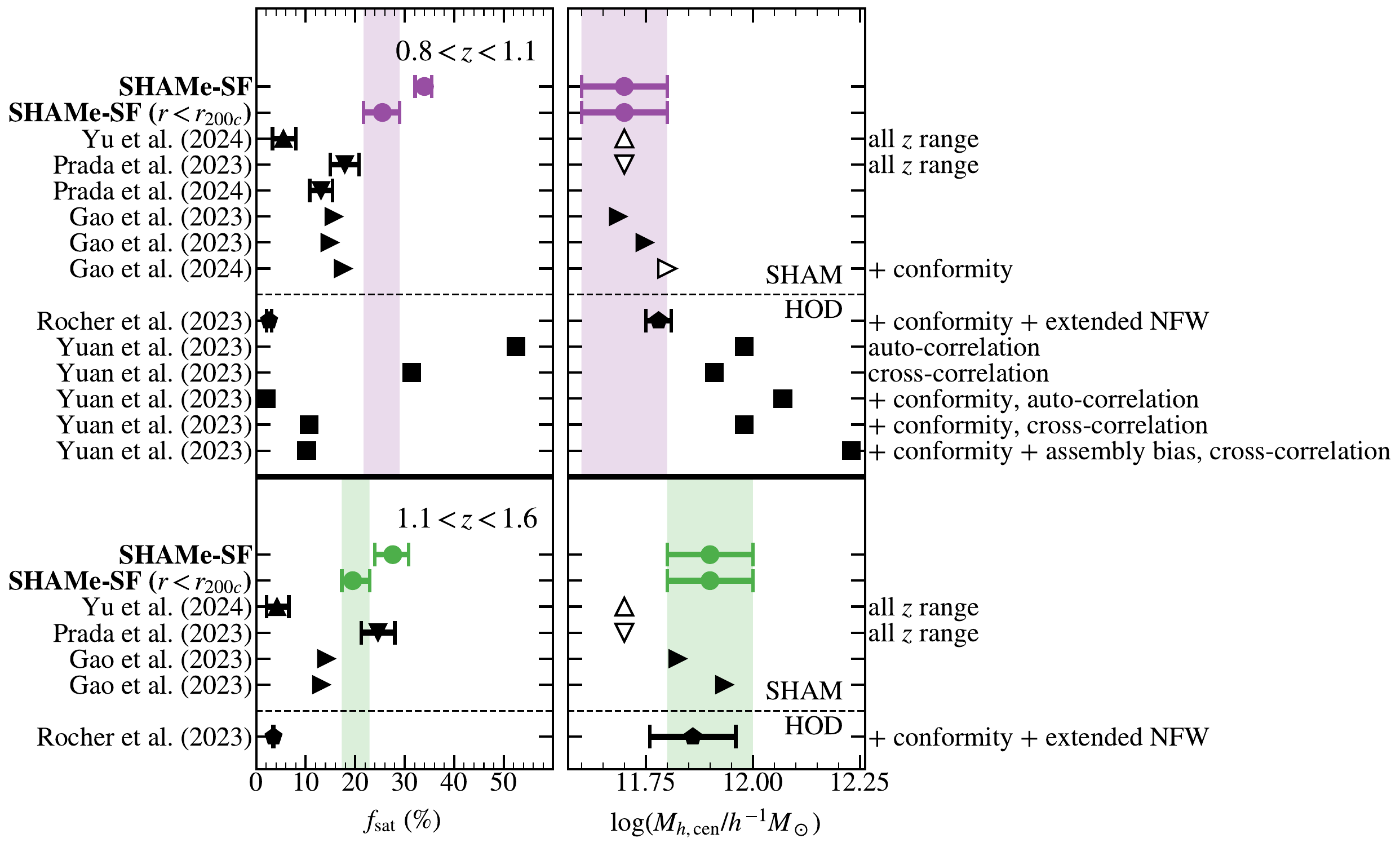}
    \caption{Satellite fractions (left) and average centrals halo mass (right) for different analyses of DESI-ELGs \citep{Rocher2023:DESI,Yu:2024_DESISHAM,Prada:2023,Gao2:2023,Gao3:2024,Yuan:2024_conformity} for the two redshift intervals analysed in this work ($0.8<z<1.1$, purple; $1.1<z<1.6$, green). The black dashed line separates HODs and SHAMs within each redshift bin. The satellite fractions are computed for subhalos as $N_{\rm{satellites}}/(N_{\rm{centrals}}+N_{\rm{satellites}})$. When the average halo mass for centrals is not provided, we add the halo mass with the highest probability (white symbols). In the case of \cite{Yu:2024_DESISHAM} and \cite{Prada:2023}, average halo occupations are computed for $0.8<z<1.6$.}
\label{fig:DESIsatandmass}
\end{figure*}

\subsection{Halo occupation number}
\label{sec:compHOD}
We begin by comparing the average mass for centrals (or mass with the maximum central probability if the average halo mass is not provided) to get a first impression of how our results compare to other DESI ELG analyses. As shown in the right column of Figure~\ref{fig:DESIsatandmass}, our results are compatible within 1$-\sigma$ with all analyses except for \cite{Yuan:2024_conformity} when including conformity and assembly bias simultaneously, where the main difference comes from the latter. In all cases, the average mass of halos hosting centrals increases with redshift.

We can also compare the shape of the central occupation. For HODs, we can directly compare the parametrization of the central occupation. Extended parametrizations for centrals in HODs are Gaussians (sometimes including decreasing power laws), error functions and log-normal distributions. All models predict a central occupation that increases to a peak probability and decreases afterwards except for \cite{Gao2:2023} and \cite{Gao3:2024}, which measure a second increase for higher halo masses on their SHAM. The width of the probability distribution also changes between models. SHAMe-SF, \cite{Rocher2023:DESI,Prada:2023} and \cite{Yuan:2024_conformity} (the latter when fitting auto correlations including only ELG conformity) find that the probability of finding central ELGs on halos above $10^{13}\hMpc$ solar masses is negligible ($<10^{-3}$), whereas \cite{Yu:2023} and \cite{Yuan:2024_conformity} (all other fits) find broader distributions.

\subsection{Assembly bias}
\label{sec:compassemblybias}
In the case of SHAMs, assembly bias appears naturally  \citep{ChavesMontero:2016} and can be further tuned using explicit parameters (e.g. \citealt{Contreras:2021shame}). However, it must be included explicitly in HODs since their functional form depends only on halo mass \citep{Hearin:2016}. 
Assembly bias can be introduced using many different secondary properties. The most extended ones are concentration and tracers of the small/large scale environment \citep{Xu:2021a,Yuan:2021a,Hadzhiyska:2022_AbacusHOD,Beltz-Mohrmann:2023}

In the case of DESI ELGs, \cite{Rocher2023:DESI} tested whether adding assembly bias through halo concentration, local halo density or local halo density anisotropy enhances the fit quality. Using the parametrization described in \cite{Hadzhiyska:2022}, the authors found little to no improvement after adding this effect. On the contrary, using the same parametrization and shear as the secondary property, \cite{Yuan:2024_conformity} found that including assembly bias to the Vanilla HOD improves the fit to the ELG autocorrelation and its cross-correlation with LRGs. The authors leave the quantification of this effect for a future analysis. In our case, we measured an assembly bias signal larger than 5\% (and non-compatible with 0 within the $1\sigma$ region) for both redshift bins.

\subsection{ELGs as satellites}
\subsubsection{Satellite fractions}
\label{sec:comsatfrac}

In this section, we compare the satellite fraction inferred from SHAMe-SF fits to the predictions of DESI 1\% analyses. All satellite fractions are summarized in the left column of Figure~\ref{fig:DESIsatandmass}. Since all of these works use \texttt{Rockstar} as their halo finder, the most direct comparison implies considering satellites only objects with $r<\rth$ where $r$ is the central-satellite distance, for which we found a satellite fraction of \DESIl{sat2}$^{+ \DESIl{satup2} }_{- \DESIl{satdown2} }$\%. 

The range of satellite fractions predicted by HODs varies drastically based on their assumptions. \cite{Yuan:2024_conformity} found that 50\% of ELGs were satellites when fitting only their projected correlation function and 30\% when adding the cross-correlation with LRGs. The satellite fraction drops to 2\% (10\% when including cross-correlations) if conformity is included. This change is related to how conformity is modelled and where satellites are placed. \cite{Rocher2023:DESI}, impose strict conformity, linking the presence of ELG satellites to having an ELG central. The authors predict satellite fractions of 2.6$\pm$0.5\% for the low-z sample and 3.5$\pm$0.1\%, similar to our fraction of conformal halos (see Fig~\ref{fig:DESIconf}).

As for SHAMs, \cite{Gao2:2023,Gao3:2024} fit scales comparable to \cite{Rocher2023:DESI} and \cite{Yuan:2024_conformity} for cross-correlation between different stellar mass bins. They measure satellite fractions of 15.7\% and 17.6\% (when including conformity), which are more similar to our inferred values. \cite{Prada:2023} and \cite{Yu:2024_DESISHAM} fit ELG clustering up to 4$\hMpc$, but find completely different satellite fractions: $17.9\pm 2.9$\% and $5.5^{+2.5}_{-2.2}$\% respectively. The main difference in the satellite fraction originates from how the subhaloes are chosen within the SHAM techniques (removing high $\vpeak$ haloes in \citeauthor{Yu:2024_DESISHAM}, or using a Gaussian selection function in \citeauthor{Prada:2023}). The first term of our functional form (eq.~\ref{eq:model}) is more similar to the choice of \citeauthor{Prada:2023}, which could explain the similarity in the satellite fractions. The satellites are downsampled in both models regardless of their host halo mass.

\subsubsection{Satellite radial phase space}
\label{sec:compphasespace}
Beyond modelling the number of satellites, halo-based models must include prescriptions for the position and velocities of satellites. In the case of subhalo-based models, the problem is reduced by adequately selecting the subhaloes that would host ELG satellites. 

For satellite positions, HOD typical choices ranged from sampling from an assumed radial profile (likely an NFW, \citealt{NFW:1995}), or randomly choosing the position of a dark matter particle within the halo. However, previous studies by \cite{Orsi:2018} on a SAM point out that ELGs are typically located at the outskirts of haloes rather than following an NFW profile. This is included on HODs using either a modified NFW profile or adding additional probability functions to select the dark matter particles \citep{Avila:2020, Hadzhiyska:2022_onehalo, Reyes-Peraza:2024}. For DESI-HODs, \cite{Rocher2023:DESI} find a significant improvement when sampling satellite positions from a scaled NFW profile with an additional exponential term for larger central-satellite distances. This extra term allows satellite placement beyond $\rth$, which would not be considered satellites by \texttt{Rockstar}. In the case of \cite{Yuan:2024_conformity}, the chosen method was randomly sampling from the dark matter particle distribution. Radial profiles inferred from our SHAMe-SF fits are more similar to \cite{Rocher2023:DESI}'s parametrization, and satellite ELGs only follow the dark matter distribution in the low-$z$ sample in the mass range $M_{\rm{h}}>10^{12.5}\hMsun$ (see Fig~\ref{fig:DESIphasespace} and Appendix~\ref{app:DESIhighmassphase}).

Likewise, it is necessary to assign velocities to the satellites. Typical approaches range from sampling from a given distribution \citep{Hadzhiyska:2022_onehalo}, taking (or rescaling) the velocity of a dark matter particle \citep{Rocher2023:DESI,Yuan:2024_conformity}, and sometimes adding an extra infall component term to a sampled velocity (\citealt{Avila:2020,Rocher2023:DESI}). SHAMe-SF predicts velocity distributions tilted towards negative radial velocities for all mass bins, not following the underlying dark matter distribution (see Figure~\ref{fig:DESIphasespace}).

\subsubsection{Conformity}
\label{sec:compconformity}
Some HODs and SHAM-based models applied to DESI's data include conformity to improve the fitting to small scales \citep{Gao3:2024} (for $r < 0.3 \, \ihMpc$) or the transition between the one-halo and the two-halo term \citep{Rocher2023:DESI, Yuan:2024_conformity}. Conformity is modelled as the probability of hosting ELG satellites depending on whether the central galaxy is also an ELG. 

As discussed in Section~\ref{sec:DESIconformity}, our fits predict that less than 5\% of halos host ELG centrals and ELG satellites simultaneously, and their typical masses are $\log( M_{\rm h}/\hMsun) =  \DESIl{m200cenwithsat}^{+\DESIl{m200cenwithsatup}}_{-\DESIl{m200cenwithsatdown}}$ and  $\DESIh{m200cenwithsat}^{+\DESIh{m200cenwithsatup}}_{-\DESIh{m200cenwithsatdown}} $, for the low-$z$ high-$z$ bins, respectively (see Figure~\ref{fig:DESIconf}). Halos hosting only ELG satellites have, on average, higher halo masses: $\log( M_{\rm h}/\hMsun) \sim  12.5$. This tendency in the average halo masses would also appear for \cite{Yuan:2024_conformity} and \cite{Gao3:2024}'s analyses since the probabilities of conformal and non-conformal satellites have an offset in host halo masses. However, the number of conformal satellites (compared to satellites hosted by non-ELGs) is larger for \cite{Gao3:2024}, as is shown in the halo occupation number they measure. For HODs that include conformity, the resultant occupation for ELG satellites will not be exactly the one parametrised by the power law. This is the case of \cite{Rocher2023:DESI} and \cite{Yuan:2024_conformity}. For the former, imposing strict conformity will remove all satellites when a central is absent regardless of the $\langle N_{\rm{sat}} \rangle$ probability given by the HOD. Conformity also affects the shape of the central HOD, which drives centrals to occupy higher halo masses to be able to host satellites.

We also look at the scenario presented in \cite{Favole:2016} for g-selected ELGs at  $z\sim 0.8$, where they distinguish between ELG centrals (with masses around $10^{12}\, \hMsun$) and ELG satellites hosted by quiescent centrals whose halo masses were larger than $10^{13}\,\hMsun$. As mentioned, we find smaller halo masses for ELGs in the lower mass bin.  \cite{Favole:2016} also find that some of the most massive halos would host more than one satellite, which happens for 1.3\% of the halos. These values align with our findings from Section~\ref{sec:DESIanisotropy} when analysing anisotropy, where we found $\DESIl{fhalos_multisat}^{+\DESIl{fhalos_multisatup}}_{-\DESIl{fhalos_multisatdown}}$\% and $\DESIh{fhalos_multisat}^{+\DESIh{fhalos_multisatup}}_{-\DESIh{fhalos_multisatdown}}$ of halos hosting more than one ELG satellite for the low-$z$ and high-$z$ samples, respectively. A combination of central-satellite conformity pairs and satellite-satellite pairs can source high clustering on smaller scales. 

\section{Summary and conclusions}
\label{sec:conclusions}
In this work, we analysed the galaxy-halo connection of DESI ELG galaxies between $0.8<z<1.1$ (low-$z$) and $1.1<z<1.6$. Our inferences are derived after fitting the galaxy clustering of these galaxies using SHAMe-SF, an extension to subhalo abundance matching specially developed for star-forming galaxies. The SHAMe-SF model can reproduce the projected correlation function and monopole and quadrupole of the correlation function in redshift space for both redshift bins between scales of 0.1 and 30 $\ihMpc$.

After fitting the DESI-ELG clustering measurements (Sect.~\ref{sec:DESIclus}), we populate a gravity-only simulation using the best-fit parameters and infer the galaxy halo connection. Our findings for DESI ELGs can be summarized as follows:
\begin{itemize}
    \item ELGs inhabit haloes of average mass $\log( M_{\rm h} \, [\hMsun]) = \DESIl{m200cen2}\,\pm\,\DESIl{m200cenerr2}$ ($\DESIh{m200cen2}\,\pm\,\DESIh{m200cenerr2}$) when they are central and $\DESIl{m200sat2}\,\pm\,\DESIl{m200saterr2}$ ($\DESIh{m200sat2}\,\pm\,\DESIh{m200saterr2}$) when they are satellites in the low-z (high-z) redshift bin, which match the values found by other DESI analyses (Sect.~\ref{sec:compHOD}). 
    \item We detect a non-zero signal of assembly bias for the high$-z$ (10\%) and low$-z$ ($\sim$15\%) samples. In the latter case, we also find that the signal is scale-dependent (Sect.~\ref{sec:DESIassembly}). 
    \item In the low$-z$ (high$-z$) sample, \DESIl{sat} $^{+ \DESIl{satup} }_{- \DESIl{satdown} }$\% (\DESIh{sat} $^{+ \DESIh{satup} }_{- \DESIh{satdown} }$\%) of the galaxies are satellites if considering \texttt{FoF}+\texttt{SUBFIND} central/satellite definition, which includes galaxies outside $\rth$. This fraction decreases to \DESIl{sat2}$^{+ \DESIl{satup2} }_{- \DESIl{satdown2} }$\% (\DESIh{sat2}$^{+ \DESIh{satup2} }_{- \DESIh{satdown2} }$\%) when counting as satellites only the galaxies within $\rth$. Both estimations are generally higher than estimations of other studies with different galaxy-halo connection models (see Sect.~\ref{sec:satfrac} and Sect.~\ref{sec:comsatfrac}). 
    \item When analysing the phase space of satellite subhaloes, we find two distinct populations of satellites. The main population extends from the inner part of the halo up to beyond $\rth$. Their velocity distribution leans towards infalling. We also find a significant number of satellites outside $\rth$ with negative radial velocity (in-falling) for all halo mass bins. As discussed by \cite{Hadzhiyska:2022_onehalo}, they can be subhaloes misclassified by the halo finder (Sect.~\ref{sec:DESIphasespace}).
    \item In both redshift bins, ELG satellites seem to be distributed almost isotropically within their haloes. We measure signatures of anisotropy in the galaxy clustering below 10\% (Sect.~\ref{sec:DESIanisotropy}).
    \item Most satellites are hosted by non-ELG centrals  (Sect.~\ref{sec:DESIconformity}). Only $\DESIl{fhalos_conformal}^{+\DESIl{fhalos_conformalup}}_{-\DESIl{fhalos_conformaldown}}$\% ($\DESIh{fhalos_conformal}^{+\DESIh{fhalos_conformalup}}_{-\DESIh{fhalos_conformaldown}}$\% ) of the haloes in the sample host both centrals and satellites. The typical halo masses of haloes hosting both ELG centrals and satellites are lower than when they only host satellites.
    
\end{itemize}

The capability of the model to accurately reproduce these statistics was validated in Appendix~\ref{app:TNG} using DESI-ELG samples from the TNG300-1 simulation and the \texttt{L-Galaxies} semi-analytical model run on TNG300-1-Dark merger trees \citep{LGalaxies_Ayromlou:2021}. We showed these tests and compared our mock samples with similar available catalogues in the same Appendix section. 

We highlight the ability of the SHAMe-SF model to infer the radial phase-space distribution of satellites within different mass bins. Given the different predictions between TNG300 and LGalaxies, we can use the SHAMe-SF model to provide further constraints on the satellite quenching and stripping modelling on hydrodynamical simulations and semi-analytical models.
After validating the SHAMe-SF model and confirming its capability to produce constraints on the galaxy-halo connection, we aim to use it to obtain cosmological constraints. 

\begin{acknowledgements}
    We thank the DESI collaboration, particularly Antoine Rocher, for making the clustering measurements of DESI 1\% available.
    SOM thanks the hospitality of Andrew Hearin, Georgios Zacharegkas and the rest of the CPAC Group at Argonne National Laboratory, where part of this work was carried out, and Mary Gerhardinger for useful discussions during that period.
    SOM is funded by the Spanish Ministry of Science and Innovation under grant number PRE2020-095788. 
    SC acknowledges the support of the `Juan de la Cierva Incorporac\'ion' fellowship (IJC2020-045705-I). 
    REA acknowledges support from project PID2021-128338NB-I00 from the Spanish Ministry of Science and support from the European Research Executive Agency HORIZON-MSCA-2021-SE-01 Research and Innovation Programme under the Marie Skłodowska-Curie grant agreement number 101086388 (LACEGAL). 
    JCM, acknowledges support from the European Union’s Horizon Europe research and innovation programme (COSMO-LYA, grant agreement 101044612). IFAE is partially funded by the CERCA program of the Generalitat de Catalunya.
    The authors also acknowledge the computer resources at MareNostrum and the technical support provided by Barcelona Supercomputing Center (RES-AECT-2024-2-0022)
    Catalogues will be public upon publication of the paper.

\end{acknowledgements}

%
%

\bibliographystyle{aa} 
\bibliography{aa.bib} 

\begin{appendix} 
\section{SHAMe-SF model validation}
\label{app:TNG}

Hydrodynamical simulations and semi-analytical models (SAMs) provide a unique opportunity to test empirical models such as HODs and SHAMs against different plausible realizations of the Universe before applying them to observational data. Since both hydrodynamical simulations and SAMs contain all the information on the galaxy-halo connection, we can use them to validate whether the assumptions made on SHAMs and HODs hold after fitting the galaxy clustering. We can also analyse different predictions of each model that are not constrained directly by the model parameters, such as galaxy assembly bias or the satellite distribution, comparing them with the "ground truth" of the simulation.

In \cite{SOM2024}, we verified that the SHAMe-SF model could reproduce the galaxy clustering of samples selected by SFR or colour cuts (without considering incompleteness). However, the selection criteria for ELGs are more complex. Through this appendix, we validate the behaviour of the SHAMe-SF model when applied to mock DESI-ELG samples (see Section~\ref{sec:thColours} for selection criteria) and specifically how accurately we can predict their galaxy-(sub)halo connection after fitting the clustering. We assumed that the redshift bin ($0.8 < z < 1.1$) is narrow enough to exclude redshift evolution, and chose our galaxies at redshift $z = 1$ from TNG300-1 and \texttt{L-Galaxies}. After applying the stellar mass and SFR cuts, the resultant number densities were $\bar{n} = 4.25\times10^{-3}\,\ihMpcC$ (DESI-TNG) and $\bar{n} = 2.73\times10^{-3}\,\ihMpcC$ (DESI-SAM). 

We dedicate Section~\ref{sec:genclus} to the galaxy clustering fits, while Section~\ref{sec:synpred} includes all the validation of the galaxy-(sub)halo connection inferences.

\subsection{Galaxy clustering}
\label{sec:genclus}
Before venturing into fitting DESI ELG's galaxy clustering, we ensured that SHAMe-SF can reproduce the galaxy clustering of both mock DESI samples. Through this section, we present two auxiliary simulations used to estimate the covariance matrices (\ref{sec:auxsims}), discuss the volume correction introduced for large scales, given the difference in size between TNG300 (Sect.~\ref{app:box}) and the simulation used to create the emulator (and DESI's survey effective volume), the covariance matrices (Sect.~\ref{sec:syncv}), and finally present our fits for DESI-TNG and DESI-SAM mock ELG samples (Sect.~\ref{sec:synfits}).

\subsubsection{Auxiliary simulations}
\label{sec:auxsims}
In addition to the simulations used to create the emulator and the posterior predictive distributions described in Sect.~\ref{sec:thBaccoPlanck}, we used two gravity-only simulations with lower resolution: TNG-mimic and LTNG. 
    
The TNG-mimic simulation has the same volume and initial conditions as TNG300-1-Dark, but employs only 625$^3$ particles (equivalent resolution to TNG300-3-Dark). LTNG also has Gaussian initial conditions and resolution, but its volume is 27 times larger. Both simulations were run with an updated version the \texttt{L-Gadget3} code \citep{Angulo:2012,Springel:2005} used on the other BACCO simulations.

\subsubsection{Box size correction}
\label{app:box}
Our DESI ELG mocks were built using TNG300-1, a relatively small simulation box, $L=205\,\hMpc$. This implies that the correlation functions on large scales are systematically underestimated due to the lack of long wavelength modes. To correct for this, we proceed as follows.

First, we find the set of SHAMe-SF parameters that best fit the clustering of either DESI-TNG or DESI-SAM mocks as described in Section~\ref{sec:clustering} for the monopole and the quadrupole (since we are using $\pi_{\rm max} = 40 \ihMpc$ on $w_{\rm{p}}$). We only consider separations below $r = 7\,\hMpc$ to ensure the finite box size does not affect the measurements. Then, we use those parameters to populate TNG-mimic and our $1\,\ihGpc$ simulation, described in Sect.~\ref{sec:thBaccoPlanck}.  We then correct the clustering measurements as:
\begin{equation}
    \overline{\vec{\xi}} \rightarrow \overline{\vec{\xi}} + (\overline{\vec{\xi}}_{1000} - \overline{\vec{\xi}}_{200})
\end{equation}

\noindent where $\overline{\vec{\xi}} = \{w_p,\xi_{\ell = 0}, \xi_{\ell=2}\}$ and the subscripts 1000 and 200 refer to quantities computed in TNG-mimic and the $1000\,\hMpc$ simulations, respectively. In Figure~\ref{fig:clusteringTNG}, we show the clustering measurements before (dashed) and after (points with errorbars) this correction (blue and orange for DESI-TNG and DESI-SAM, respectively).

\subsubsection{Covariance matrices}
\label{sec:syncv}
Given the small volume of TNG300 and the low number density of our target sample, it is not possible to have a robust estimator of the covariance matrix, especially the non-diagonal terms. We can surpass this using a larger simulation to compute the correlation matrix, and then re-scale it using the diagonal term from $3^3$ jackknife samples \citep{Zehavi:2002, Norberg:2009} directly over the mock DESI samples. 

Using the method described in Section~\ref{sec:clustering}, we obtain a preliminary fit using SHAMe-SF for each of the mock DESI catalogues using the diagonal $C_v$. Using the best-fit parameters, we populate LTNG (Appendix~\ref{sec:auxsims}), compute the correlation matrix using $9^3$ JN divisions, and re-scale them using the previously computed diagonal elements. As in the case of DESI fits, we add to the diagonal the contribution of the emulator error. 

\subsubsection{DESI-TNG and DESI-SAM clustering}
\label{sec:synfits}
\begin{figure*}
		\centering
        \includegraphics[width=0.96\textwidth]{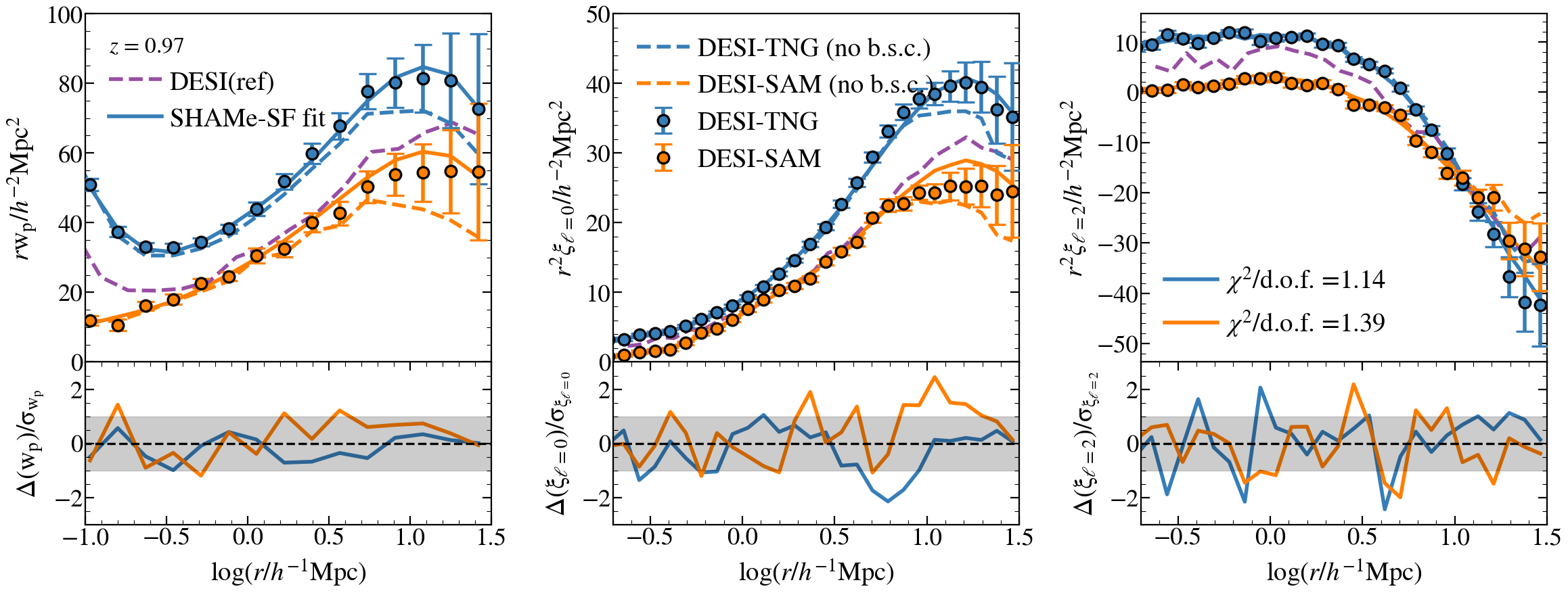}

		\caption{Projected correlation function ($w_p$), monopole and quadrupole of the DESI-TNG (blue error bars) and the DESI-SAM  (orange error bars) ELGs at $z = 1$ defined using the selection criteria explained in Sections~\ref{sec:thColours}. The solid lines show the corresponding best fit using the SHAMe-SF model. The lower panel shows the relative difference between the fits and the data normalised by the uncertainty on the measurement, with the grey shaded region indicating $1\sigma$. The blue and orange dashed lines show the measurements before the box-size correction (b.s.c.) introduced in Appendix~\ref{app:box}. The purple dashed line displays the measurements for DESI ELGs at $0.8<z<1.1$ \citep{Rocher2023:DESI} fitted in Figure~\ref{fig:clusteringDESI}.
        }	
\label{fig:clusteringTNG}
\end{figure*}

Since clustering statistics are invariant under random downsamplings to the data, we use the non-downsampled measurements of $w_p$ and $\xi_{\ell = 0,2}$. We proceed as described in Section~\ref{sec:clustering} to fit the galaxy clustering of the DESI-TNG and DESI-SAM mock samples. The best-fit results are shown in Figure~\ref{fig:clusteringTNG} for DESI-TNG (blue) and DESI-SAM (orange). As in Figure~\ref{fig:clusteringDESI}, we draw our fits using solid lines. The dashed lines represent DESI ELGs (purple) and the clustering of the DESI mocks before applying the box-size correction. We include the value of the reduced $\chi^2$  in the right panel for each sample. SHAMe-SF is able to reproduce the galaxy clustering in both cases. 

Even if the samples were defined using similar selection criteria, their behaviours differ both on large and small scales. Given the difference on the small scales, we expect DESI-SAM to have fewer satellites than DESI-TNG. Furthermore, the upturn in the projected correlation function of DESI-TNG for scales below 0.3$\hMpc$ is similar to that found for DESI's ELGs, appearing at even smaller scales, which is not found in DESI-SAM. Two possible sources are central-satellite pairs (that would manifest on the satellite distributions of conformal galaxies) or satellite-satellite pairs (producing angular anisotropies in the satellite distribution), both related to the satellite fraction of the sample.

\subsection{DESI-TNG(SAM) galaxy-subhalo connection}
\label{sec:synpred}
As described in Section~\ref{sec:DESIinference}, we compute the posterior predictive distributions for assembly bias, halo occupation number and satellite statistics. We repopulate the Planck 512$\hMpc$ simulation (Sect.~\ref{sec:thBaccoPlanck}) using 100 random elements of the MCMC chains. In this case, we compare them with the true value computed from DESI-TNG and DESI-SAM mock galaxy samples. We average the measurements over 10 random downsamplings with the same number density of the DESI low-$z$ sample ($\bar{n} = 1.047\times10^{-3}\,\ihMpcC$).

We compare our inferences with DESI-ELG mocks from other works using different selection criteria in simulations and semi-analytical models. Specifically, we compare with samples build using DESI colour selection on TNG300 from \cite{Hadzhiyska:2021} ($z = 0.8$ and $z = 1$) and \cite{Yuan:2022b} ($z\sim0.8$), $M_*$-sSFR-selected samples from the MillenniumTNG simulation at $z = 1$ \citep[][, for a number density two times larger]{Hadzhiyska:2022,Hadzhiyska:2022_onehalo}, and with the colour and [OII]-selected sample from \cite{GonzalezPerez:2018, Gonzalez-Perez:2020} in the \textsc{Galform} semi-analytical model\footnote{\textsc{Galform}: \cite{Cole:2000,Baugh:2005,Bower:2006,Lagos:2011a,Lagos:2013,Griffin:2019,Lacey:2016,GonzalezPerez:2018}}.

All the quantities computed (for the DESI mock samples and the SHAMe-SF predictions) can be found in Table~\ref{tab:mockdata}. We tested that the inferences were consistent before and after applying the box size correction.

 \subsubsection{Assembly bias}
 \begin{figure}
		\centering      \includegraphics[width=0.45\textwidth]{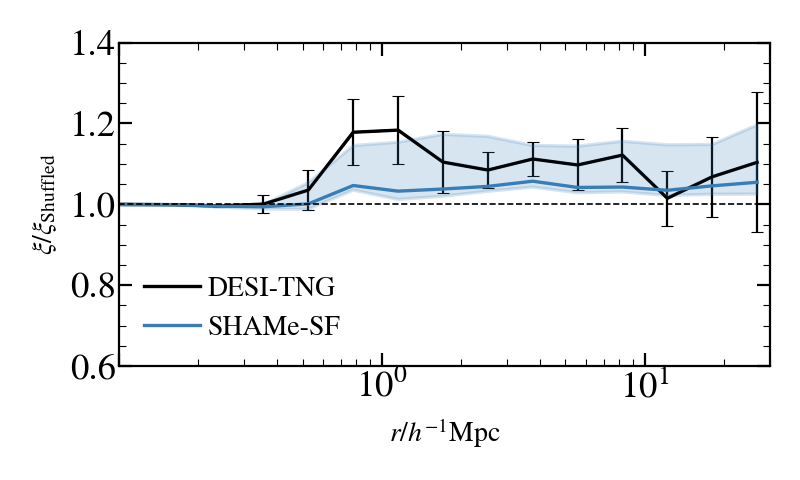}
        \includegraphics[width=0.45\textwidth]{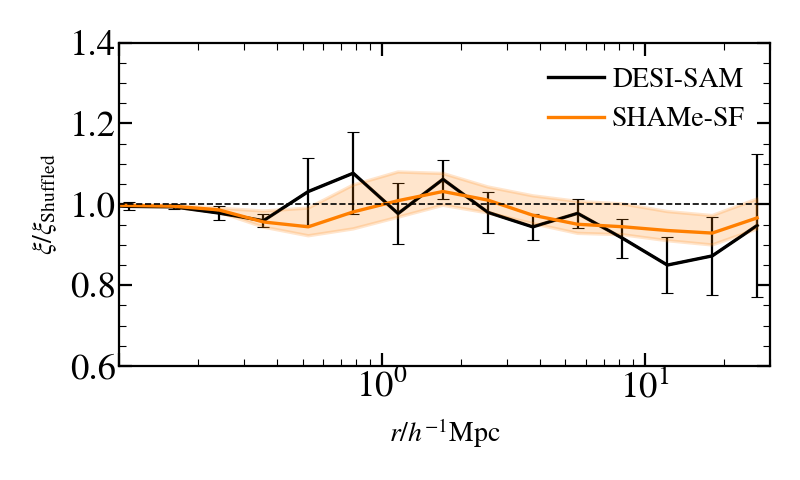}
		\caption{Same as Figure~\ref{fig:DESIassemblybias} for DESI-TNG (top, blue) and DESI-SAM (bottom, orange) mock DESI samples. The black error bars mark the measurements from the simulation/SAM, while the shaded regions indicate the predictions from the SHAMe-SF model. We can reproduce the assembly bias signal in both cases. Note the scale dependency of DESI-SAM.}	
  \label{fig:mockassemblybias}
\end{figure}
 \label{sec:synassembly}
To test whether SHAMe-SF can reproduce the dependency of galaxy clustering on properties beyond halo mass of the mock DESI samples, we follow the procedure described in Section~\ref{sec:DESIassembly}. We show the results for DESI-TNG and DESI-SAM in Figure~\ref{fig:mockassemblybias}. In each case, the prediction of the hydrodynamical simulation (or SAM) is shown with the solid black line. The error bars indicate the $1\sigma$ deviation over 10 realisations. The shaded regions in blue (for DESI-TNG) and orange (for DESI-SAM) regions show the $1\sigma$ SHAMe-SF predictions after fitting the galaxy clustering.

We find different assembly bias signatures in both samples, and good agreement between the simulation/SAM and the prediction of SHAMe-SF. This shows the potential of SHAMe-SF to constrain galaxy assembly bias from galaxy clustering. 

For DESI-TNG, the signal does not show a clear scale dependence and has a value of around 10\%. For DESI-SAM, we measured an assembly bias closer to 0, but it shows scale dependence between $\sim$5\% leaning towards negative values for larger scales. These values are in agreement with other previous works on TNG \citep{Hadzhiyska:2021}, MilleniunTNG \citep{Hadzhiyska:2022} and SAMs \cite{Gonzalez-Perez:2020}.

\subsubsection{HOD}
 \label{sec:synHOD}

\begin{figure}
		\centering
        \includegraphics[width=0.42\textwidth]{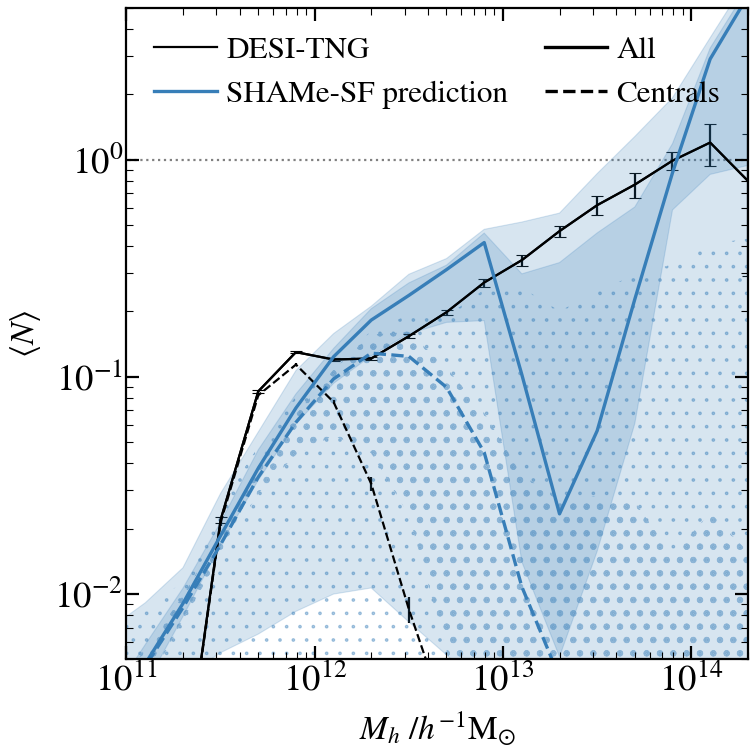}
		\includegraphics[width=0.42\textwidth]{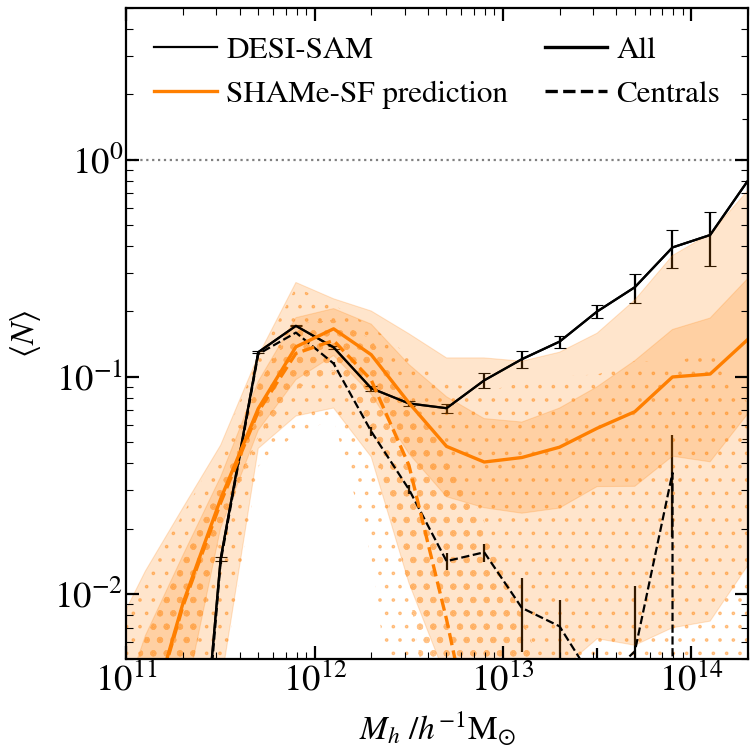}
	\caption{Similar to Figure~\ref{fig:HOD}, inferred halo occupation number for DESI-TNG (top, blue) and DESI-SAM (bottom, orange) mock DESI samples. Measurements from the simulation/SAM are shown in black. The main difference between both samples is the satellite probability.
        }		
  \label{fig:mockHOD}
\end{figure}

The halo occupation numbers of the mock DESI samples and the inferences from the SHAMe-SF model are shown in Figure~\ref{fig:mockHOD}. The reference distributions on DESI-TNG and DESI-SAM are shown with the black error bars with solid lines for all galaxies, and dashed lines for centrals. In this case, we plot the $1\sigma$ and $2\sigma$ regions for SHAMe-SF predictions (shaded region for all galaxies, big and small circle-hatched region for centrals).

SHAMe-SF predicts a broader distribution for central galaxies in DESI-TNG. A similar effect was observed by \cite{Zehavi:2011} and \cite{C24_flamingo}, where the low-mass transition is not fully constrained by galaxy clustering. The bump in the total distribution at high halo masses (dominated by satellite occupancy) is produced by the implementation of the quenching mechanism in SHAMe-SF. We underpredict the number of satellites in haloes with $M_{\rm h} > 10^{13} \, \hMsun$. Of all the satellites in the sample, 16\% are hosted by those haloes, while we predict $4^{+8}{-1}$\% (see Table~\ref{tab:mockdata}). The behaviour of the HOD for satellites is similar to \cite{GonzalezPerez:2018, Hadzhiyska:2021, Yuan:2022b} and \cite{Hadzhiyska:2022_onehalo}. For DESI-SAM, the model delivers accurate predictions for centrals and satellites hosted by haloes with masses below $10^{13} \, \hMsun$, but again underestimates the number of satellites and centrals for higher halo masses. This does not seem to affect the (normalised) predictions of the radial distance and velocity distributions.

Beyond these differences, SHAMe-SF recovers the mean halo masses of centrals for DESI-TNG ($\log( M_{\rm h} \, /\hMsun) =$ \TNG{m200cen} $^{+ \TNG{m200cenup} }_{- \TNG{m200cendown} }$) and TNGSAM ($\log( M_{\rm h}/\hMsun) =$\TNGSAM{m200cen} $^{+ \TNGSAM{m200cenup}}_{- \TNGSAM{m200cendown} }$). For satellites, we recover the average halo mass for DESI-TNG ($\log( M_{\rm h}/\hMsun) =$\TNG{m200sat} $^{+ \TNG{m200satup}}_{- \TNG{m200satdown} }$) but, as expected from the shape of the halo occupation number, we under-predict by 0.2 dex for DESI-SAM ($\log( M_{\rm h}/\hMsun)=$\TNGSAM{m200sat} $^{+ \TNGSAM{m200satup}}_{- \TNGSAM{m200satdown} }$ from SHAMe-SF, $\log( M_{\rm h}/\hMsun)=$\TNGSAMref{m200sat} as measured in \texttt{L-Galaxies}).

\subsubsection{Satellite fractions}
 \label{sec:synsatfrac}
Even if the central/satellite classification depends on the criteria used to define what a satellite is (and this definition does not impact galaxy clustering), it is a necessary check to assess the significance of the rest of the predictions in this section. If the model significantly fails to reproduce the number of satellites, all future claims about their distribution will not be accurate.

We anticipated when discussing the galaxy clustering in Appendix~\ref{sec:synfits} that the satellite fraction would be different between our mock DESI samples given the differences in the one-halo term. \TNGref{sat}\% of the galaxies in the DESI-TNG mock sample are satellites, while we find only \TNGSAMref{sat}\% in DESI-SAM. The satellite fractions of both mock samples and SHAMe-SF's inferences are shown in the top rows of Table~\ref{tab:mockdata}. SHAMe-SF is able to recover the sample fractions within the error bars for DESI-TNG (for both satellite definitions, \texttt{FoF}+\texttt{SUBFIND} and within $\rth$) and within the halo boundary for DESI-SAM. SHAMe-SF underpredicts by 3\% the number of satellites considering objects beyond $\rth$. The effect of this manifests when looking at satellite anisotropy in Appendix~\ref{sec:synanisotropy}. The satellite fractions predicted by other works also show significant deviations, ranging between 4\% \citep{Gonzalez-Perez:2020} and 36\% \citep{Hadzhiyska:2022_onehalo}. 

\begin{figure*}
		\centering
		\includegraphics[width=0.33\textwidth]{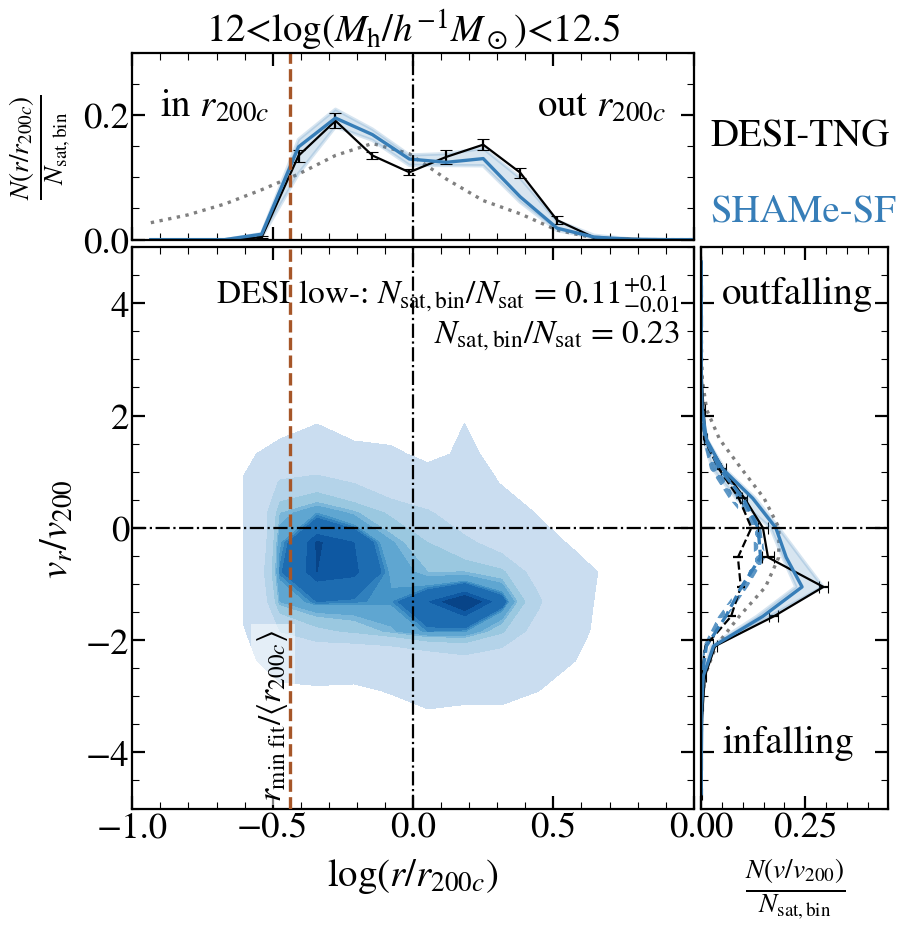}  
        \includegraphics[width=0.33\textwidth]{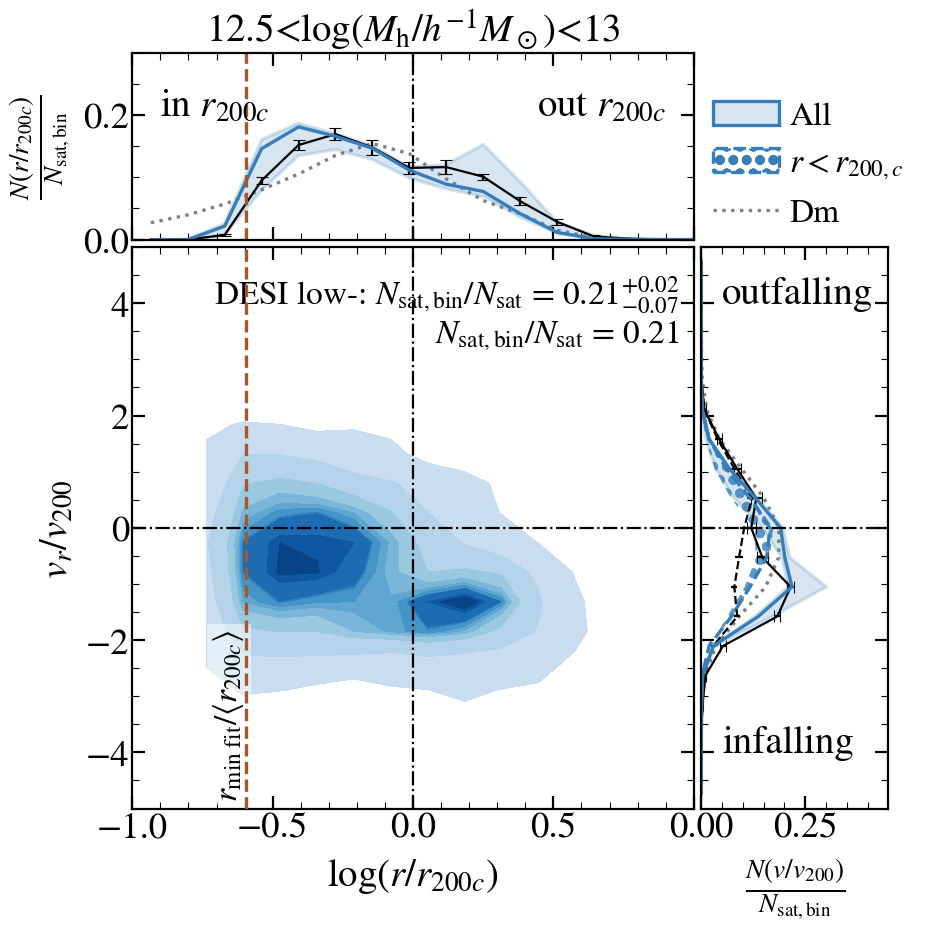}
        \includegraphics[width=0.33\textwidth]{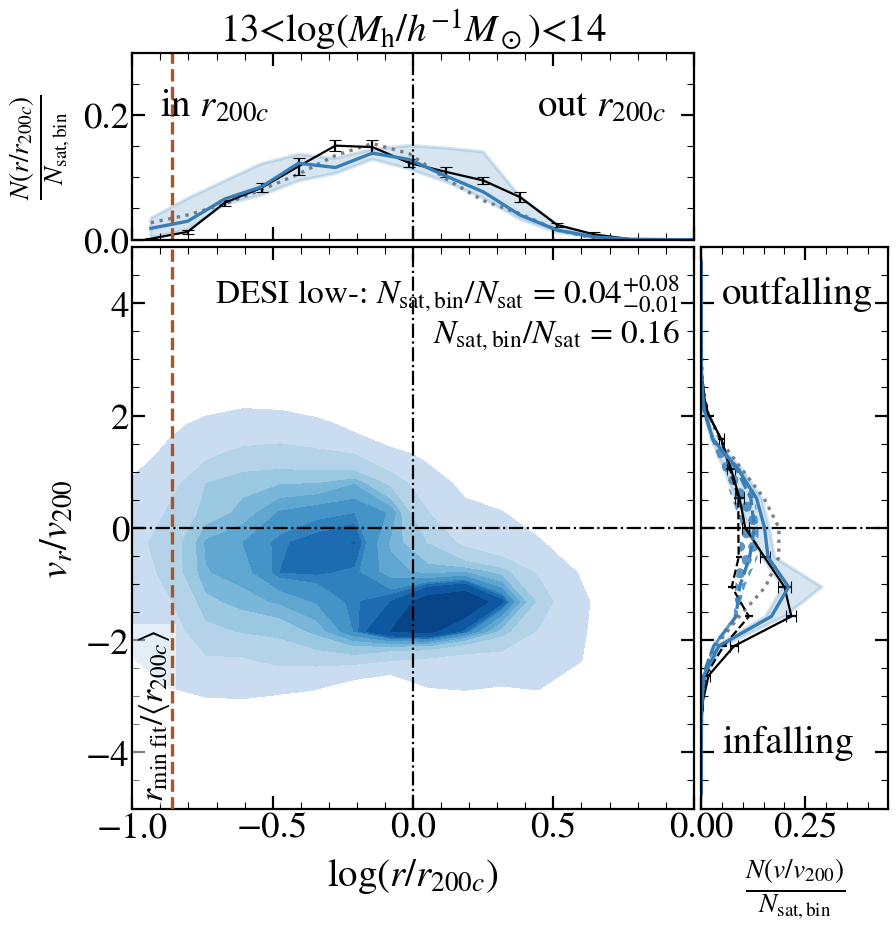}

       \includegraphics[width=0.33\textwidth]{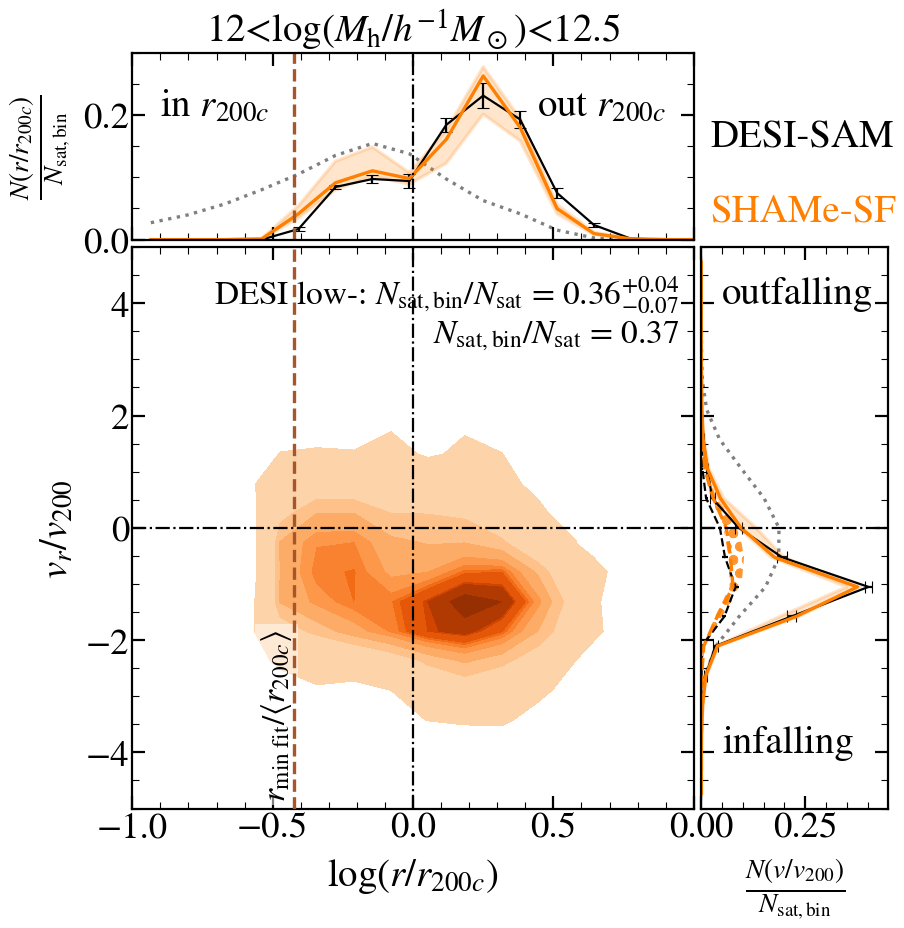} 
        \includegraphics[width=0.33\textwidth]{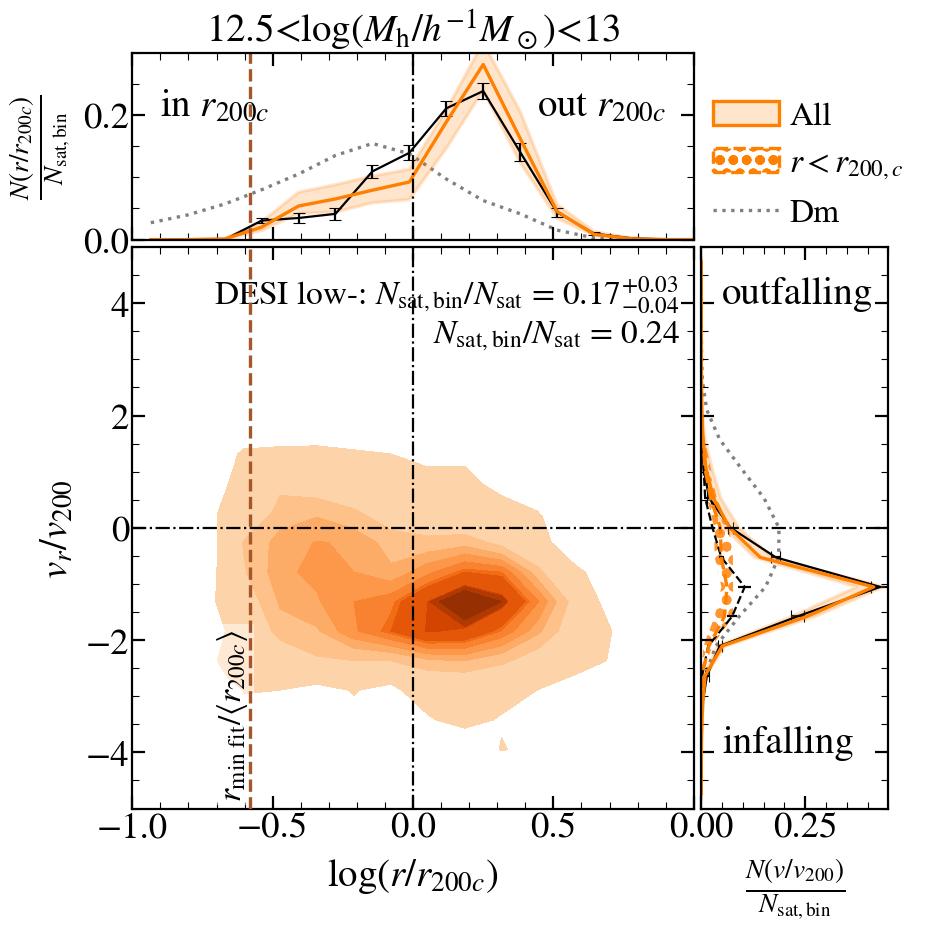} 
        \includegraphics[width=0.33\textwidth]{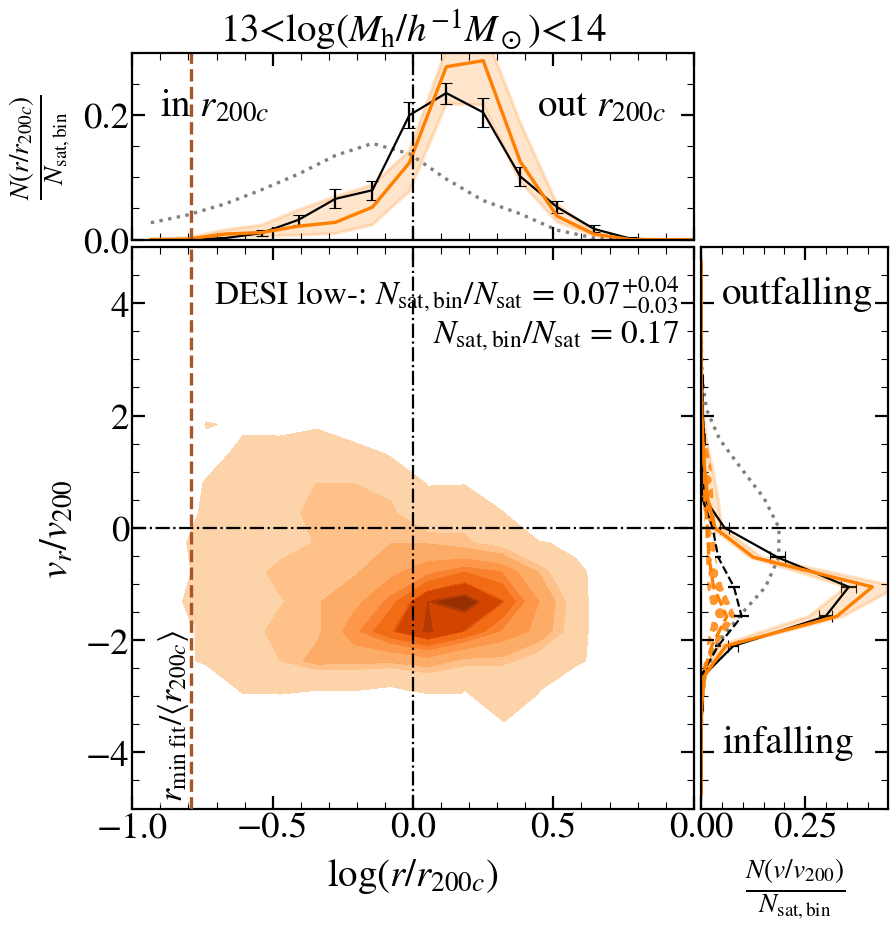}

		\caption{Similar to Figure~\ref{fig:DESIphasespace}, phase space of the central-satellite distance and radial velocity for three bins in halo mass. We only include satellites with $r>r_{\rm{min\,fit}}$. The measurements from the mock DESI catalogues are shown in black in the projected histograms, while the shaded regions mark our posterior predictive distributions. There is no difference between mass bins in DESI-SAM. For the highest mass bin in the DESI-TNG sample, the satellite density profile follows closely the distribution of randomly selected particles. This is not the case for velocities.}		
  \label{fig:mockphasespace}
\end{figure*}

\subsubsection{Satellite phase space distribution}
 \label{sec:synphasespace}
 
Satellite phase space is the main modelling challenge for HODs. Here, we analyse whether the inferences of SHAMe-SF for the mock DESI ELG samples are realistic. In Figure~\ref{fig:mockphasespace} we show the equivalent of Figure~\ref{fig:DESIphasespace} for the same mass bins, adding the comparison with the measurements on DESI-TNG/DESI-SAM on the projected histograms (black lines). We also include the fraction of satellites with $r>r_{\rm{min\,fit}}$ in each mass bin predicted by SHAMe-SF and measured in the simulation/SAM (see also Table~\ref{tab:mockdata}). SHAMe-SF reproduces the central-satellite distance and velocity distributions for the three mass bins, only finding small differences in the velocity distribution inside $\rth$ for DESI-TNG.

Looking at the 2D distribution, we find the same bimodality we pointed out for DESI galaxies (infalling and virialised subhaloes). We highlight that satellites in the high mass bin of TNG300 closely follow the randomly selected dark matter distribution (equivalent to an NFW profile). The behaviour of DESI-SAM satellites is more similar between the different mass bins, where most satellites are infalling and outside $\rth$. This is related to the implementation of the quenching mechanisms in the \texttt{L-Galaxies} version from \cite{Henriques:2015}, where tidal stripping only acts within $\rth$ (for more details on the difference between DESI-TNG and DESI-SAM quenching mechanisms, see \citealt{LGalaxies_Ayromlou:2021}). Overall, we are confident when predicting the radial distributions for DESI ELGs presented in Section~\ref{sec:DESIphasespace}.

We can compare our distance profiles with the distributions from \cite{Yuan:2022b} and \cite{Hadzhiyska:2022_onehalo}. \cite{Yuan:2022b} finds a similar mass dependence for TNG300, with a bimodality inside and outside $\rth$ that vanishes for halo masses larger than $10^{13} \, \hMsun$, where the distribution of ELG satellites coincides with the halo mass distribution. However, in the case of \cite{Hadzhiyska:2022_onehalo}, the bimodality can be found for all halo masses. The distribution of the radial velocities is similar to our DESI-TNG sample, leaning towards negative velocities with a peak associated with the infalling population outside $\rth$. 

\subsubsection{Satellite angular anisotropy}
 \label{sec:synanisotropy}
\begin{figure}
		\centering
		\includegraphics[width=0.45\textwidth]{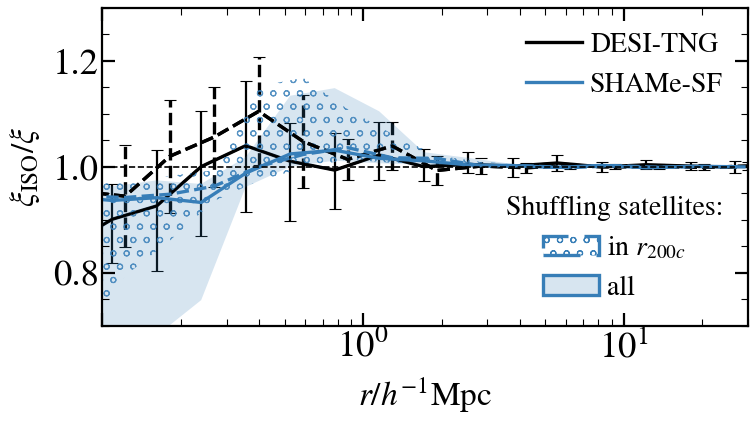}   \includegraphics[width=0.45\textwidth]{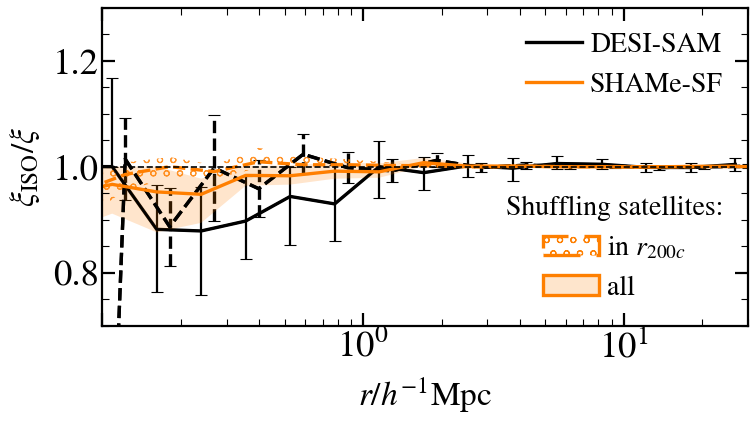}        
		\caption{Quantification of the effect of the angular satellite distribution on the galaxy clustering (similar to Figure~\ref{fig:DESIiso}). We mark the measurements from the mock DESI catalogues using the black lines: solid when all the satellites are shuffled, dashed when only changing the positions of satellites within $\rth$.}		
  \label{fig:mockiso}
\end{figure}

Here we validate our predictions from Section~\ref{sec:DESIanisotropy} when altering the angular distributions of satellites. We follow the same procedure as \cite{Hadzhiyska:2022_onehalo} and keep the central-satellite distance constant while randomly varying the angular positions of satellites within each halo. We show our results in Figure~\ref{fig:mockiso} when shuffling satellites within $\rth$ (dashed line, circle-hatched region) and all satellites (solid line and shaded region). For both shufflings, DESI-TNG presents some scale-dependent effect of anisotropy that does not appear for the DESI-SAM. The signal's shape is similar to \cite{Hadzhiyska:2022_onehalo}'s findings when comparing the isotropic HOD with the right satellite distribution, but we find a smaller amplitude.
The difference in the signal between DESI-TNG and DESI-SAM comes from the implementation of AGN feedback (anisotropic for DESI-TNG, uniform in the semi-analytical model). SHAMe-SF predicts a smaller amplitude for the small-scale anisotropy when shuffling all the satellites in DESI-SAM. This is sourced by the difference in the number of satellites outside $\rth$ measured in the semi-analytical model and predicted by SHAMe-SF. When shuffling satellites within $\rth$, we do not measure any anisotropy. We attribute this to the lack of satellite pairs measured in both the mock DESI sample and the SHAMe-SF model prediction (see Table~\ref{tab:mockdata}).

\subsubsection{Central-satellite conformity}
\label{sec:synconformity}

\begin{figure*}
		\centering
		\includegraphics[width=0.95\textwidth]{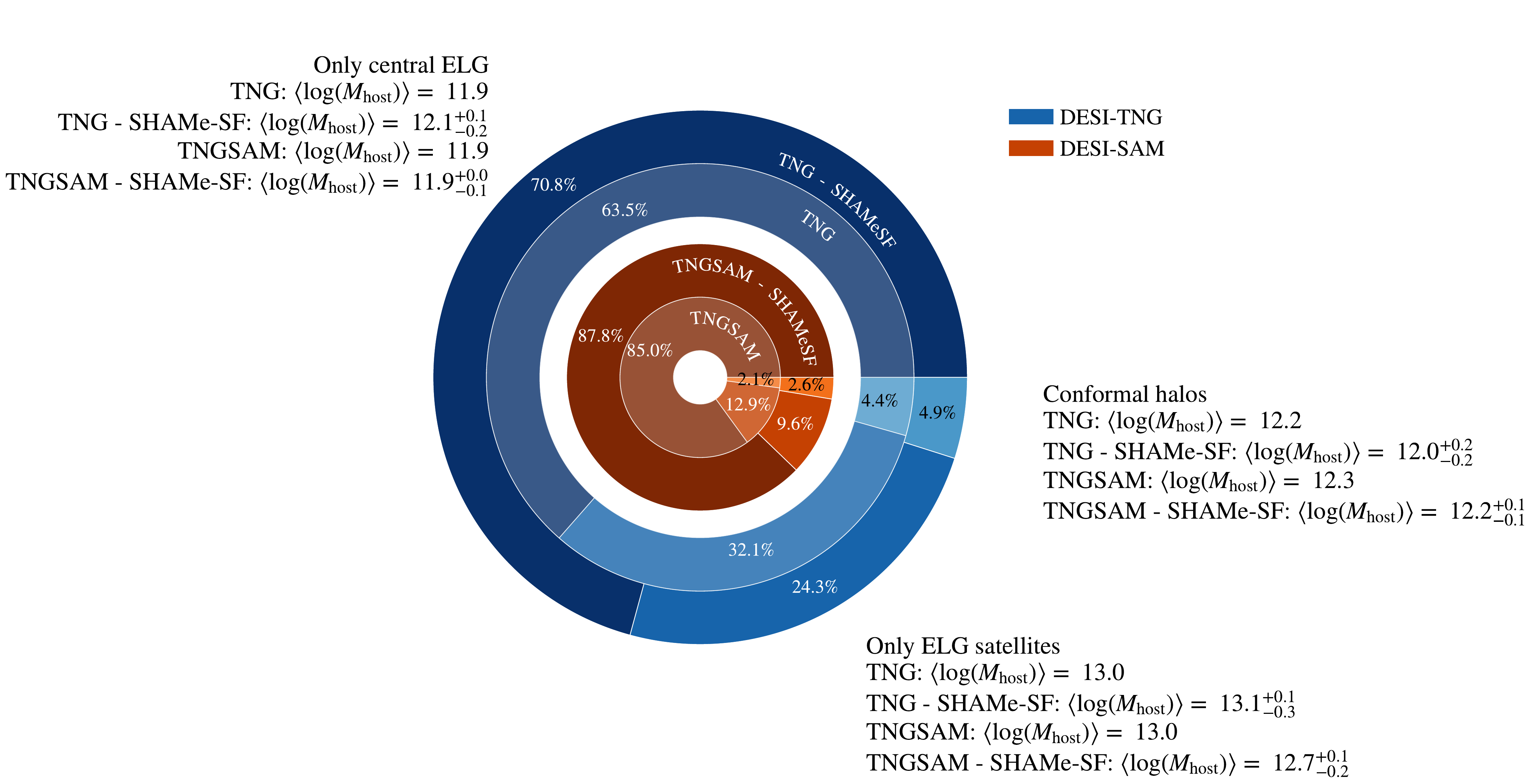}       
		\caption{Comparison between the host haloes of ELGs in the mock DESI samples (dimmer orange and palettes, inner circles for each colour) and the predictions from SHAMe-SF. As in Figure~\ref{fig:DESIconf}, we distinguish between haloes with only a central ELG, haloes with only ELG satellites and haloes with both (conformal), and provide the average halo masses measured and predicted by the model. The uncertainties on the percentage can be found in Table~\ref{tab:mockdata}. Most ELGs are centrals without other ELG satellites. haloes hosting both central and satellite ELGs have lower masses than those only containing satellites.}		
  \label{fig:mockconformity}
\end{figure*}
While discussing central-satellite conformity in Section~\ref{sec:DESIconformity}, we found that only a small percentage of haloes hosted both central and satellite ELGs. 
Even if conformity is not implemented explicitly in SHAMe-SF, it successfully recovers the percentage of conformal haloes for both DESI-TNG (\TNGref{fnoconfcentral}\%) and DESI-SAM (\TNGSAMref{fnoconfcentral}\%) mock DESI-ELG samples. The percentages and average halo masses of haloes hosting only centrals, only satellites and centrals and satellites (conformal) are summarized in Figure~\ref{fig:mockconformity} and Table~\ref{tab:mockdata} for DESI-TNG (blue) and DESI-SAM (orange) mock samples. Both quantities are accurately reproduced by SHAMe-SF.

The average masses of haloes hosting only centrals ($\log( M_{\rm h}/\hMsun) \sim 11.9$), only satellites ($\log( M_{\rm h}/\hMsun) \sim 13$) or both ($\log( M_{\rm h}/\hMsun) \sim 12.2$) are very similar between both mock DESI ELG samples. Thus, the difference in the fraction of conformal haloes is due to the difference in the satellite fraction between the samples and not necessarily to the physical processes that give rise to conformity. 

To elucidate whether the satellite placement was casual or there was a correlation with the type of central, we repeated the shuffling exercise described in Section~\ref{sec:DESIconformity}. With the SHAMe-SF model (reference) we find $\TNG{shuffled_fhalos_conformal}^{+\TNG{shuffled_fhalos_conformalup}}_{-\TNG{shuffled_fhalos_conformaldown}}$\% ($\TNGref{shuffled_fhalos_conformal}^{+\TNGref{shuffled_fhalos_conformalup}}_{-\TNGref{shuffled_fhalos_conformaldown}}$\%)
and $\TNGSAM{shuffled_fhalos_conformal}^{+\TNGSAM{shuffled_fhalos_conformalup}}_{-\TNGSAM{shuffled_fhalos_conformaldown}}$\% ($\TNGSAMref{shuffled_fhalos_conformal}^{+\TNGSAMref{shuffled_fhalos_conformalup}}_{-\TNGSAMref{shuffled_fhalos_conformaldown}}$\%). Thus, we find conformity for the DESI-TNG sample, but we cannot asses whether it is only a statistical effect in DESI-SAM given the discrepancy between our estimation and the measured percentage.

\begin{table*}
\caption{Comparison between the measured and inferred (SHAMe-SF) galaxy-(halo) connection for DESI-TNG and DESI-SAM}
\begin{tabular}{lll|ll}
\toprule 
\textbf{Property} & \textbf{DESI-TNG}& \textbf{SHAMe-SF}& \textbf{DESI-SAM}& \textbf{SHAMe-SF}\\  \midrule \midrule
\textbf{Satellite fractions (subhaloes)} & & & \\
\texttt{FoF}+\texttt{SUBFIND} definition  & \TNGref{sat}\% & \TNG{sat} $^{+ \TNG{satup} }_{- \TNG{satdown} }$\% & \TNGSAMref{sat}\% & \TNGSAM{sat} $^{+ \TNGSAM{satup}}_{- \TNGSAM{satdown} }$\% \\
Satellites only when $r<\rth$ & \TNGref{sat2}\% & \TNG{sat2} $^{+ \TNG{satup2} }_{- \TNG{satdown2} }$\% & \TNGSAMref{sat2}\% & \TNGSAM{sat2} $^{+ \TNGSAM{satup2}}_{- \TNGSAM{satdown2} }$\% \\
 \midrule 
\textbf{Halo fractions hosting:} & & & \\
Only centrals & \TNGref{fnoconfcentral}\% & \TNG{fnoconfcentral} $^{+ \TNG{fnoconfcentralup} }_{- \TNG{fnoconfcentraldown} }$\% & \TNGSAMref{fnoconfcentral}\% & \TNGSAM{fnoconfcentral} $^{+ \TNGSAM{fnoconfcentralup}}_{- \TNGSAM{fnoconfcentraldown} }$\% \\
Only satellites & \TNGref{fhalos_onlysat}\% & \TNG{fhalos_onlysat} $^{+ \TNG{fhalos_onlysatup} }_{- \TNG{fhalos_onlysatdown} }$ \%& \TNGSAMref{fhalos_onlysat}\% & \TNGSAM{fhalos_onlysat} $^{+ \TNGSAM{fhalos_onlysatup}}_{- \TNGSAM{fhalos_onlysatdown} }$ \%\\
Centrals and satellites & \TNGref{fhalos_conformal}\% & \TNG{fhalos_conformal} $^{+ \TNG{fhalos_conformalup} }_{- \TNG{fhalos_conformaldown} }$\% & \TNGSAMref{fhalos_conformal} \%& \TNGSAM{fhalos_conformal} $^{+ \TNGSAM{fhalos_conformalup}}_{- \TNGSAM{fhalos_conformaldown} }$\% \\
Centrals and satellites (sat. shuffling) & $\TNGref{shuffled_fhalos_conformal}^{+\TNGref{shuffled_fhalos_conformalup}}_{-\TNGref{shuffled_fhalos_conformaldown}}$\% & $\TNG{shuffled_fhalos_conformal}^{+\TNG{shuffled_fhalos_conformalup}}_{-\TNG{shuffled_fhalos_conformaldown}}$\% & $\TNGSAMref{shuffled_fhalos_conformal}^{+\TNGSAMref{shuffled_fhalos_conformalup}}_{-\TNGSAMref{shuffled_fhalos_conformaldown}}$\% & $\TNGSAM{shuffled_fhalos_conformal}^{+\TNGSAM{shuffled_fhalos_conformalup}}_{-\TNGSAM{shuffled_fhalos_conformaldown}}$\% \\
\\
More than one satellite & \TNGref{fhalos_multisat}\% & \TNG{fhalos_multisat} $^{+ \TNG{fhalos_multisatup} }_{- \TNG{fhalos_multisatdown} }$ \%& \TNGSAMref{fhalos_multisat}\% & \TNGSAM{fhalos_multisat} $^{+ \TNGSAM{fhalos_multisatup}}_{- \TNGSAM{fhalos_multisatdown} }$\% \\

More than one satellite  ($r<\rth$)& \TNGref{fhalos_multisat2}\% & \TNG{fhalos_multisat2} $^{+ \TNG{fhalos_multisatup2} }_{- \TNG{fhalos_multisatdown2} }$ \%& \TNGSAMref{fhalos_multisat2}\% & \TNGSAM{fhalos_multisat2} $^{+ \TNGSAM{fhalos_multisatup2}}_{- \TNGSAM{fhalos_multisatdown2} }$\% \\

 \midrule
\textbf{Average halo masses, $\langle \log(M_{h,\rm{i}}/\hMsun) \rangle$}: & & & \\
Hosting centrals  & \TNGref{m200cen} & \TNG{m200cen} $^{+ \TNG{m200cenup} }_{- \TNG{m200cendown} }$ & \TNGSAMref{m200cen} & \TNGSAM{m200cen} $^{+ \TNGSAM{m200cenup}}_{- \TNGSAM{m200cendown} }$ \\
Hosting satellites  & \TNGref{m200sat} & \TNG{m200sat} $^{+ \TNG{m200satup} }_{- \TNG{m200satdown} }$ & \TNGSAMref{m200sat} & \TNGSAM{m200sat} $^{+ \TNGSAM{m200satup}}_{- \TNGSAM{m200satdown} }$ \\
\\
Hosting only centrals  & \TNGref{m200cennosat} & \TNG{m200cennosat} $^{+ \TNG{m200cennosatup} }_{- \TNG{m200cennosatdown} }$ & \TNGSAMref{m200cennosat} & \TNGSAM{m200cennosat} $^{+ \TNGSAM{m200cennosatup}}_{- \TNGSAM{m200cennosatdown} }$ \\
Hosting only satellites  & \TNGref{m200noconsat} & \TNG{m200noconsat} $^{+ \TNG{m200noconsatup} }_{- \TNG{m200noconsatdown} }$ & \TNGSAMref{m200noconsat} & \TNGSAM{m200noconsat} $^{+ \TNGSAM{m200noconsatup}}_{- \TNGSAM{m200noconsatdown} }$ \\
Hosting centrals and satellites  & \TNGref{m200cenwithsat} & \TNG{m200cenwithsat} $^{+ \TNG{m200cenwithsatup} }_{- \TNG{m200cenwithsatdown} }$ & \TNGSAMref{m200cenwithsat} & \TNGSAM{m200cenwithsat} $^{+ \TNGSAM{m200cenwithsatup}}_{- \TNGSAM{m200cenwithsatdown} }$ 
\\ \midrule
\multicolumn{3}{l}{\textbf{Satellite distribution in host halo mass bins (100\% = all satellites)}} \\
$r<r_{\rm{min \, fit}}$  & \TNGref{fsat_inr200}\% & \TNG{fsat_inr200} $^{+ \TNG{fsat_inr200up} }_{- \TNG{fsat_inr200down} }$\% & \TNGSAMref{fsat_inr200}\% & \TNGSAM{fsat_inr200} $^{+ \TNGSAM{fsat_inr200up}}_{- \TNGSAM{fsat_inr200down} }$\% \\

$r>r_{\rm{min \, fit}}$, $M_{h}/\hMsun < 10^{12}$  & \TNGref{fsat_under12}\% & \TNG{fsat_under12} $^{+ \TNG{fsat_under12up} }_{- \TNG{fsat_under12down} }$\% & \TNGSAMref{fsat_under12}\% & \TNGSAM{fsat_under12} $^{+ \TNGSAM{fsat_under12up}}_{- \TNGSAM{fsat_under12down} }$ \%\\

$r>r_{\rm{min \, fit}}$, $10^{12}<M_{h}/\hMsun < 10^{12.5}$  & \TNGref{fsat_inm12m12p5}\% & \TNG{fsat_inm12m12p5} $^{+ \TNG{fsat_inm12m12p5up} }_{- \TNG{fsat_inm12m12p5down} }$\% & \TNGSAMref{fsat_inm12m12p5}\% & \TNGSAM{fsat_inm12m12p5} $^{+ \TNGSAM{fsat_inm12m12p5up}}_{- \TNGSAM{fsat_inm12m12p5down} }$\% \\

$r>r_{\rm{min \, fit}}$, $10^{12.5}<M_{h}/\hMsun < 10^{13}$  & \TNGref{fsat_inm12p5m13} \%& \TNG{fsat_inm12p5m13} $^{+ \TNG{fsat_inm12p5m13up} }_{- \TNG{fsat_inm12p5m13down} }$ \%& \TNGSAMref{fsat_inm12p5m13}\% & \TNGSAM{fsat_inm12p5m13} $^{+ \TNGSAM{fsat_inm12p5m13up}}_{- \TNGSAM{fsat_inm12p5m13down} }$ \%\\

$r>r_{\rm{min \, fit}}$, $10^{13}<M_{h}/\hMsun < 10^{14}$  & \TNGref{fsat_inm13m14} \%& \TNG{fsat_inm13m14} $^{+ \TNG{fsat_inm13m14up} }_{- \TNG{fsat_inm13m14down} }$\% & \TNGSAMref{fsat_inm13m14}\% & \TNGSAM{fsat_inm13m14}\% $^{+ \TNGSAM{fsat_inm13m14up}}_{- \TNGSAM{fsat_inm13m14down} }$\% \\

 \bottomrule

\end{tabular}
\label{tab:mockdata}
\end{table*}

\FloatBarrier
\section{DESI posteriors}
\subsection{Radial phase space for the higher mass bin}
\label{app:DESIhighmassphase}
\begin{figure}
		\centering

        \includegraphics[width=0.45\textwidth]{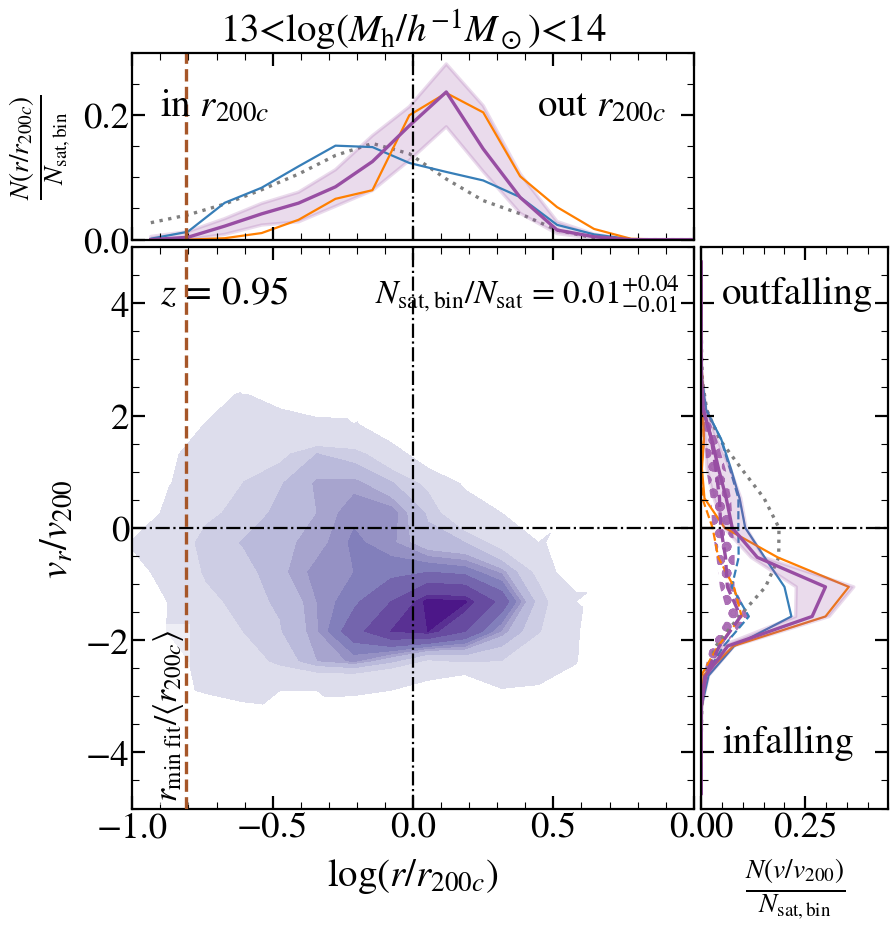}   
       \includegraphics[width=0.45\textwidth]{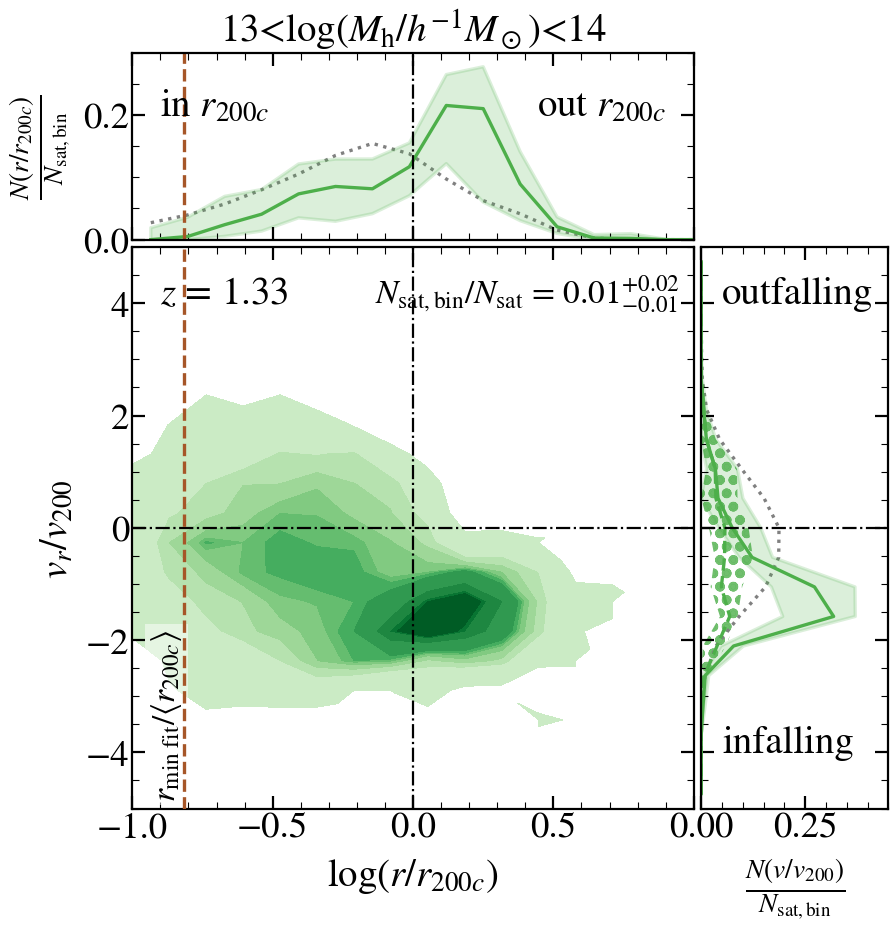}
		\caption{Same as Figure~\ref{fig:DESIphasespace}, for satellites on halos with $M_{\rm{h}}\in [10^
{13},10^{14}]\,\hMsun$. For these halo masses, the infalling population dominates for both redshift bins.
        }
  	
    \label{fig:DESIphasespacehighmass}
\end{figure}
In Figure~\ref{fig:DESIphasespacehighmass} we show the central-satellite distance and the radial velocities for the highest mass bin in halo mass, $M_{\rm{h}}\in [10^
{13},10^{14}]\,\hMsun$. The SHAMe-SF model predicts that only 1\% of all satellites are in this mass bin, fewer than in the DESI-TNG and DESI-SAM mock samples. As for the other mass bins, we distinguish two populations: a virialised population (inside and outside $\rth$) with a mostly infalling distribution of velocities and an infalling population outside $\rth$. In this mass bin, the latter dominates, similar to the case of DESI-SAM and unlike DESI-TNG. Since the radial position distribution is dominated by the infalling term, we cannot assess whether the virialised component follows the dark matter, as in the case of DESI-TNG.
\subsection{All posteriors}
\label{app:posteriors}
\begin{figure*}
		\centering
		\includegraphics[width=0.95\textwidth]{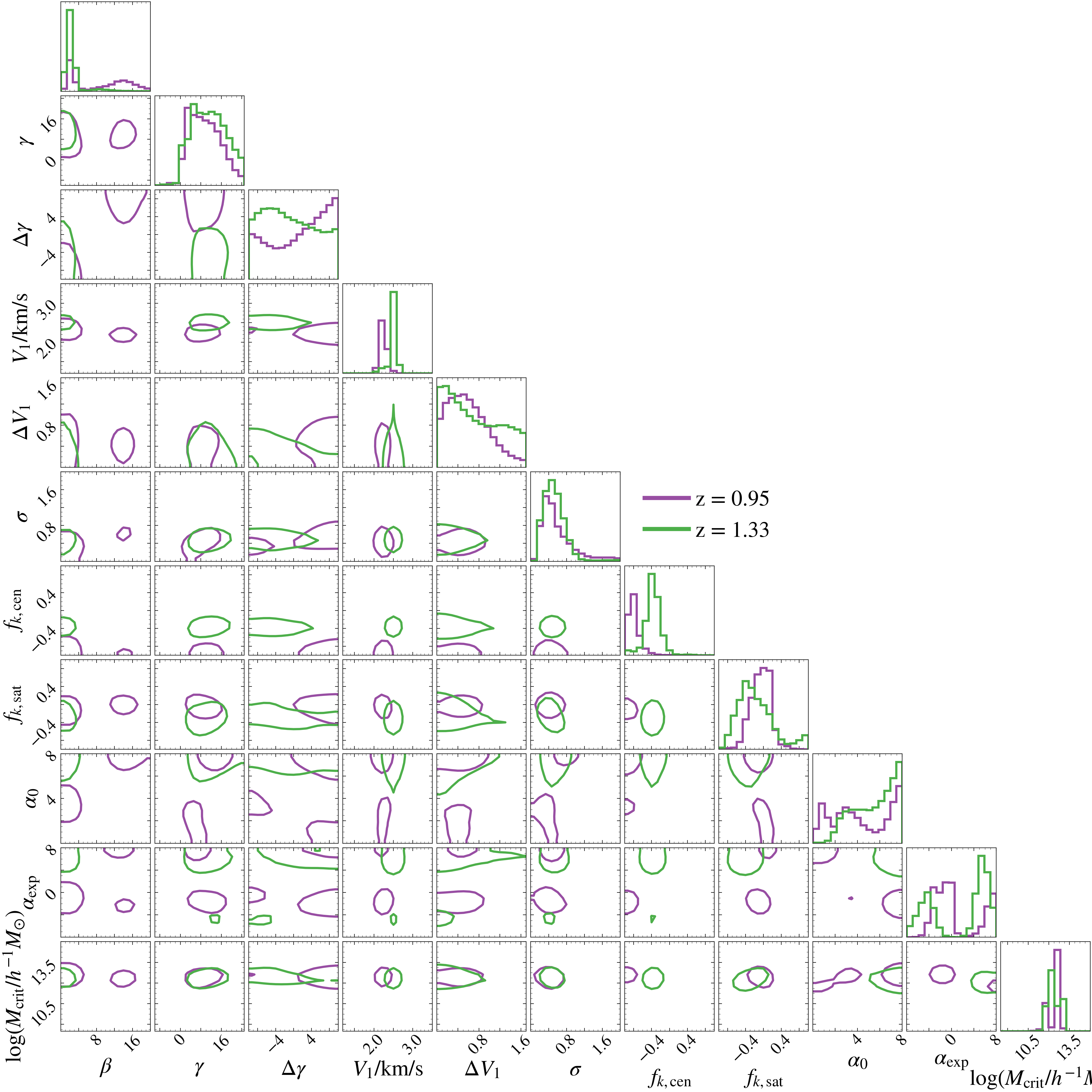}    
		\caption{Marginalised $1\sigma$ constraints all SHAMe-SF parameters, obtained from fitting the clustering of DESI-ELGs at $z=0.95$ (purple) and $1.33$ (green).}		
        \label{fig:allparams}
\end{figure*}
We discussed in Section~\ref{sec:DESIclus} the meaning of constraining some of the SHAMe-SF parameters when fitting the clustering. The full posterior distributions are shown in Figure~\ref{fig:allparams}. We use the same ranges for the parameters on the emulator training and the priors.
$\beta,\gamma,\Delta\gamma,\alpha_0$ and $\alpha_{\rm{exp}}$ control the amplitude of the score used to rank-order the subhalos before applying the cut in number density. Fixing the value of one of these parameters can be reabsorbed by the others to produce the same clustering signal. This is also true for larger values of $\sigma$. Values of $\sigma$ very close to zero would imply that the secondary dependence on concentration (parametrised by both $f_{k}$) is not important, which seems to be discarded by the posterior.

\end{appendix}

\end{document}